\documentclass[a4paper,11pt]{article}

\pdfoutput=1

\usepackage{jheppub}
\usepackage{amsmath}
\usepackage{xspace}
\usepackage{hyperref}
\usepackage{mdwlist}

\newcommand{\eq}[1]{eq.~\eqref{eq:#1}}

\renewcommand{\sec}[1]{sec.~\ref{sec:#1}}

\newcommand{\app}[1]{app.~\ref{app:#1}}
\newcommand{\fig}[1]{fig.~\ref{fig:#1}}

\newcommand{\nnu}{\nonumber\\}
\newcommand{\nn}{\nonumber}
\newcommand{\bef}{\begin{figure}[t]\centering}
\newcommand{\eef}{\end{figure}}
\def\bea#1\eea{\begin{align}#1\end{align}}
\def \be  {\begin{equation}}
\def \ee  {\end{equation}}
\def \ba  {\begin{eqnarray}}
\def \ea  {\end{eqnarray}}

\newcommand{\f}{\frac}
\newcommand{\ord}[1]{\mathcal{O}(#1)}

\newcommand{\df}{\mathrm{d}}

\newcommand{\Li}{\textrm{Li}}

\newcommand{\sdt}{\!\cdot\!}

\newcommand{\al}{\alpha}
\newcommand{\bt}{\beta}
\newcommand{\ga}{\gamma}

\newcommand{\de}{\delta}

\newcommand{\si}{\sigma}
\newcommand{\zc}{z_{\rm cut}}
\newcommand{\tg}{\theta_g}
\newcommand{\kp}{k_\perp}

\newcommand{\cG}{{\mathcal G}}
\newcommand{\cL}{{\mathcal L}}

\newcommand{\CS}{{\mathcal S}}

\newcommand{\bn}{\bar{n}}

\newcommand\as{\alpha_s}
\newcommand{\lqcd}{\Lambda_\mathrm{QCD}}

\newcommand{\Pythia}{\textsc{Pythia}\xspace}

\allowdisplaybreaks[2]

\title{Calculating the angle between jet axes}

\author[a,b]{Pedro Cal,}
\author[c]{Duff Neill,}
\author[d,e]{Felix Ringer,}
\author[a,b]{Wouter J. Waalewijn}

\affiliation[a]{Institute for Theoretical Physics Amsterdam and Delta Institute for Theoretical Physics, University of Amsterdam, Science Park 904, 1098 XH Amsterdam, The Netherlands}
\affiliation[b]{Nikhef, Theory Group, Science Park 105, 1098 XG, Amsterdam, The Netherlands}
\affiliation[c]{Theoretical Division, MS B283, Los Alamos National Laboratory, Los Alamos, NM 87545, USA}
\affiliation[d]{Physics Department, University of California, Berkeley, CA 94720, USA}
\affiliation[e]{Nuclear Science Division, Lawrence Berkeley National Laboratory, Berkeley, CA 94720, USA}
                                         
\emailAdd{p.cal@nikhef.nl}
\emailAdd{duff.neill@gmail.com}
\emailAdd{fmringer@berkeley.edu}
\emailAdd{w.j.waalewijn@uva.nl}

\abstract{We study the angle between i) the standard jet axis, ii) the axis of a jet which has been groomed using soft drop, with reduced sensitivity to soft radiation, iii) the jet axis obtained with the winner-take-all recombination scheme, which is insensitive to soft radiation at leading power. We calculate the distributions for these angles at next-to-leading logarithmic accuracy, including non-global logarithms. The angle between the standard and groomed jet axis directly probes soft wide-angle radiation, leading to a novel factorization formula. This angle is also very sensitive to nonperturbative physics, which is directly connected to nonperturbative contribution to the rapidity anomalous dimension for transverse momentum distributions. Comparing our predictions to \Pythia we find good agreement, and we foresee applications to jet substructure in proton-proton and heavy ion collisions. 
}

\preprint{NIKHEF 2019-052}

\begin{document}
\maketitle

\section{Introduction}
\label{sec:intro}

Jet substructure techniques have become important tools at the Large Hadron Collider (LHC) to study the properties of Quantum Chromodynamics (QCD) and extract the strong coupling $\al_s$, search for physics Beyond the Standard Model, and probe the quark-gluon plasma (QGP) in heavy-ion collisions. Both experimental and theoretical advancements have been crucial to enable precise comparisons between first principles calculations in QCD and LHC data. For recent reviews of jet substructure techniques and their applications, see refs.~\cite{Larkoski:2017jix,Asquith:2018igt}. 

\begin{figure}[t]
\centering
 \includegraphics[width=0.3\textwidth]{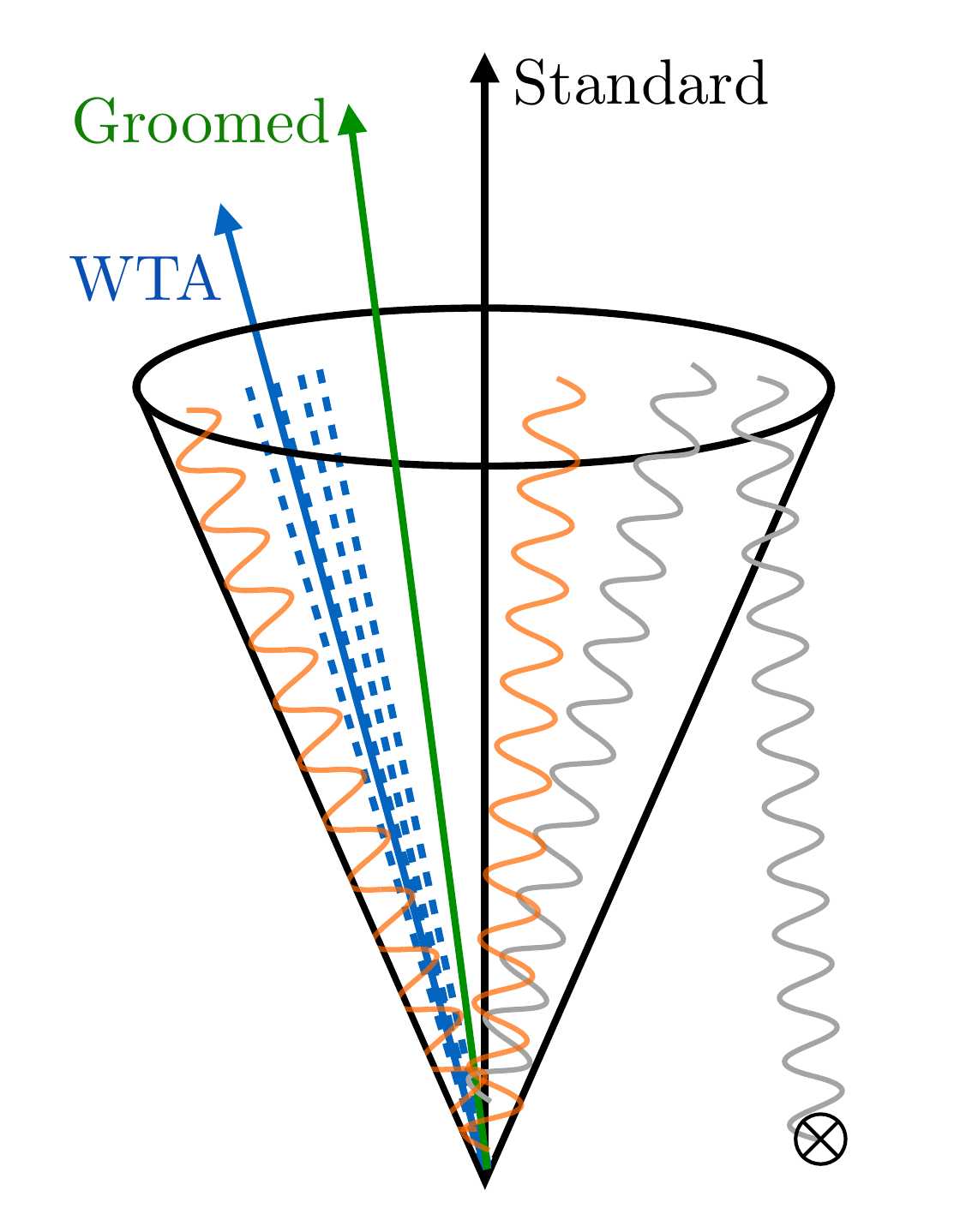}  
  \caption{The various jet axes: The standard jet axis (black) is along the total momentum of the collinear (blue) and soft (orange) radiation. For the groomed jet axis (green) groomed away soft radiation (gray) is not included. The winner-take-all axis (blue) is insensitive to soft radiation.~\label{fig:Angles-fig}} 
\end{figure}

In this paper we study three different jet axes, with varying degrees of sensitivity to soft radiation, which we now introduce:
We start from an inclusive sample of jets, obtained by using the anti-k$_T$ algorithm~\cite{Cacciari:2008gp} with jet radius parameter $R=0.8$, and refer to the resulting jet axis as the standard (ST) jet axis. Only jets with  rapidity $|\eta|<2$ are considered, and results for several bins in the jet transverse momentum $p_T$ will be reported. Applying soft drop~\cite{Larkoski:2014wba} with $\zc = 0.1$ and several values for $\beta$ to remove soft radiation, the axis of the resulting groomed jet will be called the groomed (GR) jet axis. 
Finally, we recluster the original jet using the winner-take-all (WTA) recombination scheme~\cite{Salam:WTAUnpublished,Bertolini:2013iqa}, for which the effect of soft radiation is power suppressed, yielding the WTA axis. To ensure that all radiation is inside the jet (i.e.~that the jet algorithm returns a single jet) the jet radius is increased. These various axes are pictured in~\fig{Angles-fig}, where the offset between the ST (black) and GR (green) axis is due to groomed away radiation (gray). The WTA axis (blue) clearly tracks the energetic collinear (blue) radiation. 

We will consider the distance $\theta = \sqrt{\Delta \eta^2 + \Delta \phi^2}$ in pseudorapidity $\eta$ and azimuthal angle $\phi$ between these axes, denoted with $\theta_{\rm ST,GR}$, $\theta_{\rm ST,WTA}$ and $\theta_{\rm GR,WTA}$, respectively. This distance is boost invariant and for a jet at zero rapidity, i.e.~perpendicular to the beam axis,  $\theta$ is equal to the angle between these axes (in the small angle limit, which we focus on). We therefore refer to this as an angle. These angles characterize the different sensitivity to soft radiation of these axes. In our calculations it will be convenient to express $\theta$ as a vector $\vec k_\perp$ transverse to the \emph{jet} axis, with $\theta = |\vec k_\perp|/p_T$, since vectors can easily be added. To simplify the notation, we will omit the vector symbol when referring to the norm of a vector, i.e.~$k_\perp \equiv |\vec k_\perp|$.

Given this identification between angles and transverse momenta, it should come as no surprise that all three angles between the jet axes considered here involve some form of transverse momentum resummation~\cite{Collins:1984kg,Ji:2004wu,Becher:2010tm,Collins:2011zzd,GarciaEchevarria:2011rb,Chiu:2012ir}. As such, it provides another window into the physics of transverse momentum dependent fragmentation functions and parton distribution functions, focusing on the bulk distribution of transverse momentum within jets relative to energetic collections of particles within the jet, not the energy spectrum of the underlying individual particles, similar to refs.~\cite{Moult:2018jzp,Gutierrez-Reyes:2018qez,Gao:2019ojf,Cal:2019hjc,Gutierrez-Reyes:2019vbx,Larkoski:2019fsm,Bao:2019usu,Larkoski:2019urm}.

We derive the factorization in the small-$\theta$ limit within Soft Collinear Effective Theory (SCET)~\cite{Bauer:2000ew, Bauer:2000yr, Bauer:2001ct, Bauer:2001yt,Beneke:2002ph} and carry out the resummation at next-to-leading logarithmic (NLL or NLL$'$\footnote{For NLL$'$, we simply mean the inclusion of the full matrix element of each function truncated at $O(\alpha_s)$ convolved with resummation kernel, including NGLs in the kernel when their exact structure is known. In ref.~\cite{Balsiger:2019tne}, a distinct definition of NLL$'$ was adopted, where one includes all matrix elements at the order where the logarithms generating the resummation first appear. For NGLs, this requires an $O(\alpha_s^2)$ contribution (see ref.~\cite{Becher:2016omr}), which is beyond our definition of NLL$'$. For further developments in the theory for NGLs, see refs.~\cite{Banfi:2002hw,Weigert:2003mm,Hagiwara:2015bia,Caron-Huot:2015bja,Larkoski:2015zka,Becher:2016mmh,Larkoski:2016zzc}}.) accuracy, including the contribution of non-global logarithms~\cite{Dasgupta:2001sh} in the leading-color approximation. In order to derive the factorization structure for the angle between the standard and groomed jet axis $\theta_{\rm ST,GR}$, we need to simultaneously consider the soft drop groomed jet radius $R_g$~\cite{Larkoski:2014wba,Kang:2019prh}. We perform the joint resummation of logarithms of the transverse momentum and the soft drop groomed jet radius $R_g$, and afterwards integrate over $R_g$. While for the other  observables we achieve NLL$'$ with the resummation of the non-global logs, the factorization for the angle $\theta_{\rm ST,GR}$ is considerably more complicated, limiting our accuracy to NLL in the global logarithms.
 
There are several potential applications of the observables introduced in this work: First of all, by comparing parton shower event generators to our perturbative calculations and to data from the LHC, models of nonperturbative physics can be constrained, similar to e.g.~refs.~\cite{ATLAS:2011gmi,Hoeche:2017jsi} where jet substructure results were used to constrain the hadronization and underlying event contribution. This is particularly true for the angle between the standard and groomed jet axis, which is very sensitive to nonperturbative physics as it measures the soft radiation which \emph{fails} the soft drop criterion. Interestingly, for the observables considered here the (leading) nonperturbative component is relatively well understood as it is related to the nonperturbative part of the rapidity anomalous dimension in transverse momentum resummation. Indeed, various extractions are available in the literature~\cite{Landry:2002ix,Su:2014wpa,Bacchetta:2017gcc,Bertone:2019nxa}, and it was recently proposed that this evolution kernel can also be extracted from lattice QCD~\cite{Ebert:2018gzl}.  Explicitly, our analysis predicts how the grooming parameters affect the nonperturbative part of the rapidity anomalous dimension on our observables, which can be compared to hadronization models in Monte Carlo parton showers.
Moreover, we expect that the observables considered here will have important applications in heavy-ion collisions where jets are used as probes of the created hot and dense QCD medium~\cite{Andrews:2018jcm}. For example, the measurement of the angle between the groomed and WTA axis provides a handle on the soft radiation contained in identified jets in the large background produced in heavy-ion collisions. Since the different jet axes considered in this work exhibit a very different sensitivity to soft physics, these observables are ideally suited to quantitatively explore the effects of parton energy loss and jet broadening. 

The remainder of this paper is organized as follows. In~\sec{fac}, we introduce the different factorization and resummation formulae in physical terms, which are necessary for the calculation of the angles between the three jet axes. We present all relevant expressions and technical details for the three cases in~\sec{ST_WTA} to~\sec{ST_GR}. In~\sec{implement}, we discuss different aspects of the implementation in impact parameter space and describe our nonperturbative model (which we take from transverse momentum resummation). Numerical results for LHC kinematics are presented in~\sec{results}, and in~\sec{conclusions} we draw conclusions and present an outlook.

\section{Factorization and resummation of angles between axes}
\label{sec:fac}

In this section we describe in physical terms the structure of the cross sections that encode the angles between the different jet axes described in the Introduction. The necessary factorization and resummation techniques are introduced but technical details and expressions of the various functions that appear in these factorization formulae are relegated to subsequent sections.

\subsection{Jet production}
\label{sec:fac_jet}

As a first step, common to the calculation of all these angles, we factorize the hard scattering from the jet production, by working in the narrow jet approximation, $R\ll 1$. In several examples it has been observed that the $\ord{R^2}$ power corrections to this limit are still small for values of $R$ up to 0.7~\cite{Mukherjee:2012uz,Dasgupta:2014yra}. The cross section differential in jet transverse momentum $p_T$, (pseudo)rapidity $\eta$, and  $\vec k_\perp$ (describing the angle between axes) is given by
\begin{align} \label{eq:hard_fact}
  \frac{\df \si}{\df \eta\, \df p_T\, \df^2 \vec k_\perp}
  &= \sum_{i,j,k} 
  \int\! \frac{\df x_i}{x_i}\, f_i(x_i,\mu) \int\! \frac{\df x_j}{x_j}\, f_j(x_j,\mu)
  \int\! \frac{\df z}{z}\, {\cal H}_{ij}^k(x_i, x_j, \eta, p_T/z, \mu)
  \nn \\ & \quad\times
   \cG_k(z, k_\perp, p_T R, \mu) \big[1+\mathcal{O}(R^2)\big]
\,.\end{align}
Here, the parton distribution functions (PDFs) $f_{i,j}$ describe extracting a parton with flavor $i, j$  and momentum fraction $x_{i,j}$ from a colliding proton. The hard function ${\cal H}_{ij}^k$~\cite{Aversa:1988vb,Jager:2002xm,Mukherjee:2012uz}  encodes the hard scattering of the incoming partons $i$ and $j$, and the distribution of the resulting partons with flavor $k$, rapidity $\eta$ and transverse momentum $p_T/z$. The jet function $\cG_k$ describes collinear final-state radiation, i.e.~how parton $k$ (inclusively) produces jets, with transverse momentum $z \times p_T/z = p_T$, as well as the angle between the axes of the jet, encoded in $k_\perp$. We will indicate which axes we are considering by including the appropriate superscript on the cross section $\sigma$ and the jet function $\cG_k$.

In the rest of this paper we will focus on the jet, and therefore denote the parton that initiates it with $i$, rather than $k$. We note that \eq{hard_fact} is simply a more differential version of the factorization for inclusive jet production~\cite{Kaufmann:2015hma,Kang:2016mcy,Dai:2016hzf}, since integrating our jet function over $\vec k_\perp$ yields the semi-inclusive jet function $J_i$ of ref.~\cite{Kang:2016mcy} 
\begin{align}
     \int \! \df^2 \vec k_\perp\, \cG_i(z, k_\perp, p_T R, \mu) = J_i(z,p_T R,\mu)
\,.\end{align}
  We will exploit this sum rule to factor out the jet production from the jet measurement~\cite{Kaufmann:2015hma,Cal:2019hjc}, writing:
\begin{align} \label{eq:reshuffle}
     \cG_i(z, k_\perp, p_T R, \mu) &\equiv  \sum_j J_{ij}(z, p_T R, \mu) \tilde \cG_j\bigl(k_\perp, p_T R, \al_s(\mu)\bigr) .
\end{align}
The coefficient $J_{ij}$ is akin to a jet flavor-tagged fragmentation function: it describes how a parton of flavor $i$ fragments into a jet of radius $R$ of flavor $j$. Note that this is not a factorization of physics at different scales, and requires keeping track of the jet flavor $j$. As indicated, the new jet function $\tilde \cG$ only depends on $\mu$ through the strong coupling. These functions satisfy
\begin{align}
  \int \! \df^2 \vec k_\perp\,\tilde \cG_j\bigl(k_\perp, p_T R, \al_s(\mu)\bigr)=1, \qquad
  \sum_j J_{ij}(z, p_T R, \mu) =  J_i(z,p_T R,\mu)
\,,\end{align}  
which are sufficient to determine $J_{ij}$ and $\tilde \cG_j$  from the full expressions for $\cG_i$ and $J_i$. To the order we calculate, we further introduce
\begin{align}\label{eq:Delta_iG}
     \tilde \cG_i\bigl(k_\perp, p_T R, \al_s(\mu)\bigr) = \frac{1}{\pi}\,\de(k_\perp^2) +  \de(1-z) \Delta \cG_i(k_\perp, p_T R, \mu)
\,,\end{align}
reporting only on the (simpler) $\Delta \cG_i$. This equation defines the function $\Delta \cG_i$. Note that we do not need any flavor indicies on the $\delta(k_{\perp}^2)$ term. 

The natural scales of the ingredients in \eq{hard_fact} are the same as for inclusive jet production
\begin{align}
    \mu_f &\sim \lqcd
    \,, &
    \mu_{\cal H} &\sim p_T
    \,, &
    \mu_{\cal G} &\sim p_T R
\,,\end{align}
and the resummation of logarithms of $R = \mu_\cG/\mu_{\cal H}$ is achieved by using the DGLAP~\cite{Gribov:1972ri,Altarelli:1977zs,Dokshitzer:1977sg} equation
\begin{align} \label{eq:G_evo}
\mu\f{\df}{\df\mu}\,\cG_i(z, k_\perp, p_T R , \mu)=\sum_j\int_z^1\f{\df z'}{z'}\,\f{\as}{\pi} P_{ji}(z/z') \,\cG_j(z', k_\perp, p_T R , \mu)
\,,\end{align}
to evolve the jet function $\cG$ from the jet scale $\mu_\cG$ to the hard scale $\mu_{\cal H}$. The splitting functions $P_{ji}$ are at one-loop order given in \eq{split}.

If $k_\perp \sim p_T R$, there are no additional logarithms to resum as $\cG$ involves a single scale. In this case one simply has to calculate $\Delta \cG$ for the angle between each pair of axes. However, these distributions are in fact dominated by $k_\perp \ll p_T R$, in which case $\cG$ contains two hierarchical scales and needs to be factorized. We will outline this below for each of the angles, relegating much of the details to subsequent sections.  To keep the discussion concise, we have focused on the mode structure contributing to each factorization formula we present, which dictates its form. For an introduction to factorization within the SCET formalism, we refer the reader to refs.~\cite{Stewart:notes,Becher:2014oda}. For specific examples in the literature on the technical derivation of the factorized cross section for jet observables from the underlying modes of the effective theory, see e.g.~refs.~\cite{Fleming:2007xt,Bauer:2008dt,Stewart:2009yx,Ellis:2010rwa,Becher:2011pf,Chiu:2012ir}.

\subsection{Angle between the standard and WTA axis}
\label{sec:fac_ST_WTA}

\begin{figure}
        \hfill \includegraphics[width=0.49\textwidth]{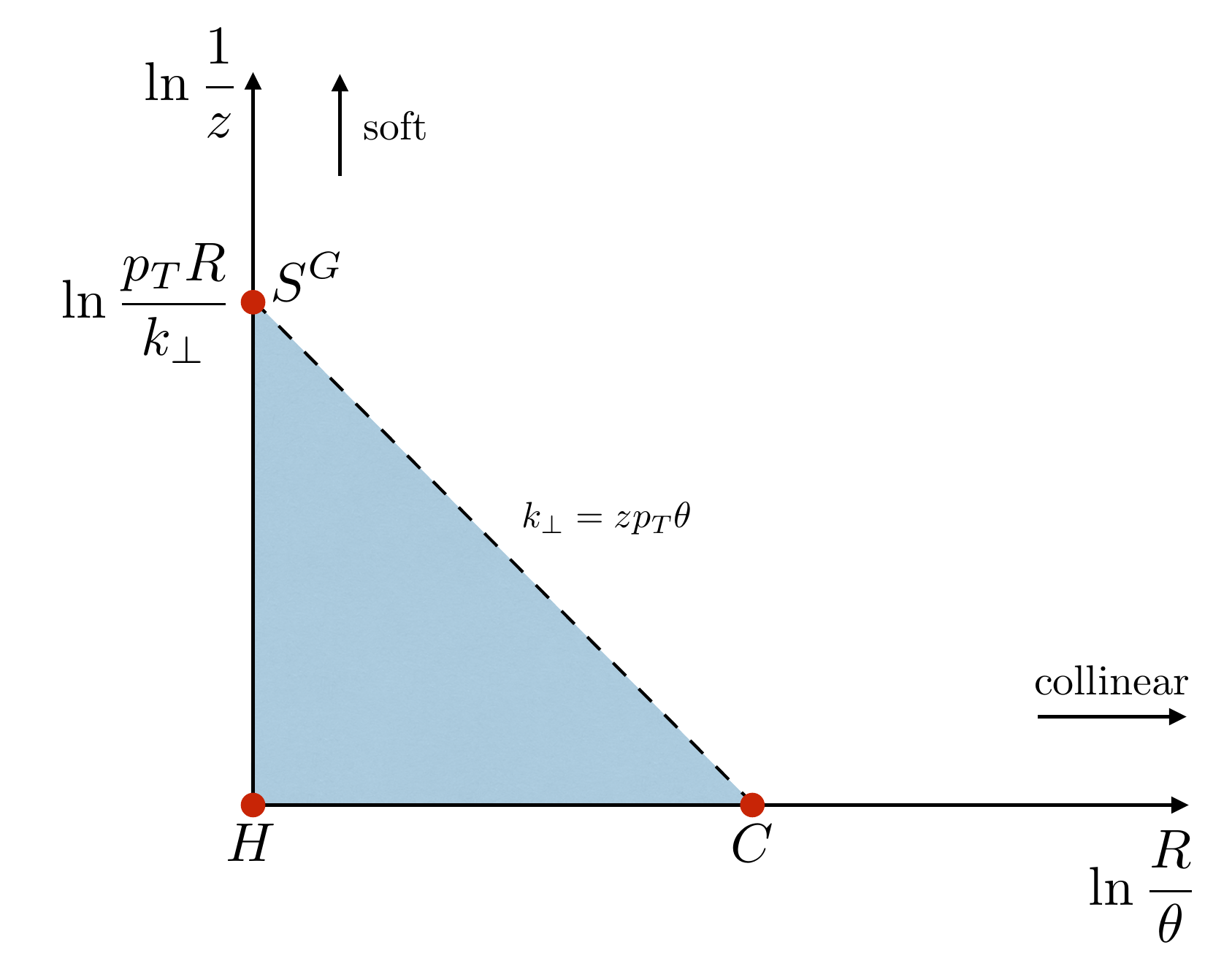} \hfill
         \includegraphics[width=0.49\textwidth]{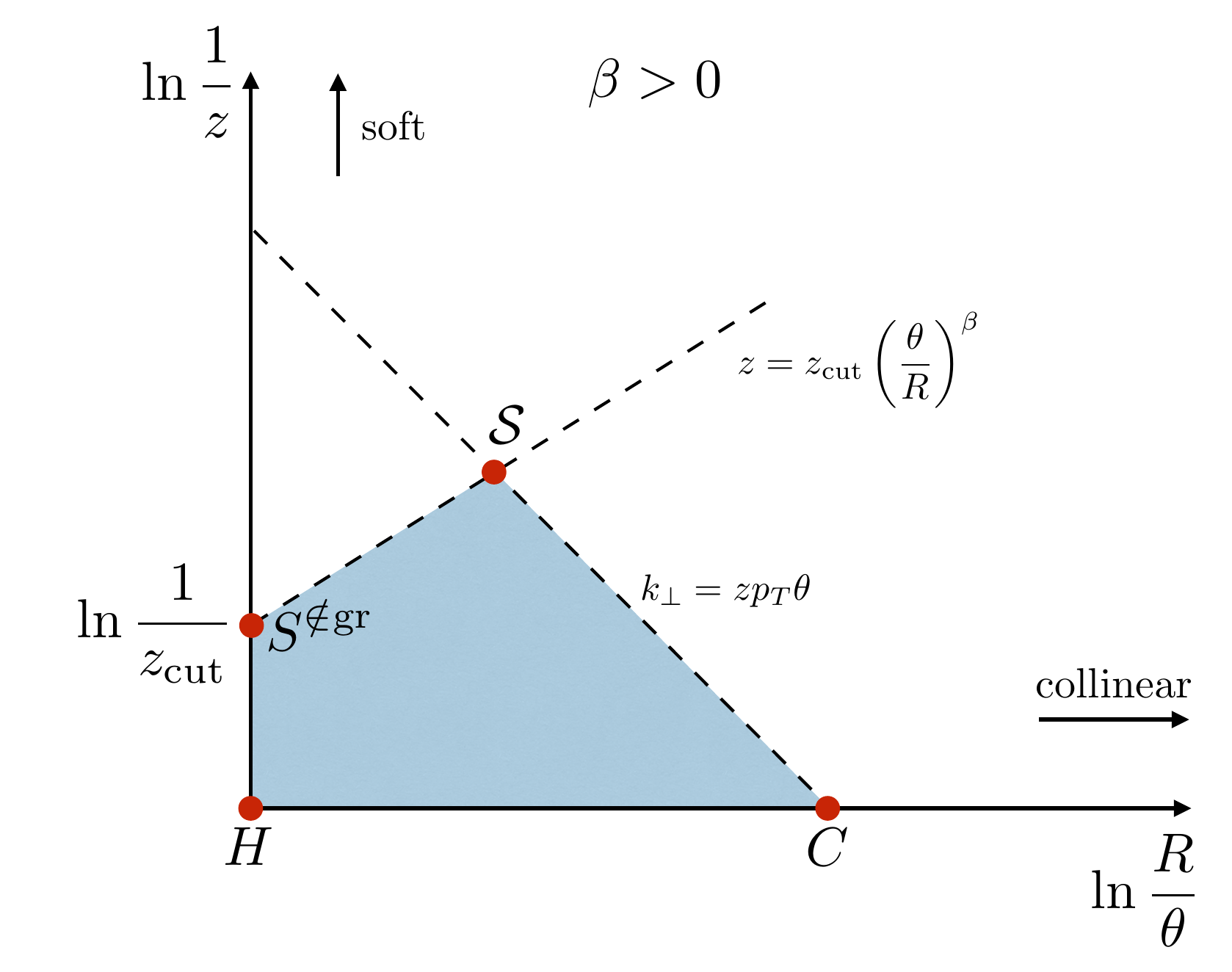} \hfill \phantom{.}
\caption{Lund diagrams for the angle between the standard and the WTA axis (left) and the groomed and the WTA axis (right). The modes that appear in the factorization for these two observables are indicated with red dots, whose parametric scaling can be read off from this diagram.~\label{fig:Lund1and2}}
\end{figure}

We start with a leading logarithmic (LL) analysis of the angle between the standard and WTA axes using the Lund diagram~\cite{Andersson:1988gp} shown in the left panel of \fig{Lund1and2}. The horizontal axis corresponds to the angle $\theta$ and the vertical axis to the momentum fraction $z$ of emissions in the jet. By using the indicated logarithmic coordinates, emissions have a uniform emission probability at LL accuracy, allowing one to directly determine the Sudakov factor from the area of the region where emissions are not allowed. In this LL picture the WTA axis is along the single hard parton in the jet, and the offset of the ST axis is due to the transverse momentum $k_\perp$ of the dominant emission. Thus cutting on the transverse momentum $k_\perp < k_\perp^c$ prohibits emissions in the red region. This leads to the following expression for the jet function $\tilde  \cG$ that describes the angle between the standard and WTA axes, 
\begin{align}
 \tilde \cG_{i}^{\rm ST, WTA}\bigl(k_\perp, p_T R, \al_s(\mu)\bigr) &\stackrel{{\rm LL}}{=} \frac{1}{2\pi\, k_\perp}\, \frac{\df}{\df k_\perp} 
 \exp\bigg[- \frac{\al_s C_i}{\pi} 
 \ln^2 \Big(\frac{p_T R}{k_\perp}\Big) \bigg]
\,.\end{align}
This is differential in $\vec k_\perp$, and the color factor $C_i = C_F$ ($C_A$) for $i=q$ ($i=g$).

Extending this analysis beyond LL using SCET, the corners of this region correspond to the degrees of freedom (or modes) in SCET, as encoded in the factorization of  $\tilde \cG$ in the limit $k_\perp \ll p_T R$: 
\begin{align} \label{eq:refact_ST_WTA}
\tilde \cG_i^{\rm ST, WTA}\bigl(k_\perp, p_T R, \al_s(\mu)\bigr) &\stackrel{{\rm NLL}'}{=} \tilde H_i(p_{T} R,\mu)\,
\int\! \df^2 \vec k_\perp' 
 C_i(k_\perp',\mu,\nu) 
  \int\! \df^2 \vec k_\perp''\, S_i^{\rm G}(\vec k_\perp \!-\! \vec k_\perp' \!-\! \vec k_\perp'' ,\mu, \nu R)
\nn\\ & \qquad \times
 S_i^{\rm NG}\Big(\frac{k_\perp''}{p_T R}\Big)
\bigg[1+ \mathcal{O}\bigg(\frac{k_\perp^2}{p_T^2R^2} \bigg)\bigg]\, .
\end{align}
The WTA axis is only sensitive to collinear radiation inside the jet, and the argument $k_\perp'$ of the collinear function $C_i$ simply encodes the angle of the WTA axis with respect to the initial collinear parton. 
The ST axis is along the total jet momentum, so the collinear and soft radiation inside the jet must balance each other, $\vec k_{\perp,c} + \vec k_{\perp,s} = 0$ (see e.g.~fig.~2 of ref.~\cite{Cal:2019hjc}). Consequently, the offset of the initial collinear parton with respect to the standard axis is given by $k_{\perp,s}$, which is the argument of the soft function, and $\vec k_\perp = \vec k_\perp' + \vec k_{\perp,s}$. 
Because only soft radiation inside the jet contributes to $k_{\perp,s}$, and radiation outside the jet is completely unconstrained, non-global logarithms~\cite{Dasgupta:2001sh} appear. The function $S_i^{\rm NG}$ encodes the leading non-global logarithms, which in the small $R$ limit are the same as for the hemisphere soft function in $e^+e^-$~\cite{Banfi:2010pa}. This is the reason why \eq{refact_ST_WTA} is limited to NLL$'$ accuracy. The hard function $\tilde H_i$ in \eq{refact_ST_WTA} describes collinear radiation with energies $p_T$ at angles of order $R$. Such emissions are not allowed inside the jet, since they would displace the WTA axis but not the ST axis, leading to $k_\perp \sim p_T R$. Following \eq{reshuffle}, we have removed the contribution from jet production, i.e.~this is the factorization for $\tilde \cG$ not $\cG$. 

 \begin{table}
   \centering
   \begin{tabular}{l|c}
     \hline \hline
     Mode: & Scaling $(\bn \sdt p ,n \sdt p ,p_\perp)$  \\ \hline
     hard & $ p_T(1,R^2,R)$ \\
     collinear & $ (p_T, k_\perp^{\,2}/p_T,  k_\perp)$ \\
     soft & $ k_\perp/R (1, R^2, R)$ \\ 
     \hline \hline
   \end{tabular}
   \caption{For the angle between the ST and WTA axis, the modes that enter in the factorization of the jet function $\cG$ are listed above.}
   \label{tab:modes_ST_WTA}
   \end{table} 
 
The power counting of the modes that underpin the factorization in \eq{refact_ST_WTA} are summarized in table~\ref{tab:modes_ST_WTA}, in terms of light-cone coordinates 
\begin{align}
  p^\mu = \bn \sdt p\, \frac{n^\mu}{2} + n \sdt p\, \frac{\bn^\mu}{2} + p_\perp^\mu
\,,\end{align}
where $n^\mu = (1,0,0,1)$ is along the jet axis, $\bn^\mu = (1,0,0,-1)$, and $p_\perp^\mu$ denotes the transverse components.
The corresponding scales are\footnote{In our numerical implementation we choose scales in terms of the impact parameter $b_\perp$, instead of its Fourier conjugate to $k_\perp$, to avoid a well-known problem in transverse momentum resummation~\cite{Frixione:1998dw}. See \sec{implement} for details.}
\begin{align} \label{eq:can_ST_WTA}
    \mu_{H} & = p_T R\,, &
    \mu_C &= k_\perp\,, &
    \mu_S &= k_\perp\,, 
    \nn \\   & &
    \nu_C &=2 p_T\,, &
    \nu_S &= 2 k_\perp/R
\,,\end{align}
where $\mu$ corresponds to invariant mass scales and $\nu$ to rapidity scales. By evaluating each of the ingredients at their natural scale and evolving them to a common one using the RG equations
\begin{align} \label{eq:RGE}
\mu\f{\df}{\df\mu}\,\tilde H_i(p_T R,\mu) &= \gamma^{\tilde H}_i(p_T R,\mu)\,\tilde H_i(p_T R,\mu)
\,, \nn \\
\mu\f{\df}{\df\mu}\,C_i(k_\perp,\mu,\nu/p_T) &= \gamma^C_i(\mu, \nu / p_T)\,C_i(k_\perp,\mu,\nu/p_T)
\,, \nn \\
\mu\f{\df}{\df\mu}\,S_i^{\rm G}(k_\perp, \mu,\nu R) &= \gamma^S_i(\mu, \nu R)\,S_i^{\rm G}(k_\perp,\mu,\nu R)
\,, \nn \\
\nu\f{\df}{\df\nu}\,C_i(k_\perp,\mu,\nu/p_T) &= -\int\! \frac{\df^2 \vec k_\perp'}{(2\pi)^2}\,
\gamma^\nu_i(\vec k_\perp-\vec k_\perp',\mu)\,C_i(k_\perp',\mu,\nu/p_T)
\,, \nn \\
\nu\f{\df}{\df\nu}\,S_i^{\rm G}(k_\perp,\mu,\nu R) &= 
\int\! \frac{\df^2 \vec k_\perp'}{(2\pi)^2}\,
\gamma^\nu_i(\vec k_\perp-\vec k_\perp',\mu)\,S_i^{\rm G}(k_\perp',\mu,\nu R)
\,, \end{align}
we resum the global logarithms of $k_\perp/(p_T R)$. The anomalous dimensions are collected in \app{anom}.

\subsection{Angle between the groomed and WTA axis}
\label{sec:fac_GR_WTA}

  \begin{table}
   \centering
   \begin{tabular}{l|c}
     \hline \hline
     Mode: & Scaling $(\bn \sdt p ,n \sdt p ,p_\perp)$  \\ \hline
     hard & $ p_T(1,R^2,R)$ \\
     soft & $\zc p_T (1, R^2, R)$ \\ 
     collinear & $ (p_T, k_\perp^{\,2}/p_T,  k_\perp)$ \\
     collinear-soft & $ (a=\zc k_\perp^\bt p_T^{1-\bt} R^{-\bt}, k_\perp^2/a,  k_\perp)$ \\
     \hline \hline
   \end{tabular}
   \caption{For the angle between the GR and WTA axis, the modes that enter in the factorization of the jet function $\cG$ when $k_\perp \ll \zc p_T R$ are listed above.}
   \label{tab:modes_GR_WTA}
   \end{table}

We start by briefly reviewing the soft drop grooming procedure~\cite{Larkoski:2014wba}. First, the identified jets are reclustered with the Cambridge/Aachen (C/A) algorithm~\cite{Dokshitzer:1997in,Wobisch:1998wt}. The C/A clustering metric only depends on the geometric distance between particles in the $\eta$-$\phi$ plane and yields a jet with an angular ordered clustering tree. Second, the obtained jets are declustered recursively where at each step of the declustering procedure, the soft drop criterion is checked
\begin{equation}
\frac{\min \left[p_{T 1}, p_{T 2}\right]}{p_{T 1}+p_{T 2}}>z_{\mathrm{cut}}\left(\frac{\Delta R_{12}}{R}\right)^{\beta} \,.
\end{equation}
Soft branches that fail this criterion are removed from the jet. Here, the $p_{T,i}$ denote the transverse momenta of the two branches and $\Delta R_{12}$ denotes their geometric distance in the $\eta$-$\phi$ plane. The soft threshold parameter $z_{\rm cut}$ and the angular exponent $\beta$ are free parameters that determine how aggressively the grooming algorithm removes soft radiation from the jet. For $\beta=0$, the soft drop criterion reduces to the modified mass drop tagger (mMDT) of ref.~\cite{Dasgupta:2013ihk}. The groomed radius $R_g=\Delta R_{12}$ is determined by the geometric distance between the two branches that satisfy the soft drop criterion, after which the soft drop procedure terminates. See refs.~\cite{Aaboud:2017qwh,Sirunyan:2017bsd,Kauder:2017cvz,Sirunyan:2018gct,Sirunyan:2018xdh,Acharya:2019djg,ATLAS:2019sol} for experimental measurements of groomed jet substructure observables. Related recent theoretical calculations of groomed jet substructure observables can be found in refs.~\cite{Frye:2016aiz,Salam:2016yht,Marzani:2017mva,Larkoski:2017cqq,Larkoski:2017iuy,Baron:2018nfz,Kang:2018vgn,Makris:2018npl,Kang:2018jwa,Kardos:2018kth,Chay:2018pvp,Napoletano:2018ohv,Lee:2019lge,Hoang:2019ceu,Mehtar-Tani:2019rrk}.

For the angle between the groomed and WTA axis, we will again start with a LL analysis, using the Lund planes shown in \fig{Lund1and2}. Soft drop removes wide-angle soft radiation in a jet, so the dominant emission that sets $k_\perp$ must now pass the soft-drop grooming condition, $z > \zc (\theta/R)^\bt$, shown in the right panel. This leads to 
\begin{align}
 \tilde \cG_{i}^{\rm GR, WTA}\bigl(k_\perp, p_T R, \zc, \beta, \al_s(\mu)\bigr) &\stackrel{{\rm LL}}{=} \frac{1}{2\pi\, k_\perp}\,\frac{\df}{\df k_\perp} \bigg\{
 \exp\bigg\{- \frac{\al_s C_i}{\pi} 
 \bigg[\ln^2 \Big(\frac{p_T R}{k_\perp}\Big) 
 \nn \\ & \quad
 - \Theta(\zc p_T R-k_\perp) \frac{1}{1+\bt} \ln^2 \Big(\frac{\zc p_T R}{k_\perp}\Big)\bigg] \bigg\}
\,.\end{align}
Note that if $k_\perp > \zc p_T R$ the soft-drop grooming condition is always satisfied and the Lund diagram is in fact the same as for the ungroomed case, shown in the left panel of  \fig{Lund1and2}. 

Extending our analysis beyond LL accuracy, if $k_\perp \gg \zc p_T R$, the transverse momentum of the radiation that is removed by grooming is negligible compared to the measured $k_\perp$, and the formula for the cross section is the same as in \eq{refact_ST_WTA}. We will therefore focus on the case $k_\perp \ll \zc p_T R$, and interpolate between these regimes in our numerical results\footnote{One could also separately consider the regime $k_\perp \sim \zc p_T R$, constraining this interpolation.}. The modes in SCET correspond to the red dots at the corners of the blue region in the Lund plane and are summarized in table~\ref{tab:modes_GR_WTA}.  The factorization formula for $\tilde \cG$ is given by
\begin{align} \label{eq:refact_GR_WTA}
\tilde \cG_i^{\rm GR, WTA}\bigl(k_\perp, p_T R, \zc,\bt, \al_s(\mu)\bigr) &\stackrel{{\rm NLL}'}{=} \tilde H_i(p_{T} R,\mu)\, S_i^{\notin {\rm gr}}(\zc p_T R, \beta, \mu) 
  \\ & \quad \times
 \int\! \df^2 \vec k_\perp' 
 C_i(k_\perp',\mu,\nu/p_T) 
 \CS_i(\vec k_\perp - \vec k_\perp' , p_T R, \zc, \bt, \mu, \nu/p_T)
 \nn \\ & \quad \times
 S_i^{\rm NG}(\zc)
\bigg[1+ \mathcal{O}\bigg(\frac{k_\perp^2}{\zc^2 p_T^2R^2}, \zc, \Big(\frac{\zc k_\perp^\bt}{(p_T R)^\bt}\Big)^{1/(1+\bt)} \bigg)\bigg]\, .
\nn\end{align}
The hard and collinear radiation are unaffected by the grooming and the corresponding functions are thus the same as in \eq{refact_ST_WTA}. The collinear-soft function $\CS_i$ describes the contribution from collinear-soft radiation that \emph{passes} the grooming condition to the angle between the jet axes. The soft function $S_i^{\notin {\rm gr}}$ encodes the wide-angle soft radiation, which is always groomed away, and therefore does not affect the shape of the $k_\perp$ distribution but only the total rate. In this case non-global logarithms arise because the hard mode and soft mode have the same angular scale, but the soft radiation inside the jet must be much less energetic in order to fail the grooming condition. Fortunately, the leading NGLs are again described by the same function as in \eq{refact_ST_WTA}, but now with $\zc$ as argument. 

By evaluating each function in \eq{refact_GR_WTA} at its natural scale
\begin{align} \label{eq:can_GR_WTA}
    \mu_H & = p_T R\,, &
    \mu_{S^{\notin {\rm gr}}} &= \zc p_T R\,, &
    \mu_C &= k_\perp\,, &
    \mu_{\CS} &= k_\perp\,, 
    \nn \\   & & & &
    \nu_C &= 2 p_T\,, &
    \nu_{\CS} &= 2 p_T \zc^{\frac{1}{1+\bt}} \Big(\frac{k_\perp}{p_TR}\Big)^{\f{\bt}{1+\bt}} 
\,,\end{align}
and evolving them to a common $\mu$ and $\nu$ scale, the global logarithms of $k_\perp/(p_T R)$ and $\zc$ are resummed. The renormalization group equations are partially the same as in \eq{RGE}, except for:
\begin{align} 
\mu\f{\df}{\df\mu}\,S_i^{\notin {\rm gr}}(\zc p_T R, \beta, \mu) &=  \gamma^{S^{\notin {\rm gr}}}_i(\zc p_T R, \bt, \mu) \,S_i^{\notin {\rm gr}}(\zc p_T R, \beta, \mu)
\,, \\
\mu\f{\df}{\df\mu}\,\CS_i(k_\perp, p_T R, \zc, \bt,\mu,\nu/p_T) &= \gamma^{\CS}_i(p_T R, \zc, \bt, \mu, \nu/p_T)\,\CS_i(k_\perp,p_T R, \zc, \bt,\mu,\nu/p_T)
\,, \nn \\
\nu\f{\df}{\df\nu}\,\CS_i(k_\perp,p_T R, \zc, \bt,\mu,\nu/p_T) &= 
\int\! \frac{\df^2 \vec k_\perp'}{(2\pi)^2}\,
\gamma^\nu_i(\vec k_\perp-\vec k_\perp',\mu)\,\CS_i(k_\perp',p_T R, \zc, \bt,\mu,\nu/p_T)
\,. \nn \end{align}

\subsection{Angle between the standard and groomed axis}
\label{sec:fac_ST_GR}

The factorization structure for the angle between the standard and groomed jet axis is rather different. Since  collinear radiation is never groomed away, this observable is highly sensitive to soft radiation, similar to collinear drop~\cite{Chien:2019osu}. The effect of grooming is power suppressed when $k_\perp \gg \zc p_T R$, so the cross section is power suppressed in this region. We therefore focus on the opposite limit $k_\perp \ll \zc p_T R$, smoothly matching the cross section to 0 when crossing $k_\perp \sim \zc p_T R$. It turns out that we must consider the cross section differential in $k_\perp$ and the groomed jet radius $R_g$, since there are two different cases that need to be considered, and then integrate the combined resummed result over $R_g$. It is  convenient to work in terms of $\theta_g \equiv R_g/R$. 

\begin{figure}
        \hfill \includegraphics[width=0.49\textwidth]{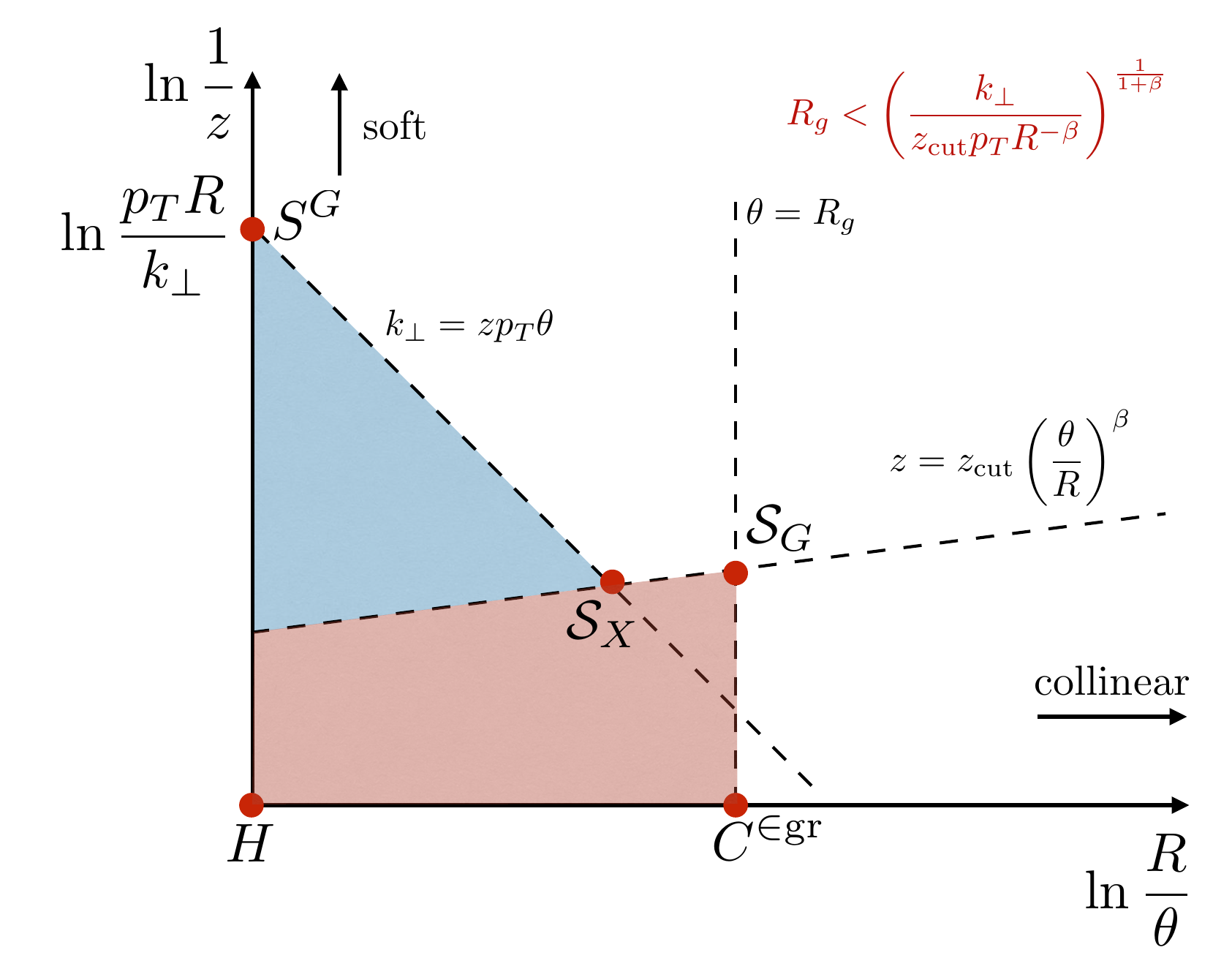} \hfill
         \includegraphics[width=0.49\textwidth]{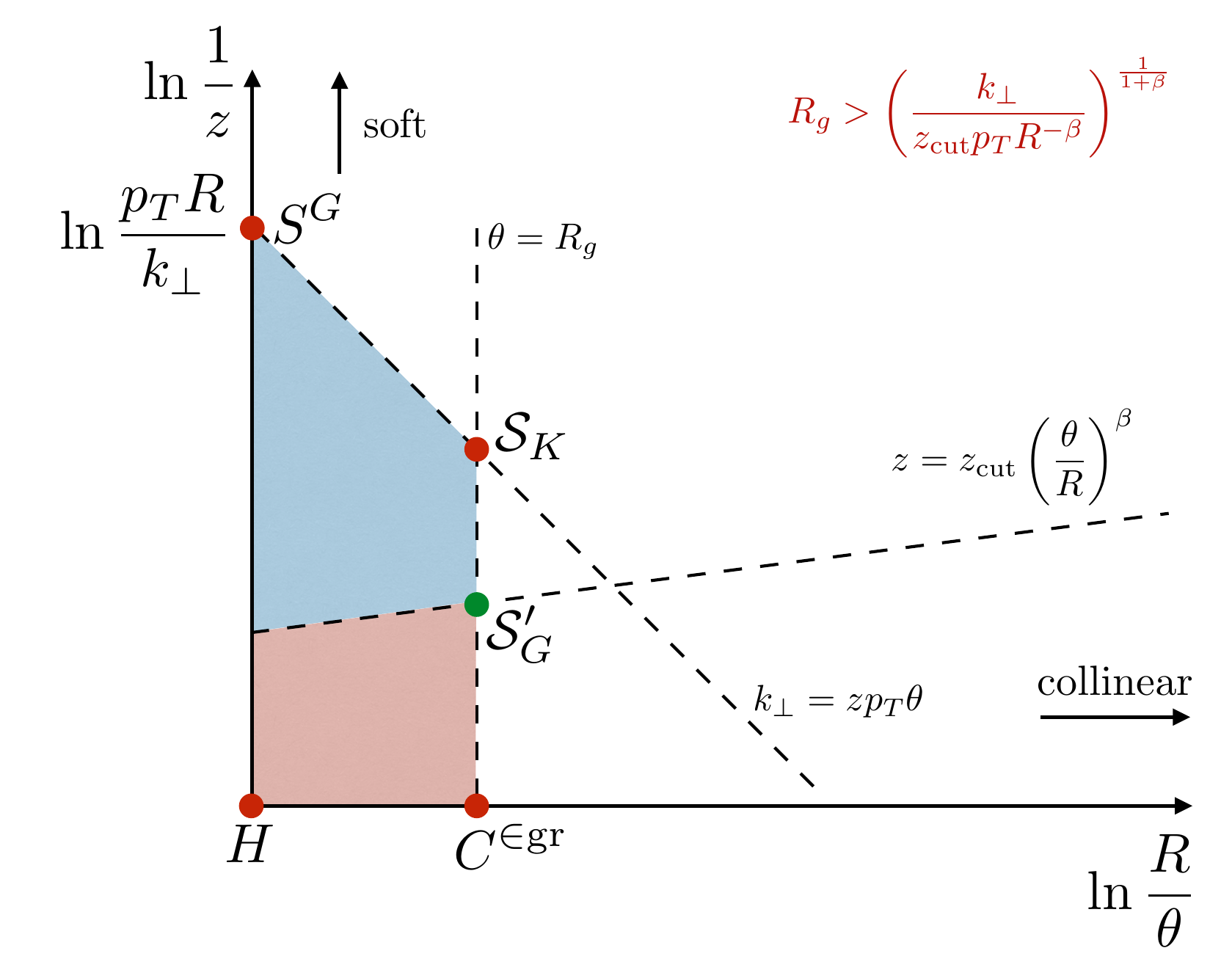} \hfill \phantom{.}
    \caption{Lund diagrams for the angle between the standard and the groomed jet axis when $k_\perp \ll \zc p_T R$, for regime $A$ (left) and $B$ (right). The relevant SCET modes are indicated by red and green dots, and their power counting can be read off, see table~\ref{tab:modes_ST_GR}.~\label{fig:Lund}}
\end{figure}

The Lund planes for the LL analysis are shown in \fig{Lund} and will now be discussed. We start by considering the measurement of the groomed jet radius. The cross section with a cut $\theta_g < \theta_g^c$ prohibits emissions in the red region with $z > \zc (\theta/R)^\bt$ and $\theta/R > \theta_g^c$~\cite{Larkoski:2014wba}. Ref.~\cite{Kang:2019prh} extends this analysis to NLL using SCET, exploiting that for the cumulative distribution in $\theta_g^c$ one can separately impose $\theta/R < \theta_g^c$ on the collinear and collinear-soft radiation (analogous to the factorization of the measurement in jet veto resummation).
If in addition one requires $k_\perp < k_\perp^c$, emissions with $z < \zc (\theta/R)^\bt$ and $\theta/R > \theta_g$ (\emph{not} $\theta_g^c$) are groomed away by soft drop and thus forbidden for $k_\perp = z \theta p_T > k_\perp^c$, corresponding to the blue regions in \fig{Lund}. There are now two cases depending on whether $\theta_g$ is smaller or larger than $[k_\perp^c/(\zc p_T R)]^{1/(1+\bt)}$, which we will refer to as regime $A$ and $B$.\footnote{In principle one can also consider $\theta_g \sim [k_\perp^c/(\zc p_T R)]^{1/(1+\bt)}$, but this only enters beyond NLL.} The Lund diagrams for these regimes are displayed in the left and right panel of \fig{Lund}, and the power counting of the momenta corresponding to the modes is summarized in table~\ref{tab:modes_ST_GR}.

 \begin{table}
   \centering
   \begin{tabular}{l|ll}
     \hline \hline
     Mode: & Regime $A$ & Regime $B$ \\ \hline
     hard & \multicolumn{2}{c}{$p_T(1,R^2,R)$} \\
     soft & \multicolumn{2}{c}{$k_\perp/R (1, R^2, R)$} \\ 
     collinear & \multicolumn{2}{c}{$p_T (1,R_g^2, R_g)$} \\
     collinear-soft$_G$ &  \multicolumn{2}{c}{$\zc \theta_g^\bt p_T (1,R_g^2, R_g)$} \\
     collinear-soft$_X$ &  $(b=(\zc p_T)^{1/(1+\bt)}(k_\perp/R)^{\bt/(1+\bt)},k_\perp^2/b, k_\perp)$ \quad \phantom{a} & \\
     collinear-soft$_K$ &  & $k_\perp/R_g (1,R_g^2, R_g)$ \\
     \hline \hline
   \end{tabular}
   \caption{The modes that enter in the factorization of $\cG$ for the angle between the standard and groomed jet axis, when $k_\perp \ll \zc p_T R$. Regime $A$ ($B$) correspond to $R_g$ smaller (larger) than $[k_\perp R^\bt/(\zc p_T)]^{1/(1+\bt)}$.}
   \label{tab:modes_ST_GR}
   \end{table}

The situation is the simplest for regime $A$, since the blue region is insensitive to the precise value of $\theta_g$, i.e.~the resummation of $\theta_g$ and $k_\perp$ are independent of each other. At LL accuracy, we simply calculate the sum of the area of the red and blue regions to obtain 
\begin{align}
 \tilde \cG_{i,A}^{\rm ST, GR}\bigl(k_\perp, p_T R, \zc,\bt, \theta_g, \al_s(\mu)\bigr) &\stackrel{{\rm LL}}{=} \frac{1}{2\pi\, k_\perp}\,\frac{\df}{\df k_\perp}\,\frac{\df}{\df \theta_g}\,\exp\bigg\{- \frac{\al_s C_i}{\pi} \bigg[
 \bt \ln^2 \theta_g + 2\ln \theta_g \ln \zc 
 \nn \\ & \quad
 + \frac{1}{1+\bt} \ln^2 \Big(\frac{k_\perp}{\zc p_T R}\Big)
  \bigg] \bigg\}
\end{align}
This can be extended beyond LL, using the factorization formula for the jet function $\tilde \cG$ differential in $\theta_g$, 
\begin{align} \label{eq:refact_ST_GR_A}
\tilde \cG_{i,A}^{\rm ST, GR}\bigl(k_\perp, p_T R, \zc,\bt, \theta_g, \al_s(\mu)\bigr)  &\stackrel{{\rm NLL}}{=} \frac{\df}{\df \theta_g} \bigg[\tilde H_i(p_{T} R,\mu)\, 
 C_i^{\in \rm gr}(\theta_g p_T R,\mu) \,  \CS_{G,i}\bigl(\zc \theta_g^{1+\bt} p_T R, \bt, \mu\bigr)
\nn \\ & \quad\hspace*{-3cm} \times
 \int\! \df^2 \vec k_\perp'\, 
 \CS_{X,i}(k_\perp' , p_T R, \zc, \bt, \mu, \nu/p_T)
  \int\! \df^2 \vec k_\perp''\, 
 S_i^{\rm G}(\vec k_\perp - \vec k_\perp' - \vec k_\perp'' ,\mu, \nu R) 
 \nn \\ & \quad\hspace*{-3cm} \times
 S_i^{\rm NG}\Big(\frac{k_\perp''}{p_T R}\Big)\,
 \CS_i^{\rm NG+AC}(\zc \theta_g^\bt)\bigg]
\,.\end{align}
For this angle we restrict ourselves to NLL accuracy, which is why the power corrections to the factorization formula have been omitted.
The collinear function $C_i^{\in \rm gr}$ describes collinear radiation, which is never groomed away. It can provide the emission that sets $\theta_g$, and any other collinear emissions must be at smaller angles. Similarly, for the collinear-soft$_G$ mode, emissions must satisfy $\theta/R < \theta_g^c$ if they pass the soft drop condition, as described by the collinear-soft function $\CS_{G,i}$. Wide-angle soft radiation is always groomed away, and its total transverse momentum is described by the same soft function $S_i^{\rm G}$ that we encountered before. Finally, $\CS_{X,i}$ describes collinear-soft$_X$ emissions, which only contribute to the transverse momentum if they fail the soft drop condition. There are also non-global logarithms and clustering logarithms associated with the collinear function $C_i$ and collinear-soft function $\CS_{G,i}$. The non-global logarithms from correlated hard and soft emissions are the same as for the standard vs.~WTA axis and described by $S_i^{\rm G}$. These are the same as in the resummation of $\theta_g$ and described by $\CS_i^{\rm NG+AC}$~\cite{Kang:2019prh}. 
The renormalization group equations for the new ingredients in \eq{refact_ST_GR_A} are given by
\begin{align}
   \mu \frac{\df}{\df \mu}\, C_i^{\in \rm gr}(\theta_g^c p_T R,\mu) &=  \ga^{C^{\in \rm gr}}_i(\theta_g^c p_T R,\mu)\, C_i^{\in \rm gr}(\theta_g^c p_T R,\mu)
   \,, \\   
   \mu \frac{\df}{\df \mu}\, \CS_{G,i}\bigl(\zc (\theta_g^c)^{1+\bt} p_T R,\bt, \mu\bigr) &=  \ga^{\CS_G}_i\bigl(\zc (\theta_g^c)^{1+\bt} p_T R,\bt, \mu\bigr)\, \CS_{G,i}\bigl(\zc (\theta_g^c)^{1+\bt} p_T R,\bt, \mu\bigr)   
   \,,\nn \\
\mu\f{\df}{\df\mu}\,\CS_{X,i}(k_\perp, \zc p_T R, \bt, \mu, \nu R) &= \gamma^{\CS_X}_i(\zc p_T R, \bt, \mu, \nu R)\,\CS_{X,i}(k_\perp, \zc p_T R, \bt, \mu, \nu R)
\,, \nn \\
\nu\f{\df}{\df\nu}\,\CS_{X,i}(k_\perp, \zc p_T R, \bt, \mu, \nu R) &= 
\!-\! \int\! \frac{\df^2 \vec k_\perp'}{(2\pi)^2}\,
\gamma^\nu_i(\vec k_\perp-\vec k_\perp',\mu)\,\CS_{X,i}(k_\perp', \zc p_T R, \bt, \mu, \nu R)
\,.\nn\end{align}
We wrote $\theta_g^c$ to explicitly indicate that the functions are cumulative distributions, i.e.~integrated over $\theta_g$ up to $\theta_g^c$. The natural scales of these modes are
\begin{align}\label{eq:can-ST-GR-A}
    \mu_H & = p_T R\,, &
   \mu_{C^{\in \rm gr}} &= p_T R\, \tg \, , &
   \mu_{\CS_G} &= \zc p_T R \, \tg^{1+\bt} \,, &
   \mu_{S^G} &= \kp \,  ,&
   \mu_{\CS_X} &= k_\perp\,,
   \nn \\ 
     & & & & & &
   \nu_{S^G} &= \f{2 \kp}{R}\,, &
   \nu_{\CS_X} &= \f{2}{R}\bigl(\zc p_T R \, \kp^\bt \bigr)^{\f{1}{1+\bt}}.
\end{align}

For regime $B$ we find it clarifying to work differentially in $\theta_g$. 
In the LL analysis there must a single emission that sets $\theta_g$, i.e.~$\theta/R = \theta_g$ and $z>\zc \theta_g^\bt$. Any other emissions must be outside the colored area, leading to 
\begin{align}
  \tilde \cG_{i,B}^{\rm ST, GR}\bigl(k_\perp, p_T R, \zc,\bt, \theta_g, \al_s(\mu)\bigr) &\stackrel{{\rm LL}}{=} -\frac{\alpha_s C_i}{\pi} \frac{2}{\theta_g} \ln (\zc \theta_g^\bt) 
  \\  & \hspace{-5ex} \times
 \frac{1}{2\pi\, k_\perp}\, \frac{\df}{\df k_\perp} \exp\bigg\{- \frac{\al_s C_i}{\pi} \bigg[\ln^2 \theta_g + 2 \ln \theta_g \ln \Big(\frac{k_\perp}{\theta_g p_T R}\Big) \bigg] \bigg\}
\,.\nn\end{align}
In the SCET analysis there are the four modes corresponding to the outer corners of the colored region, as one might anticipate. However, the emission that sets $\theta_g$ (and only this emission) is sensitive to the soft drop condition, which is why there is also a collinear-soft$_G$ mode contribution in this case. This leads to unusual behavior: the corresponding collinear-soft function only contributes if it has a single emission that sets $\theta_g$. In particular, if it has (independent) emissions that do not set $\theta_g$, these emissions only have the constraint $\theta/R < \theta_g$, resulting in scaleless integrals. The factorization formula in SCET for $\tilde \cG$ is given by 
\begin{align} \label{eq:refact_ST_GR_B}
\tilde \cG_{i,B}^{\rm ST, GR}\bigl(k_\perp, p_T R, \zc,\bt, \theta_g, \al_s(\mu)\bigr) &\stackrel{{\rm NLL}}{=} \tilde H_i(p_{T} R,\mu)\,
 \bigg[\frac{\df}{\df \theta_g}\, C_i^{\in \rm gr}(\theta_g p_T R,\mu) 
 \nn  \\ & \hspace{-15ex} 
 +  \CS_{G,i}'\bigl(\theta_g, \zc \theta_g^\bt p_T, \bt, \mu\bigr)  C_i^{\in \rm gr}(\theta_g p_T R,\mu) \bigg]
 \nn \\ &  \hspace{-15ex} \times
 \int\! \df^2 \vec k_\perp' 
 \CS_{K,i}(k_\perp' ,  \mu, \nu \tg R)
 S_i^{\rm G}(\vec k_\perp - \vec k_\perp' ,\mu, \nu R) 
 S_i^{\rm NG}\Big(\frac{k_\perp}{p_T R}\Big) 
\,.\end{align}
The two terms in the square brackets correspond to the collinear mode setting $\theta_g$ (when the derivative acts on $C_i^{\in \rm gr}$) or the collinear-soft$_G$ mode setting $\theta_g$. While the collinear-soft$_G$ mode has the same power counting as in regime A, the function is very different. This is also reflected in the unusual nature of the renormalization group equation of $\CS_{G,i}'$,\footnote{Note that only $R_g > 0$ contributions need to be considered here, since we are differential in $R_g$ and assume regime B. In particular, no (plus) distributions at $R_g=0$ are required.} 
\begin{align}
 \mu \frac{\df}{\df \mu}\, \CS_{G,i}'\bigl(\theta_g, \zc \theta_g^\bt p_T,\bt, \mu\bigr)
  = -\frac{\df}{\df \theta_g}\,  \ga^{C^{\in \rm gr}}_i(\theta_g p_T R,\mu)
\,,\end{align}
which follows from consistency of the factorization theorem and agrees with a direct calculation at order $\al_s$. The terms in the square brackets of eq.~\eqref{eq:refact_ST_GR_B} can be interpreted as a matching coefficient where we integrated out modes at the scales $\theta_gp_TR$ and $\zc \theta_g^{1+\bt} p_T R$. Its running down to the $k_\perp$ scale will generate non-global logarithms which we conjecture is also described by the function $\CS_i^{\rm NG+AC}$ of ref.~\cite{Kang:2019prh}, but with the different argument $k_\perp/(\theta_gp_T R)$. These NGLs require further study and are not included in our numerical analysis. The function $\CS_{K,i}$ describes the contribution to $k_\perp$ from collinear-soft$_K$ emissions that are groomed away, which for this mode simply means $\theta/R > \theta_g$. It's renormalization group equation is given by  
\begin{align}
\mu\f{\df}{\df\mu}\,\CS_{K,i}(k_\perp,\mu, \nu \theta_g R) &= \gamma^{\CS_K}_i(\mu, \nu \theta_g R)\,\CS_{K,i}(k_\perp, \mu, \nu \theta_g R)
\nn \,, \\
\nu\f{\df}{\df\nu}\,\CS_{K,i}(k_\perp,\mu,\nu \theta_g R) &= 
\!-\! \int\! \frac{\df^2 \vec k_\perp'}{(2\pi)^2}\,
\gamma^\nu_i(\vec k_\perp-\vec k_\perp',\mu)\,\CS_{K,i}(k_\perp', \mu, \nu \theta_g R).
\,.\end{align}
The natural scales for the modes in this regime are 
\begin{align}\label{eq:can-ST-GR-B}
    \mu_H & = p_T R\,, &
   \mu_{C^{\in \rm gr}} &= p_T R\, \tg \, , &
   \mu_{\CS_G'} &= \zc p_T R \, \tg^{1+\bt} \,, &
   \mu_{S^G} &= \kp \,  ,&
   \mu_{\CS_K} &= k_\perp\,,
   \nn \\ 
     & & & & & &
   \nu_{S^G} &= \f{2\kp}{R}\,, &
   \nu_{\CS_K} &= \f{2\kp}{ \tg R} .
\end{align}

\section{Standard vs.~winner-take-all axis}
\label{sec:ST_WTA}

In this section we provide further details describing our calculation of the angle between the standard and winner-take-all axis. In \sec{cG_ST_WTA} we discuss the calculation of jet function $\tilde \cG^{\rm ST,WTA}$ when $k_\perp \sim p_T R$, and the resummation of jet radius logarithms. In \sec{refact_ST_WTA} we consider $k_\perp \ll p_T R$, presenting the ingredients that enter the refactorization of $\tilde \cG$ and checking the singular/nonsingular decomposition. Details on our numerical implementation are give in \sec{implement}, with anomalous dimensions relegated to \app{anom}.

\subsection{Jet function for $k_\perp \sim p_T R$}
\label{sec:cG_ST_WTA}

We are working in the collinear limit $R\ll1$ and can therefore use the collinear phase-space and squared matrix element~\cite{Giele:1991vf},
\begin{align} \label{eq:coll}
\int\! \df \Phi_2\, \si_{2,q}^c &= 
\frac{\al_s}{\pi} \f{(e^{\gamma_E}\mu^2)^\epsilon}{\Gamma(1-\epsilon)}\int_0^1 \df x\,
C_F\bigg[\frac{1+x^2}{1-x} - \epsilon\, (1-x) \bigg]
\int\f{\df q_\perp}{q_\perp^{1+2\epsilon}}
\,,\nn \\
\int\! \df \Phi_2\, \si_{2,g}^c &= 
\frac{\al_s}{\pi} \f{(e^{\gamma_E}\mu^2)^\epsilon}{\Gamma(1-\epsilon)}\int_0^1 \df x\,
\bigg\{C_A \Big[\frac{x}{1-x}+\frac{1-x}{x}+x(1-x)\Big] 
\nn \\ & \qquad
+ n_f T_F \Big[1-2x(1-x)-2\epsilon\, x (1-x)\Big]\bigg\}
\int\f{\df q_\perp}{q_\perp^{1+2\epsilon}}
\,,\end{align}
to calculate the jet function for the angle between the standard and winner-take-all axis
\begin{align} \label{eq:cG_ST_WTA}
  \Delta \cG_i^{\rm ST,WTA}\bigl(k_\perp, p_T R, \al_s(\mu)\bigr) &=  \int\! \df \Phi_2\, \si_{2,i}^c\, \Theta\Big(\frac{ q_\perp}{x(1\!-\!x)p_T} < R\Big) \bigg[ \Theta\Big(x > \frac12\Big) \frac{1}{\pi} \delta\Big(k_\perp^2 \!-\! \frac{q_\perp^2}{x^2}\Big) 
  \nn \\ & \quad
  + \Theta\Big(x < \frac12\Big) \frac{1}{\pi} \delta\Big(k_\perp^2 - \frac{q_\perp^2}{(1-x)^2}\Big) 
  - \frac{1}{\pi} \de(k_\perp^2)  \bigg] 
\,.\end{align}
The final term in this expression subtracts off the semi-inclusive jet function, see \eq{Delta_iG}. This leads to
\begin{align} \label{eq:G_i}  
    \Delta \cG_q^{\rm ST,WTA}\bigl(k_\perp, p_T R, \al_s(\mu)\bigr)&=
     \frac{\al_s C_F}{\pi^2}\, \Theta\Bigl(k_\perp< \frac{p_T R}{2}\Bigr) \bigg\{ -\frac{1}{2\mu^2} \cL_1\Big(\frac{k_\perp^2}{\mu^2}\Big) 
   \\ & \quad \hspace*{-2cm}
  + \frac{1}{\mu^2} \cL_0\Big(\frac{k_\perp^2}{\mu^2}\Big) 
  \bigg[ \ln \Big(\frac{p_T R}{\mu}\Big) + \ln\Big(1 - \frac{k_\perp}{p_T R}\Big) +\frac32\, \frac{k_\perp}{p_T R} - \frac34\bigg]
  \nn \\ & \quad\hspace*{-2cm}
  + \de(k_\perp^2) \bigg[- \ln^2 \Big(\frac{p_T R}{\mu}\Big) + \frac32 \ln \Big(\frac{p_T R}{\mu}\Big) - \frac32 \ln 2 + \frac{\pi^2}{6} - \frac32\bigg]  \bigg\}
 \,, \nn \\ 
\Delta {\cal G}_g^{\rm ST, WTA}\bigl(k_\perp, p_T R, \al_s(\mu)\bigr) &= \frac{\al_s}{\pi^2}\, \Theta\Bigl(k_\perp< \frac{p_T R}{2}\Bigr) \bigg\{ -\frac{C_A}{2}\frac{1}{\mu^2} \cL_1\Big(\frac{k_\perp^2}{\mu^2}\Big) 
  \nn \\ & \quad \hspace*{-2cm}
  + \frac{1}{\mu^2} \cL_0\Big(\frac{k_\perp^2}{\mu^2}\Big) 
  \bigg[ \frac{\beta_0}{2} \bigg( \frac{k_\perp^{3}}{p_T^{3}R^{3}}-\frac32 \frac{k_\perp^{2}}{p_T^{2}R^{2}}+\frac32 \frac{k_\perp}{p_T R} -\frac12 \bigg)  \nn \\ & \quad \hspace*{-2cm}
  + C_A \bigg(-\frac32 \frac{k_\perp^{3}}{ p_T^{3}R^{3}} +\frac94 \frac{k_\perp^{2}}{p_T^{2}R^{2}}-\frac34 \frac{k_\perp}{p_T R} + \ln\Big(1- \frac{k_\perp}{ p_T R} \Big) + \ln \Big( \frac{p_T R}{\mu} \Big) \bigg)  \bigg]
  \nn \\ & \quad\hspace*{-2cm}
  + \de(k_\perp^2) \bigg[ C_A \bigg( -\ln^2\Big(\frac{p_T R}{\mu} \Big) +\frac{\pi^2}{6}+\frac{5}{16} \bigg) +\frac{\beta_0}{2}\bigg(\ln\Big(\frac{p_T R}{2\mu}\Big) -\frac{29}{24} \bigg) \bigg]  \bigg\}    
\,,\nn\end{align}
where $\beta_0$ is given in \eq{beta_0} and the plus distributions are defined as
\begin{align}
  \mathcal{L}_n(x) = \bigg[\frac{\ln^n x}{x}\bigg]_+
\,.\end{align}
Note that the terms where the plus distribution multiplies a power of $k_\perp$ do not require a plus prescription. It is worth emphasizing that in \eq{G_i} the explicit $\mu$ dependence cancels between the various terms, so the only $\mu$ dependence is in the scale $\al_s$.

\subsection{Refactorization for $k_\perp \ll p_T R$}
\label{sec:refact_ST_WTA}

We now present expressions for the various ingredients entering in the refactorization of the jet function in \eq{refact_ST_WTA}. First of all, we give the coefficients $J_{ij}$ in \eq{reshuffle}, that are independent of the angle we consider~\cite{Cal:2019hjc}  
\begin{align}
 J_{qq}(z,p_{T} R,\mu) 
  &= \de(1-z) + \f{\as}{2\pi} \bigg\{\ln\Big(\f{\mu^2}{p_T^2 R^2} \Big) P_{qq}(z) 
\\
 & \quad
+C_F\bigg[-2(1+z^2)\cL_1(1-z) + \Big(\frac{13}{2}- \f{2\pi^2}{3} \Big)\delta(1-z) - 1+z  \bigg]\bigg\}
, \nnu
 J_{qg}(z,p_{T} R,\mu) 
 &=\f{\as}{2\pi}\bigg[\Big(\ln\Big(\f{\mu^2}{p_T^2 R^2} \Big) - 2 \ln(1-z) \Big) P_{gq}(z) - C_Fz \bigg]
, \nnu
J_{gq}(z,p_{T} R, \mu) 
 & =  \f{\as}{2\pi}\bigg[\Big(\ln\Big(\f{\mu^2}{p_T^2 R^2} \Big) - 2\ln(1-z) \Big)  P_{qg}(z) - T_F 2z(1-z) \bigg]
, \nnu
J_{gg}(z, p_{T} R, \mu) 
& = \de(1-z) + \f{\as}{2\pi}\bigg\{\ln\Big(\f{\mu^2}{p_T^2 R^2} \Big)  P_{gg}(z)
\nnu 
& \quad 
  - \frac{4C_A (1-z+z^2)^2}{z}\, \cL_1(1-z)
  + \Big[C_A \Big(\frac{5}{12} - \frac{2\pi^2}{3}\Big) + \frac{23}{12} \bt_0\Big]  \delta(1-z)  \bigg\}
,\nn\end{align}
where the splitting functions are given in \eq{split}.

The hard function is up to one-loop order given by~\cite{Kang:2017mda,Cal:2019hjc} 
\begin{align}
\tilde H_q(p_{T} R,\mu) &= 1 + \frac{\al_s C_F}{2\pi} \biggl(- \frac12 \ln^2\Big(\f{\mu^2}{p_T^2 R^2} \Big)  - \frac32 \ln\Big(\f{\mu^2}{p_T^2 R^2} \Big)  - \frac{13}{2} + \frac{3\pi^2}{4} \biggr) + \ord{\al_s^2}
 \,, \nn \\
\tilde H_g(p_{T} R,\mu) &= 1 + \frac{\al_s}{2\pi} \biggl[ C_A\bigg( - \frac12 \ln^2\Big(\f{\mu^2}{p_T^2 R^2} \Big)  - \frac{5}{12} + \frac{3\pi^2}{4}\bigg) + \beta_0 \bigg(-\frac12 \ln\Big(\f{\mu^2}{p_T^2 R^2} \Big)  - \frac{23}{12}\bigg) \biggr] 
\nn \\ & \quad 
+ \ord{\al_s^2}
.\end{align}
The global soft function is at one-loop order given by~\cite{Kang:2017mda,Kang:2017glf}
\begin{align} \label{eq:S_nlo}
  S_i^{\rm G}(k_\perp,\mu,\nu R) = \frac{1}{\pi} \de(k_\perp^2) \!+\! \frac{\al_s C_i}{2\pi^2} \bigg[- \frac{1}{\mu^2} {\cal L}_1 \Big(\frac{k_\perp^2}{\mu^2}\Big)+\! \frac{1}{\mu^2} {\cal L}_0 \Big(\frac{k_\perp^2}{\mu^2}\Big) \!\ln \frac{\nu^2 R^2}{4\mu^2}
 \!-\! \frac{\pi^2}{12} \de(k_\perp^2)\bigg]
\,,\end{align}
where the color factor $C_i = C_A$ ($C_F$) for $i=g$ ($q$). The leading NGLs in the large $N_c$ approximation are taken from the solution to the BMS equation~\cite{Banfi:2002hw} up to five-loop order~\cite{Schwartz:2014wha} 
\begin{align} \label{eq:S_NG}
 S_{q}^{\rm NG}(\widehat L) &= 1 - \frac{\pi^2}{24} \widehat L^2 + \frac{\zeta_3}{12} \widehat L^3 + \frac{\pi^4}{34560} \widehat L^4+ \Big(-\frac{\pi^2 \zeta_3}{360} + \frac{17\zeta_5}{480}\Big) \widehat L^5 + \ord{L^6}
 \,, \nn \\
 S_{g}^{\rm NG}(\widehat L) &=  \bigl[S_{q}^{\rm NG}(\widehat L)\bigr]^2
\,.\end{align}
The argument of this $S_i^{\rm NG}$ is in impact parameter space given by 
\begin{align}
 \widehat L = \frac{\al_s N_c}{\pi} \ln (b_\perp p_T R)
\,.\end{align}
While \eq{S_NG} technically does not resum the NGLs, we have checked that the effect beyond the cubic term on our numerical results is less than a percent.

\begin{figure}[t]
     \hfill \includegraphics[width=0.48\textwidth]{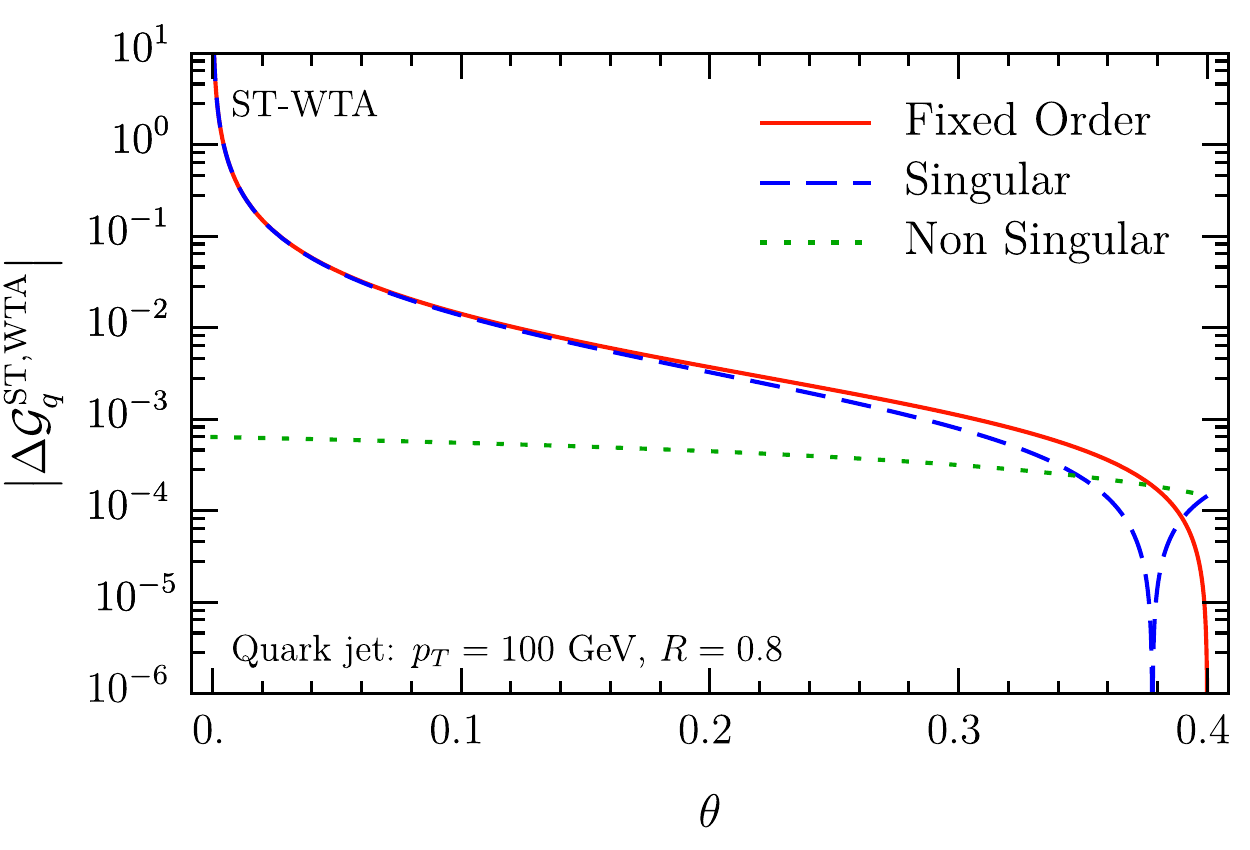} \hfill 
     \includegraphics[width=0.48\textwidth]{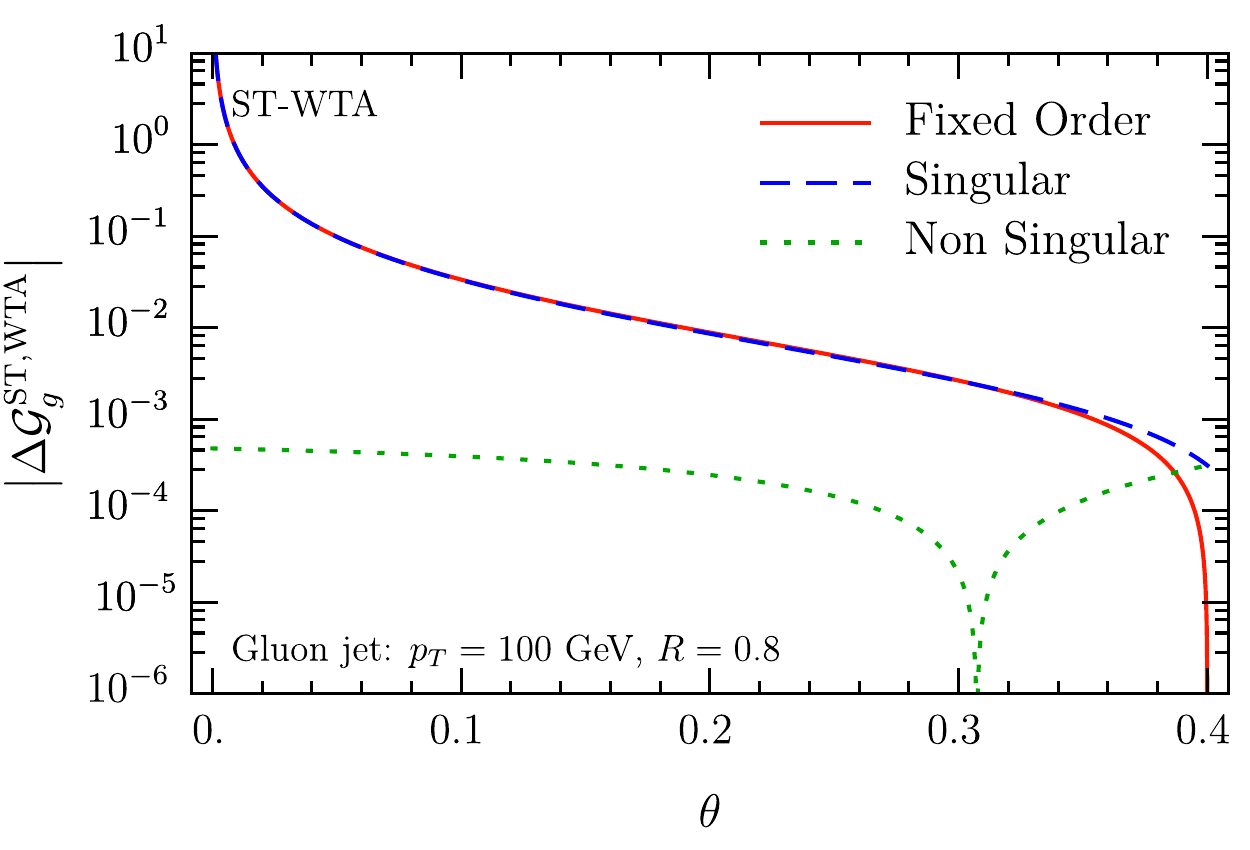} \hfill \phantom{.} \\
    \caption{Testing the refactorization (blue dashed) of the jet function $\Delta \cG^{\rm ST,WTA}$ (red solid), for the angle between the standard and WTA axis in \eq{refact_ST_WTA} at $\ord{\al_s}$. The nonsingular (green dotted) is the difference. The left (right) panel corresponds to quark (gluon) jets.~\label{fig:refact_ST_WTA}}
\end{figure}

The only new ingredient is the collinear function, whose calculation is very similar to \eq{cG_ST_WTA}, except that we can drop the restriction that the two particles are inside the jet since this is automatically the case for $k_\perp \ll p_T R$. Explicitly,
\begin{align}
  C_i(k_\perp, \mu,\nu/p_T) 
  &= \int\! \df \Phi_2\, \si_{2,i}^c\, \Big(\frac{\nu}{2(1-x)p_T}\Big)^\eta
    \bigg[ \Theta\Big(x > \frac12\Big) \frac{1}{\pi} \delta\Big(k_\perp^2 - \frac{q_\perp^2}{x^2}\Big) 
   \nn \\ & \quad
  + \Theta\Big(x < \frac12\Big) \frac{1}{\pi} \delta\Big(k_\perp^2 - \frac{q_\perp^2}{(1-x)^2}\Big)  \bigg] 
\,,\end{align}
where we use a rapidity regulator $\eta$ with associated scale $\nu$~\cite{Chiu:2011qc,Chiu:2012ir}. This leads to
\begin{align} 
  C_q(k_\perp, \mu,\nu/p_T)  &=  \frac{1}{\pi} \de(k_\perp^2) +
     \frac{\al_s C_F}{\pi^2} \bigg[
     \frac{1}{\mu^2} \mathcal{L}_0\Big(\frac{k_\perp^2}{\mu^2}\Big) \Big(- \frac34 - \ln \Bigl(\frac{\nu}{2p_T}\Bigr)\Big) 
   \nn \\ & \quad  
     +
     \de(k_\perp^2) \Big(-\frac32 \ln 2 - \frac{\pi^2}{6} + \frac{7}{4}\Big) \bigg]
   \,, \nn \\ 
C_g(k_\perp, \mu,\nu/p_T)  &=  \frac{1}{\pi}\delta(k_\perp^2)+\frac{\alpha_s}{\pi^2}\bigg\{\frac{1}{\mu^2} {\cal L}_0 \Big(\frac{k_\perp^2}{\mu^2}\Big) \Big(-\frac{\beta_0}{4}-C_A \ln \Bigl(\frac{\nu}{2p_T}\Bigr) \Big) 
\nn\\& \quad
+ \delta(k_\perp^2) \Big[C_A \Big(\frac{25}{48}-\frac{\pi^2}{6}\Big)+\frac{\beta_0}{2}\Big(\frac{17}{24}-\ln 2\Big) \Big] \bigg\}
\,.
\end{align}
Alternatively, since at this order there are at most two partons and the WTA is along the most energetic one, we can extract this collinear function from the TMD fragmentation function. In impact parameter $b_\perp$ space, using the conventions of ref.~\cite{Kang:2017glf}, this relationship is given by 
\begin{align}
   C_i(b_\perp,\mu,\nu)  = \sum_j \int\! \df z\, \Theta\Big(z - \frac12\Big)\, \,\tilde C_{j/i}(z, b_\perp, \mu,\nu)
\,.\end{align}
See also~\cite{Bain:2016rrv}. We conclude this section by verifying the validity of the refactorization in \eq{refact_ST_WTA} in \fig{refact_ST_WTA}, by plotting the jet function $\Delta \cG$, the singular expression obtained from the right-hand side and their nonsingular difference. We have converted the transverse momentum to an angle, including the appropriate Jacobian $\df^2 k_\perp = 2\pi \theta p_T^2\, \df \theta$, and taken $\mu = p_T R$, which only affects the scale in $\al_s$.
The nonsingular indeed vanishes for small angles. In fact, the refactorization seems to hold over most of the range, i.e.~$\theta = k_\perp/p_T \lesssim 0.4 R$. 

\section{Groomed vs.~winner-take-all axis}
\label{sec:GR_WTA}

In \sec{cG_GR_WTA} we calculate the jet function $\cG^{\rm GR,WTA}$ when $k_\perp \sim z_{\rm cut}p_T R\sim p_T R$. In \sec{refact_ST_WTA} the ingredients that enter its refactorization for $k_\perp \ll z_{\rm cut} p_T R\ll p_T R$ are given, and the singular/nonsingular decomposition is checked.

\subsection{Jet function for $k_\perp \sim p_T R$}
\label{sec:cG_GR_WTA}

The calculation is similar to the one in \sec{cG_ST_WTA}. In terms of the collinear phase-space and matrix element squared in \eq{coll}, 
\begin{align} 
\Delta \cG_i^{\rm ST,GR}\bigl(k_\perp,p_TR, \zc, \bt, \tg, \al_s(\mu)\bigr) &=
\int \df \Phi_2\, \si_{2,i}^c \Theta (\theta < R)  \nn \\
&\hspace*{-3cm} \times \biggl\{ \Theta \biggl(\frac12 > x> \zc \Bigl(\f{\theta}{R}\Bigr)^\beta \biggr)
\frac{1}{\pi} \biggl[\delta\Big(k_\perp^2 - \frac{q_\perp^2}{(1-x)^2}\Big) - \delta(k_\perp^2)\biggr]
 \nn \\
&\hspace*{-3cm} +  \Theta \biggl(1- \zc \Bigl(\f{\theta}{R} \Bigr)^\beta > x> \f12\biggr)\, \frac{1}{\pi} \biggl[ \delta\Big(k_\perp^2 - \frac{q_\perp^2}{x^2}\Big) - \delta(k_\perp^2)\biggr] \biggr\}
\,,\end{align}
using the shorthand $\theta = q_\perp/[x (1-x)p_T]$. The additional condition compared to  \eq{cG_ST_WTA} is that the partons must pass soft drop. This leads to
\begin{align} \label{eq:cG_GR_WTA}
\Delta {\cal G}_q^{\rm GR,WTA}\bigl(k_\perp,p_TR, \zc, \bt, \tg, \al_s(\mu)\bigr) &= \frac{\al_s C_F}{\pi^2} \bigg( \Theta\Bigl(\zc p_T R<k_\perp< \frac{p_T R}{2}\Bigr) 
  \nn \\ &  \quad \hspace*{-5cm}
 \times\bigg[ - \frac{1}{2\mu^2} \cL_1\Big(\frac{k_\perp^2}{\mu^2}\Big)  + \frac{1}{\mu^2} \cL_0\Big(\frac{k_\perp^2}{\mu^2}\Big) 
  \bigg( \ln\Big(\frac{p_TR}{\mu}\Big)+\ln\Big(1-\frac{k_\perp}{p_T R}\Big)+\frac32\frac{k_\perp}{p_T R}-\frac34  \bigg) \bigg]
  \nn \\ & \quad \hspace*{-5cm}
  + \Theta(k_\perp< \zc p_T R) \,  \bigg\{ -\frac{\beta}{2(1+\beta)} \f{1}{\mu^2} \cL_1 \Big(\frac{k_\perp^2}{\mu^2}\Big)  + \f{1}{\mu^2}\cL_0 \Big(\frac{k_\perp^2}{\mu^2}\Big) \bigg[\frac{\beta}{1+\beta}\ln\Big(\frac{p_T R}{\mu} \Big)
  \nn \\ & \quad \hspace*{-5cm}
+\frac32 \zc ^{\frac{1}{1+\beta}}\Big(\frac{k_\perp}{p_T R}\Big)^{\frac{\beta}{1+\beta}} +\ln\bigg(1-\zc ^{\f{1}{1+\beta}}\Big(\frac{k_\perp}{p_T R} \Big)^{\f{\beta}{1+\beta}}\bigg)-\frac{1}{1+\beta}\ln \zc -\frac34 \bigg]  \bigg\}
 \nn\\ & \quad \hspace*{-5cm}
  +\delta(k_\perp^2) \bigg[-\frac{\beta}{1+\beta}\ln^2\Big(\f{p_T R}{\mu}\Big)+\f{1}{1+\beta}\ln^2 \zc +\frac{2}{1+\beta}\ln \zc  \ln\Big(\f{p_T R}{\mu}\Big)
  \nn\\ & \quad \hspace*{-5cm}
  +\f32 \ln\Big(\frac{p_T R}{2\mu}\Big)+ \frac{2}{\beta} {\rm Li}_2(\zc )-\frac{3 \zc }{\beta}+\frac{\pi^2}{6}-\frac32\bigg] \bigg)
\,, \nn \\
\Delta {\cal G}_g^{\rm GR,WTA}\bigl(k_\perp,p_TR, \zc, \bt, \tg, \al_s(\mu)\bigr) &=  \frac{\al_s}{\pi^2}\, \bigg(\Theta\Bigl(\zc  p_T R<  k_\perp< \frac{p_T R}{2}\Bigr) \bigg\{ -\frac{C_A}{2}\frac{1}{\mu^2} \cL_1\Big(\frac{ k_\perp^{\,2}}{\mu^2}\Big) 
  \nn \\ & \hspace*{-5cm}
  + \frac{1}{\mu^2} \cL_0\Big(\frac{ k_\perp^{\,2}}{\mu^2}\Big) 
  \bigg[   \f{\bt_0}{2}\bigg( \frac{ k_\perp^{3}}{p_T^{3}R^{3}}-\f32 \frac{ k_\perp^{2}}{p_T^{2}R^{2}}+\f32 \frac{ k_\perp}{p_T R}-\f12 \bigg)
   \nn \\ &  \hspace*{-5cm}
 + C_A \bigg( -\f32 \frac{ k_\perp^{3}}{ p_T^{3}R^{3}} +\f94 \frac{ k_\perp^{2}}{ p_T^{2}R^{2}}-\f34  \frac{ k_\perp}{p_T R} + \ln\Big(1- \frac{ k_\perp}{ p_T R} \Big) + \ln \Big( \frac{p_T R}{\mu} \Big) \bigg) \bigg] \bigg\}
  \nn \\ &  \hspace*{-5cm}
+\Theta(k_\perp<\zc  p_T R) \bigg\{-C_A\frac{\beta}{2(1+\beta)}\f{1}{\mu^2}\cL_1 \Big(\frac{ k_\perp^{\,2}}{\mu^2}\Big) +  \f{1}{\mu^2} \cL_0\Big(\frac{ k_\perp^{\,2}}{\mu^2}\Big) 
  \nn \\ &  \hspace*{-5cm} \times
 \bigg[ \f{\beta_0}{2}  \bigg( \zc ^{\frac{3}{1+\beta}}\Big(\frac{ k_\perp}{p_T R}\Big)^{\frac{3\beta}{1+\beta}}  -\f32 \zc ^{\frac{2}{1+\beta}}\Big(\frac{ k_\perp}{p_T R}\Big)^{\frac{2\beta}{1+\beta}} +\f32 \zc ^{\frac{1}{1+\beta}}\Big(\frac{ k_\perp}{p_T R}\Big)^{\frac{\beta}{1+\beta}} -\f12 \bigg)
  \nn \\ &  \hspace*{-5cm}
+ C_A \bigg( \frac{\beta}{1+\beta} \ln\Big(\frac{p_T R}{\mu}\Big)-\f32 \zc ^{\frac{3}{1+\beta}}\Big(\frac{ k_\perp}{p_T R}\Big)^{\frac{3\beta}{1+\beta}}  +\f94  \zc ^{\frac{2}{1+\beta}}\Big(\frac{ k_\perp}{p_T R}\Big)^{\frac{2\beta}{1+\beta}} 
  \nn \\ &  \hspace*{-5cm}
-\f34  \zc ^{\frac{1}{1+\beta}}\Big(\frac{ k_\perp}{p_T R}\Big)^{\frac{\beta}{1+\beta}}  
+\ln\Big(1-\zc ^{\f{1}{1+\beta}}\Big(\frac{ k_\perp}{p_T R}\Big)^{\f{\beta}{1+\beta}}\Big)-\frac{1}{1+\beta}\ln \zc  \bigg) \bigg] \bigg\}
  \nn \\ &  \hspace*{-5cm}
  + \de( k_\perp^{\,2}) \bigg\{ C_A \bigg[ -\frac{\beta}{1+\beta}\ln^2\Big(\frac{p_T R}{\mu} \Big) +\frac{1}{1+\beta}\ln^2 \zc +\frac{2}{1+\beta}\ln \zc \ln\Big(\frac{p_T R}{\mu}\Big)
  \nn \\ &  \hspace*{-5cm}
 +\frac{2}{\beta}\text{Li}_2(\zc ) +\frac{\pi^2}{6}+\frac{5}{16} + \frac{1}{\beta}\Big(\zc ^3-\f94 \zc ^2+ \f32 \zc \Big) \bigg] 
  \nn \\ &  \hspace*{-5cm}
 +\frac{\beta_0}{2}\bigg[\ln\Big(\frac{p_T R}{2\mu}\Big) -\frac{29}{24} +\frac{1}{\beta}\Big(-\f23 \zc ^3+\f32 \zc ^2 - 3 \zc \Big)\bigg] \bigg\}  \bigg)
\,.\end{align}
Taking $\bt \to \infty$ we recover $\Delta \cG^{\rm ST,WTA}$. We can obtain the result for $\bt=0$ by taking the limit of the above expressions. Two particular types of terms that require care are
\begin{align}\label{eq:betatozero}
  \frac{1}{k_\perp^2}\, \zc ^{\f{n}{1+\beta}} \Bigl(\f{k_\perp}{p_T R}\Bigr)^{\f{n\,\beta}{1+\beta}} &= \delta(k_\perp) \biggl[\f{2(1+\bt)}{n\,\beta} \zc^n -2\zc^n \ln \zc -2 \zc^n \ln \Bigl(\f{p_T R}{\mu} \Bigr) \biggr] \nn \\
 & \quad + \f{1}{\mu^2} \cL_0\Big(\frac{k_\perp^2}{\mu^2}\Big) \, \zc^n +{\cal O}(\beta) 
\,, \nn \\
\frac{1}{k_\perp^2}\, \ln \bigg( 1- \zc ^{\f{1}{1+\beta}} \Big(\f{k_\perp}{p_T R}\Big)^{\f{\beta}{1+\beta}} \bigg) &= \de(k_\perp^2) \bigg[-\f{2}{\beta} {\rm Li}_2(\zc )-2{\rm Li}_2(\zc ) 
\nn \\ & \quad
-2 \ln \Bigl( \f{p_T R}{\mu} \Bigr)\ln (1-\zc )-2\ln (1-\zc )\ln \zc  \bigg]
\nn \\ & \quad
+ \f{1}{\mu^2} \cL_0\Big(\frac{ k_\perp^{\,2}}{\mu^2}\Big) \ln (1-\zc )+\mathcal{O}(\beta) \,.
\end{align}
The $1/\beta$ poles that appear in these expressions exactly cancel those already present in \eq{cG_GR_WTA}.
This leads to the following expression for $\bt=0$ 
\begin{align}\label{eq:Gbetazeroquark}
\Delta {\cal G}_q^{\rm GR,WTA}\bigl(k_\perp,p_T R,\zc, \beta=0, \al_s(\mu)\bigr)= \, & \frac{\al_s}{\pi^2} C_F\bigg( \Theta\Bigl(\zc p_T R< k_\perp < \frac{p_T R}{2}\Bigr) 
  \nn \\ &  \hspace*{-5cm}
\times \bigg\{ - \frac{1}{2\mu^2} \cL_1\Big(\frac{k_\perp^{\,2}}{\mu^2}\Big)  + \frac{1}{\mu^2} \cL_0\Big(\frac{ k_\perp^{\,2}}{\mu^2}\Big) 
  \bigg[ \ln\Big(\frac{p_TR}{\mu}\Big)+\ln\Big(1-\frac{ k_\perp}{p_T R}\Big)+\frac32\frac{ k_\perp}{p_T R}-\frac34  \bigg] \bigg\}
  \nn \\ & \hspace*{-5cm}
  + \Theta\Bigl( k_\perp< \zc p_T R\Bigr) \f{1}{\mu^2} \cL_0 \Big(\frac{ k_\perp^{\,2}}{\mu^2}\Big) \bigg[ \ln(1-\zc )-\ln \zc +\frac32 \zc -\frac34 \bigg]
 \nn\\ & \hspace*{-5cm}
  +\delta( k_\perp^2) \bigg[\ln^2 \zc +\ln \zc  \left(2\ln\Big(\frac{p_T R}{\mu}\Big)-2\ln(1-\zc )-3 \zc \right)   + \ln\Big(\frac{p_T R}{\mu}\Big)
  \nn\\ & \hspace*{-5cm} 
  \times \Bigl(-2\ln(1-\zc )-3 \zc +\frac32 \Bigr)-2\, {\rm Li}_2(\zc )+3 \zc +\frac{\pi^2}{6}-\frac32-\frac32 \ln 2\bigg] \bigg) \,,
  \nn\\
 \Delta {\cal G}_g^{\rm GR,WTA}\bigl(k_\perp,p_T R,\zc, \beta=0, \al_s(\mu)\bigr)=\, & \frac{\al_s}{\pi^2}\, \bigg(\Theta\Bigl(\zc  p_T R<  k_\perp< \frac{p_T R}{2}\Bigr) \bigg\{ -\frac{C_A}{2}\frac{1}{\mu^2} \cL_1\Big(\frac{ k_\perp^{\,2}}{\mu^2}\Big) 
  \nn \\ & \hspace*{-5cm}
  + \frac{1}{\mu^2} \cL_0\Big(\frac{ k_\perp^{\,2}}{\mu^2}\Big) 
  \bigg[ \f{\bt_0}{2}\bigg( \frac{ k_\perp^{3}}{p_T^{3}R^{3}}-\f32 \frac{ k_\perp^{2}}{p_T^{2}R^{2}}+\f32 \frac{ k_\perp}{p_T R}-\f12 \bigg)
   \nn \\ &  \hspace*{-5cm}
 + C_A \bigg( -\f32 \frac{ k_\perp^{3}}{ p_T^{3}R^{3}} +\f94 \frac{ k_\perp^{2}}{ p_T^{2}R^{2}}-\f34  \frac{ k_\perp}{p_T R} + \ln\Big(1- \frac{ k_\perp}{ p_T R} \Big) + \ln \Big( \frac{p_T R}{\mu} \Big) \bigg) \bigg] \bigg\}
  \nn \\ &  \hspace*{-5cm}
+\Theta( k_\perp<\zc  p_T R)\, \f{1}{\mu^2} \cL_0\Big(\frac{ k_\perp^{\,2}}{\mu^2}\Big)   \bigg[ \f{\bt_0}{2} \bigg( \zc^3 -\f32 \zc^2 +\f32 \zc -\f12 \bigg)
  \nn \\ &  \hspace*{-5cm}
+ C_A \bigg(-\f32 \zc^3 +\f94 \zc^2 -\f34 \zc +\ln(1-\zc )-\ln \zc  \bigg) \bigg]  
  \nn \\ &  \hspace*{-5cm}
  + \de( k_\perp^{\,2})  \bigg\{ C_A \bigg[\ln^2 \zc +2 \ln \zc \ln\Big(\f{p_T R}{\mu}\Big)+\Bigl(\ln \zc +\ln \Big(\f{p_T R}{\mu}\Big)\Bigr) \Big(3 \zc ^3 -\f92 z^2_{\rm cut}  \nn \\ &  \hspace*{-5cm}
  + \f32 \zc  -2\ln (1-\zc ) \Big)
  -z^3_{\rm cut} +\f94 z^2_{\rm cut}-\f32 \zc  -2{\rm Li}_2(\zc ) +\f{\pi^2}{6} +\f{5}{16}\bigg] \nn\\ & \hspace*{-5cm}
  +\beta_0\bigg[\Bigl(\ln \zc +\ln \Big(\f{p_T R}{\mu}\Big)\Bigr) \left(-z^3_{\rm cut}+\f32 z^2_{\rm cut}  -\f32 \zc  \right) \nn \\ &  \hspace*{-5cm}
+\f13 z^3_{\rm cut}-\f34 z^2_{\rm cut} +\f32 \zc  +\f12 \ln \left( \f{p_T R}{2\mu} \right) -\f{29}{48} \bigg]\bigg\}   \bigg) 
\,,\end{align}
which we checked by also calculating it directly.

\subsection{Refactorization for $k_\perp \ll p_T R$}
\label{sec:refact_GR_WTA}

The two new ingredients compared to \eq{refact_ST_WTA} are the soft function~\cite{Kang:2018jwa}
\begin{equation}
S_i^{\notin {\rm gr}}(\zc p_T R, \beta, \mu) = 1+\frac{\alpha_s C_i}{\pi} \frac{1}{1+\beta}\bigg[\ln^2 \Big(\frac{z_{\rm cut}p_T R}{\mu} \Big) - \frac{\pi^2}{24}\bigg] \,,
\end{equation}
with color factor $C_i = C_A$ ($C_F$) for $i=g$ ($q$),
and the collinear-soft function\footnote{We thank Z.~B.~Kang and K.~Lee for corroborating this result.}
\begin{align}\label{eq:ktSD}
\CS_i^{\rm G}(k_\perp, p_T R, \zc, \bt,\mu,\nu/p_T) &=   \frac{1}{\pi}\delta(k_\perp^2)+\frac{\alpha_s C_i}{\pi^2}\bigg[-\f{\beta}{2(1+\beta)} \frac{1}{\mu^2}\cL_1\Big(\frac{k_\perp^{\,2}}{\mu^2}\Big) 
 \\ & \hspace{-5ex}
+  \frac{1}{\mu^2}\cL_0\Big(\frac{ k_\perp^{\,2}}{\mu^2}\Big) \, \ln\bigg(\frac{\nu}{2p_T}\, z_{\rm cut}^{-1/(1+\beta)}\Big(\frac{p_T R}{\mu}\Big)^{\frac{\beta}{1+\beta}}\bigg) 
-\f{\beta}{1+\beta}\delta(k_\perp^2) \f{\pi^2}{24}\bigg] \,.
\nn \end{align}
Taking the $\bt \to \infty$ limit, $S_i^{\notin {\rm gr}} \to 1$ and $\CS_i^{\rm G}(k_\perp, p_T R, \zc, \bt,\mu,\nu/p_T) \to   S_i^{\rm G}(k_\perp,\mu,\nu R)$ in \eq{S_nlo}, recovering the result without grooming.
The NGLs are encoded by the same expression in \eq{S_NG}, but now involve the following logarithm
\begin{align}
 \widehat L = -\frac{\al_s N_c}{\pi} \ln \zc
\,.\end{align}

\begin{figure}[t]
     \hfill \includegraphics[width=0.48\textwidth]{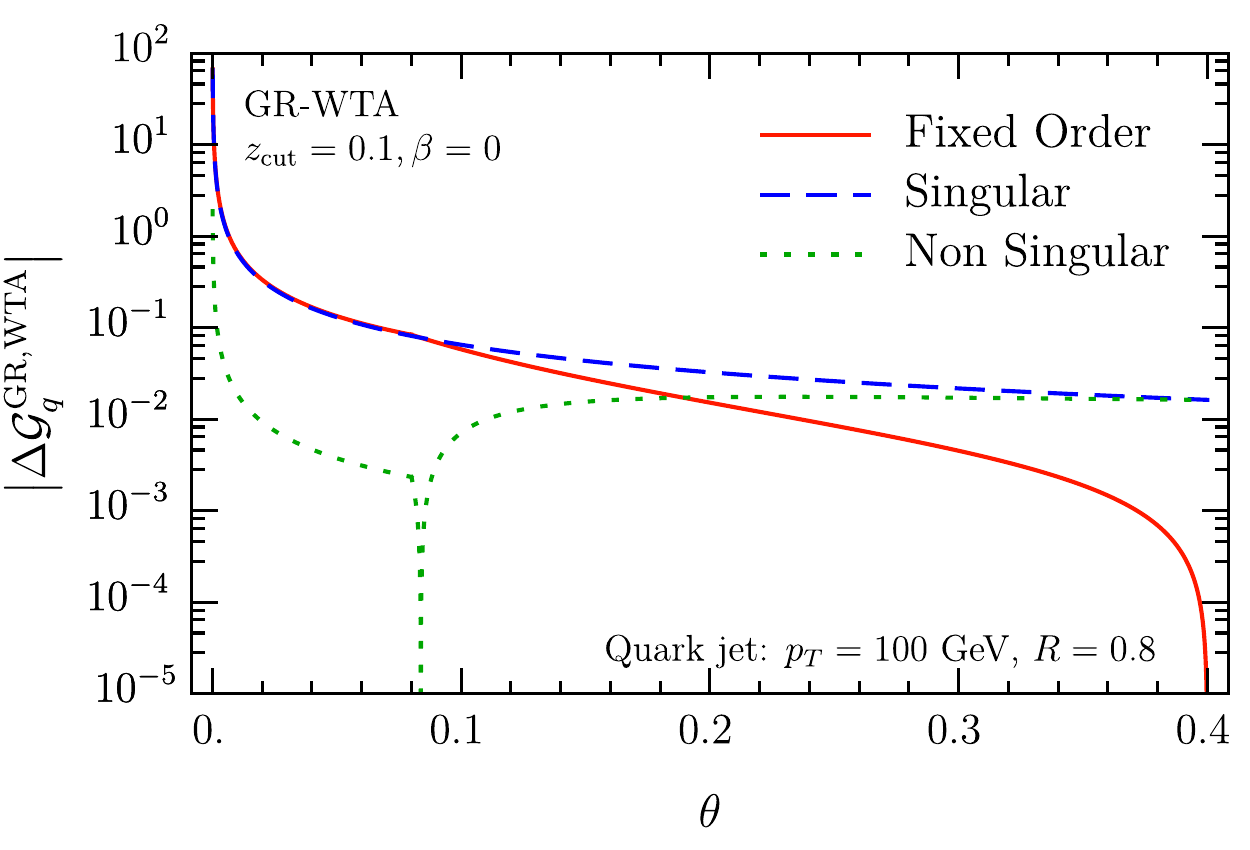} \hfill 
     \includegraphics[width=0.48\textwidth]{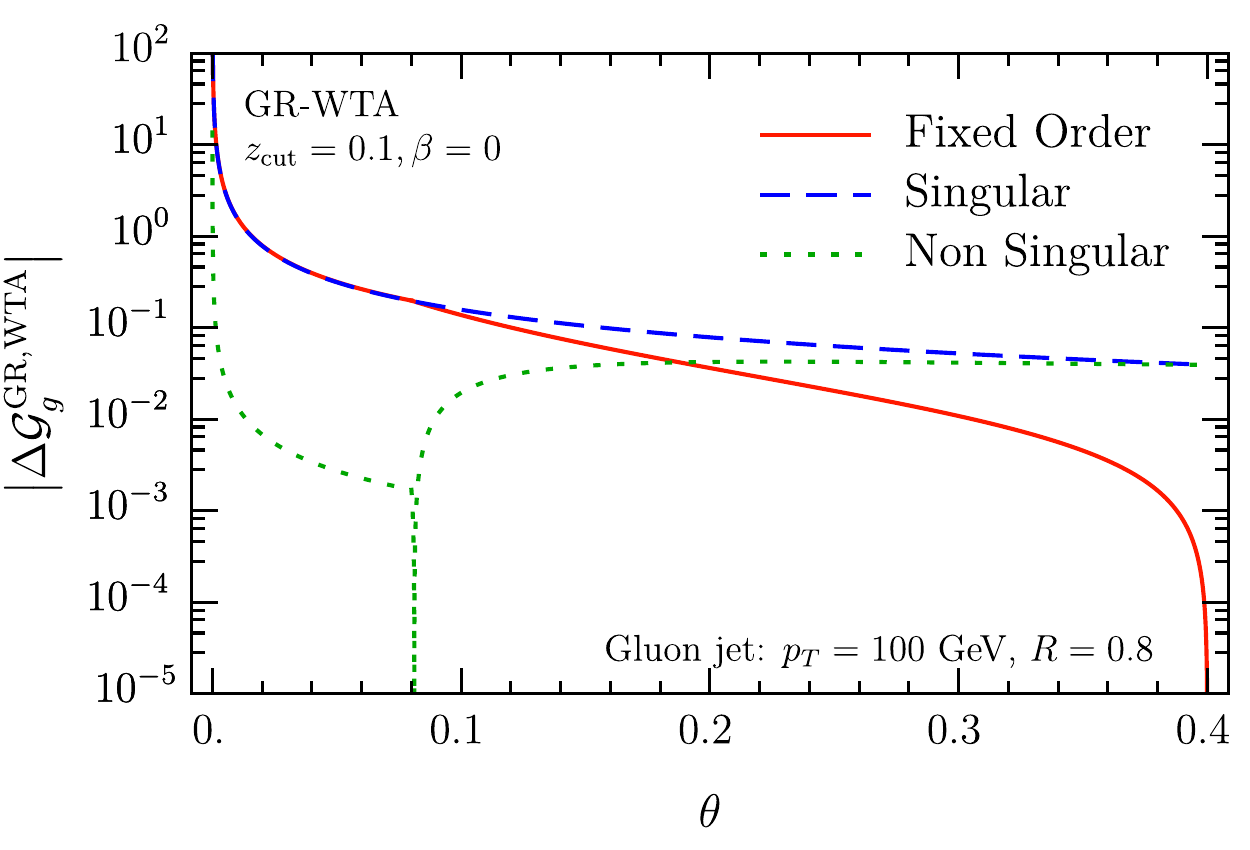} \hfill \phantom{.} \\
    \caption{Testing the refactorization (blue dashed) of the jet function $\Delta \cG^{\rm GR,WTA}$ (red solid), for the angle between the groomed axis (with $\zc=0.1$, $\bt=0$) and WTA axis in \eq{refact_ST_WTA} at $\ord{\al_s}$. The nonsingular (green dotted) is the difference. The left (right) panel corresponds to quark (gluon) jets.~\label{fig:refact_GR_WTA}}
\end{figure}

We numerically test the size of the power corrections to the refactorization in \eq{refact_GR_WTA}, by plotting the jet function $\Delta \cG$, the singular expression obtained from the right-hand side and their nonsingular difference. As in \fig{refact_ST_WTA}, we show the distribution differential in $\theta$ and take $\mu = p_T R$. The nonsingular encodes the size of power corrections and is very small at small angles. There is a transition at $\theta = \zc R$, since for larger angles there is no effect of grooming, and the plot coincides with that in \fig{refact_ST_WTA}. Since  this transition is still in the resummation region, this does not imply that there is no effect of grooming for larger angles in the \emph{resummed} cross section.

\section{Standard vs. groomed jet axis~\label{sec:ST_GR}}

We present results for the jet function $\cG^{\rm ST, GR}$ at fixed order, in the kinematic regime where $k_\perp \sim z_{\rm cut}p_T R\sim p_T R$, in sec.~\ref{sec:cG_ST_GR}. The ingredients entering its refactorization are given in \sec{refact_ST_GR}.

\subsection{Jet function for $k_\perp \sim p_T R$}
\label{sec:cG_ST_GR}

The measurement for the jet function differential in the groomed radius $\theta_g$ and the angle between the standard and groomed jet axis, encoded in $k_\perp$, is given by 
\begin{flalign}
\Delta \cG_i^{\rm ST,GR}\bigl(k_\perp,p_TR, \zc, \bt, \tg, \al_s(\mu)\bigr) &=
\int \df \Phi_2\, \si_{2,i}^c \Theta (\theta < R) \nn \\
&\hspace*{-3cm} \bigg[ \Theta \bigl(x> \zc (\theta/R)^\beta \bigr)\, \Theta \bigl(1-x> \zc (\theta/R)^\beta \bigr)\,\frac{1}{\pi}  \delta (k_\perp^2)\,  \delta (\tg - \theta/R)\nn \\
&\hspace*{-3cm}+\Theta \bigl(1-x < \zc (\theta/R)^\beta \bigr)\, \f{1}{\pi}\delta \Bigl(k_\perp^2-\f{q_\perp^2}{x^2} \Bigr)\, \delta (\tg)\nn \\
&\hspace*{-3cm}+ \Theta \bigl(x < \zc (\theta/R)^\beta \bigr)\,  \f{1}{\pi}\delta \Bigl(k_\perp^2-\f{q_\perp^2}{(1-x)^2} \Bigr) \delta (\tg) - \frac{1}{\pi}\delta(\kp^2) \delta (\tg) \bigg]
\,,\end{flalign}
using the shorthand $\theta = q_\perp/[x (1-x)p_T]$.
When the softest particle is groomed away, the groomed axis lies on top of the remaining parton, whereas the axes coincide when the parton passes grooming. We find
\begin{align}
\Delta \cG_q^{\rm ST,GR}\bigl(k_\perp,p_TR, \zc, \bt, \tg, \al_s(\mu)\bigr) &= \frac{\alpha_s C_F}{\pi^2}\bigg(\delta(\kp^2)\Theta(\tg<1)\Big[ (3 \zc - 2)\bt \cL_1(\tg)
 \nn \\ &\hspace*{-4cm}
+ \Bigl(3\zc -\f32 -2 \ln \zc \Bigr) \cL_0(\tg) +\f{2}{\tg} \ln (1-\zc \tg^\bt)\Big] 
  +\delta(\tg)\Theta(k_\perp < \zc p_T R)  
  \nn  \\&\hspace*{-4cm} \times  
  \bigg\{-\f{1}{2(1+\beta)}\,\f{1}{\mu^2} \cL_1 \Big(\frac{k_\perp^{\, 2}}{\mu^2}\Big)
  + \f{1}{\mu^2} \cL_0 \Big(\frac{k_\perp^{\,2}}{\mu^2}\Big) \bigg[\f32 \f{k_\perp}{p_TR}-\f32 \Big( \f{k_\perp}{p_TR} \Big)^{\f{\beta}{1+\beta}}\zc^{\f{1}{1+\beta}}
  \nn \\ &\hspace*{-4cm}
  + \ln \Big(1-\f{k_\perp}{p_T R}\Big)
   -\ln \bigg(1-\Big( \f{k_\perp}{p_TR} \Big)^{\f{\beta}{1+\beta}}\zc^{\f{1}{1+\beta}}\bigg)+ \f{1}{1+\beta}\ln\Big(\f{\zc p_T R}{\mu}\Big) \bigg]\bigg\}
 \nn \\ &\hspace*{-4cm}
  + \delta(\tg)\delta(k_\perp^2) \bigg[-\f{1}{1+\bt} \ln^2 \Big(\f{\zc p_T R}{\mu} \Big)
+ \f{3}{\bt} \zc \bigg]\bigg) \,, \nn \\
\Delta {\cal G}_g^{\rm ST,GR}\bigl(k_\perp,p_TR, \zc, \bt, \tg, \al_s(\mu)\bigr) &= \frac{\al_s}{\pi^2} \bigg(\delta(\kp^2)\Theta(\tg<1)\bigg\{ \f{\bt_0}{2} \big[(2\zc^3 -3\zc^2  \nn \\ 
&\hspace*{-4cm} +3\zc -1)\cL_0(\tg) + (6\zc^3 -6\zc^2 +3\zc)\bt \cL_1(\tg) \big]+ C_A \Big[ \Big(-3\zc^3 +\f92 \zc^2 \nn \\ 
&\hspace*{-4cm} -\f32 \zc  -2 \ln \zc \Big)\cL_0(\tg) + \Big(-9\zc^3 +9\zc^2-\f32\zc -2 \Big)\bt  \cL_1(\tg)   \nn \\
 &\hspace*{-4cm}  + \f{2}{\tg} \ln \Big(1-\zc \tg^\bt \Big)\Big]  \bigg\} + \de(\tg) \Theta\bigl( k_\perp< \zc p_T R\bigr)  \bigg\{-\f{C_A}{2(1+\beta)} \f{1}{\mu^2} \cL_1 \Big(\frac{ k_\perp^{\,2}}{\mu^2}\Big)\nn  \\
 &\hspace*{-4cm} + \f{1}{\mu^2} \cL_0 \Big(\frac{ k_\perp^{\,2}}{\mu^2}\Big) \bigg[\f{\bt_0}{2}\bigg( \f{k^3_\perp}{p^3_T R^3}- \Big( \f{k_\perp}{p_T R}\Big)^{\f{3\beta}{1+\beta}} \zc ^\f{3}{1+\beta} +\f32 \f{k^2_\perp}{p^2_T R^2}-\f32 \Big( \f{k_\perp}{p_T R}\Big)^{\f{2\beta}{1+\beta}} \zc ^\f{2}{1+\beta} \nn \\
&\hspace*{-4cm}  -\f32 \f{k_\perp}{p_T R}+ \f32 \Big( \f{k_\perp}{p_T R}\Big)^{\f{\beta}{1+\beta}} \zc ^\f{1}{1+\beta}  \bigg)  +C_A \bigg(-\f32 \f{k^3_\perp}{p^3_T R^3}+\f32 \Big( \f{k_\perp}{p_T R}\Big)^{\f{3\beta}{1+\beta}} \zc ^\f{3}{1+\beta} 
\nn \\ &\hspace*{-4cm}
 +\f94 \f{k^2_\perp}{p^2_T R^2}-\f94 \Big( \f{k_\perp}{p_T R}\Big)^{\f{2\beta}{1+\beta}} \zc ^\f{2}{1+\beta}
  -\f34 \f{k_\perp}{p_T R}+\f34\Big( \f{k_\perp}{p_T R}\Big)^{\f{\beta}{1+\beta}} \zc ^\f{1}{1+\beta} \nn \\
&\hspace*{-4cm}  + \f{1}{1+\beta}\ln \Big(\f{\zc p_T R}{\mu} \Big) + \ln \Big(1-\f{k_\perp}{p_T R} \Big)-\ln \Big(1- \Big(\f{k_\perp}{p_T R}\Big)^{\f{\beta}{1+\beta}} \zc ^\f{1}{1+\beta}\Big)  \bigg) \bigg] \bigg\} \nn  \\
&\hspace*{-4cm} + \de(\tg) \delta( k_\perp^2) \bigg\{ C_A \bigg[ \f{1}{\bt} \Big(-\zc^3 +\f94\zc^2-\f32\zc \Big)
- \f{1}{1+\bt}\, \ln^2 \Big(\f{\zc p_T R}{\mu} \Big) \bigg]\nn \\ 
&\hspace*{-4cm} +\frac{\beta_0}{\bt} \Big(\f13 \zc^3 -\f34 \zc^2 +\f32 \zc\Big) \bigg\} \bigg)
.\end{align}
When taking the limit $\beta\to 0$, similar care needs to be taken as for the angle between the groomed and the WTA axis, see \eq{betatozero}. The corresponding results for quarks and gluons with $\beta=0$ are given by
\begin{align}
\Delta \cG^{\rm ST,GR}_q\bigl(k_\perp,p_TR,z_{\rm cut},\beta=0,\tg,\al_s(\mu)\bigr) &= \\
&\hspace*{-4cm}\frac{\alpha_s C_F}{\pi^2}\bigg\{\delta(\kp^2)\Theta(\tg<1) \Big(3\zc -\f32 -2 \ln \zc + 2 \ln (1-\zc) \Big)\cL_0(\tg) \nn \\
&\hspace*{-4cm}  +\delta(\tg)\Theta( k_\perp< \zc p_T R)  \bigg[-\f{1}{2\mu^2} \cL_1 \Big(\frac{ k_\perp^{\,2}}{\mu^2}\Big)+ \f{1}{\mu^2} \cL_0 \Big(\frac{ k_\perp^{\,2}}{\mu^2}\Big) \bigg(\ln \Big(1-\f{k_\perp}{p_T R}\Big)\nn \\
  &\hspace*{-4cm}+\ln \Big( \f{\zc p_T R}{\mu}\Big)+\f32 \f{k_\perp}{p_T R}-\ln (1-\zc ) -\f32 \zc  \bigg)\bigg]\nn  \\
   &\hspace*{-4cm}+ \delta(\tg) \delta( k_\perp^2) \bigg[ -\ln ^2 \Big(\f{\zc p_T R}{\mu}\Big) + (3z_{\rm cut }+2\ln (1-\zc )) \ln \Big(\f{\zc p_T R}{\mu}\Big)  \nn \\
  &\hspace*{-4cm}-3\zc +2\rm{Li}_2 (\zc )\bigg] \bigg\}, 
  \nn \\
\Delta {\cal G}_g^{\rm ST,GR}\bigl(k_\perp,p_TR,z_{\rm cut},\beta=0,\tg,\al_s(\mu)\bigr) &= 
\nn \\ &\hspace*{-4cm}
\frac{\al_s}{\pi^2} \bigg(\delta(\kp^2) \Theta (\tg<1)\bigg[\f{\bt_0}{2}  (2\zc^3 -3\zc^2 +3\zc -1)\cL_0(\tg) \nn \\ 
&\hspace*{-4cm} +C_A  \Big(-3\zc^3 +\f92 \zc^2 -\f32 \zc -2 \ln \zc +2 \ln(1-\zc)\Big)\cL_0(\tg) \bigg]    \nn \\
& \hspace*{-4cm} + \de(\tg) \Theta( k_\perp< \zc p_T R)  \bigg\{-\f{C_A}{2\mu^2} \cL_1 \Big(\frac{ k_\perp^{\,2}}{\mu^2}\Big)+ \f{1}{\mu^2} \cL_0 \Big(\frac{ k_\perp^{\,2}}{\mu^2}\Big)  \nn  \\
&\hspace*{-4cm}\times \bigg[\f{\bt_0}{2}\bigg( \f{k^3_\perp}{p^3_T R^3} +\f32 \f{k^2_\perp}{p^2_T R^2}-\f32 \f{k_\perp}{p_T R}- \zc ^3-\f32\zc ^2+\f32 \zc \bigg)\nn \\
&\hspace*{-4cm}  +C_A \bigg(-\f32 \f{k^3_\perp}{p^3_T R^3} +\f94 \f{k^2_\perp}{p^2_T R^2}-\f34 \f{k_\perp}{p_T R}+ \ln \Big(1-\f{k_\perp}{p_T R} \Big)+ \ln \Big(\f{\zc p_T R}{\mu} \Big)\nn \\
&\hspace*{-4cm}  -\ln (1-\zc)+\f32 \zc^3 -\f94\zc^2 +\f34 \zc \bigg) \bigg] \bigg\} \nn  \\
&\hspace*{-4cm} + \de(\tg) \delta( k_\perp^2) \bigg\{ C_A \bigg[- \ln^2 \Big(\f{\zc p_T R}{\mu}\Big) +\Big(-3z^3_{\rm cut}+\f92 z^2_{\rm cut}-\f32 \zc  \nn \\
& \hspace*{-4cm} + 2\ln(1-\zc) \Big)  \ln \Big(\f{\zc p_T R}{\mu} \Big)  +\zc^3\nn -\f94 \zc^2 +\f32 \zc +\Li_2 (\zc)   \bigg] \nn \\ 
&\hspace*{-4cm} +\beta_0 \bigg[(\zc^3 -\f32 \zc^2 +\f32 \zc) \ln \Big(\f{\zc p_T R}{\mu} \Big) -\f13 \zc^3 +\f34 \zc^2 -\f32 \zc  \bigg] \bigg\} \bigg).
\nn \end{align}

\subsection{Refactorization for $k_\perp \ll p_T R$}
\label{sec:refact_ST_GR}

We start with the new functions that enter in the factorization for regime $A$. The same collinear function appeared in the factorization of the soft drop groomed jet radius, and is up to one-loop order given by~\cite{Kang:2019prh} 
\begin{align}
C_q^{\in {\rm gr}}(\tg^c p_T R,\mu)=&\,1+\f{\as C_F}{\pi} \bigg[\ln^2\Big(\frac{\mu}{\theta_g^c p_T R}\Big) + \frac32 \ln\Big(\frac{\mu}{\theta_g^c p_T R}\Big) + \frac{13}{4}-\frac{3\pi^2}{8} \bigg] \,,\\
C_g^{\in {\rm gr}}(\tg^c p_T R,\mu)=&\, 1+\f{\as}{\pi} \bigg[C_A  \ln^2\Big(\frac{\mu }{\theta_g^c p_T R }\Big) + \frac{\beta_0}{2} \ln\Big(\frac{\mu}{\theta_g^c p_T R}\Big) +C_A \Big(\frac{5}{24}-\frac{3\pi^2}{8}\Big)-\bt_0 \frac{23}{24} \bigg] \,.
\nn \end{align}
It is related to the unmeasured jet function of ref.~\cite{Ellis:2010rwa}. Next, we consider the collinear-soft functions. $\CS_{G,i}$ also appeared in the factorization of the soft drop groomed jet radius, and is given by~\cite{Kang:2019prh}
\begin{align}
\CS_{G,i}\bigl(\zc \theta_g^{1+\beta} p_T R,\bt, \mu\bigr) = 1 + \f{\alpha_s C_i}{ \pi (1+\beta)} \bigg[-\ln ^2 \Big( \f{\mu}{\zc \tg^{1+\beta} p_T R } \Big) +\f{\pi^2}{24} \bigg]
\end{align}
The collinear-soft function $\CS_{X,i}$ is sensitive to $k_\perp$ and the soft drop grooming condition, and is up to one-loop order given by
\begin{align}
 \CS_{X,i}(k_\perp, p_T R, \zc, \bt, \mu, \nu/p_T) = & \,  \frac{1}{\pi}\delta(k_\perp^2)+\frac{\alpha_s C_i}{\pi^2}\bigg[\f{\beta}{2(1+\beta)} \frac{1}{\mu^2}\cL_1\Big(\frac{k_\perp^{\,2}}{\mu^2}\Big) 
\nn \\ &\hspace*{-3cm}
-  \frac{1}{\mu^2}\cL_0\Big(\frac{k_\perp^{\,2}}{\mu^2}\Big) \, \ln\bigg(\frac{\nu}{2p_T}\, z_{\rm cut}^{-1/(1+\beta)}\Big(\frac{p_T R}{\mu}\Big)^{\frac{\beta}{1+\beta}}\bigg) 
+\f{\beta}{1+\beta}\delta(k_\perp^2) \f{\pi^2}{24}\bigg] \,.
 \end{align}
Note that the ${\cal O}(\alpha_s)$ term here is the same as the collinear-soft function ${\cal S}_i^{\rm G}$ in \eq{ktSD}, but with opposite sign.

In regime $B$, the $\CS_{G,i}'$ is in fact simply the derivative of $\CS_{G,i}$ with respect to $\theta_g$. The only additional result needed is the collinear-soft $\CS_{K,i}$, which is differential both in $k_\perp$ and $\tg$. A one-loop calculation yields,\footnote{Note that the one-loop contribution to $\CS_{K,i}$ is equal to that of the global soft function $S^G_i$ in \eq{S_nlo} with $R \to \tg R$ and a different overall sign.} 
\begin{align}
\CS_{K,i}(k_\perp, \mu, \nu \theta_g R)&= \f{1}{\pi} \delta(\kp^2)+\f{\as C_i}{\pi^2} \bigg[\f{1}{2\mu^2}\mathcal{L}_1 \left(\f{k^2_\perp}{\mu^2}\right)-\f{1}{\mu^2}\mathcal{L}_0 \left(\f{k^2_\perp}{\mu^2}\right) \ln \left( \f{\tg  R \nu}{2 \mu} \right) \nn \\
&+\f{\pi^2}{24} \delta(\kp^2) \bigg] \,.
\end{align}

\begin{figure}[t]
     \hfill \includegraphics[width=0.48\textwidth]{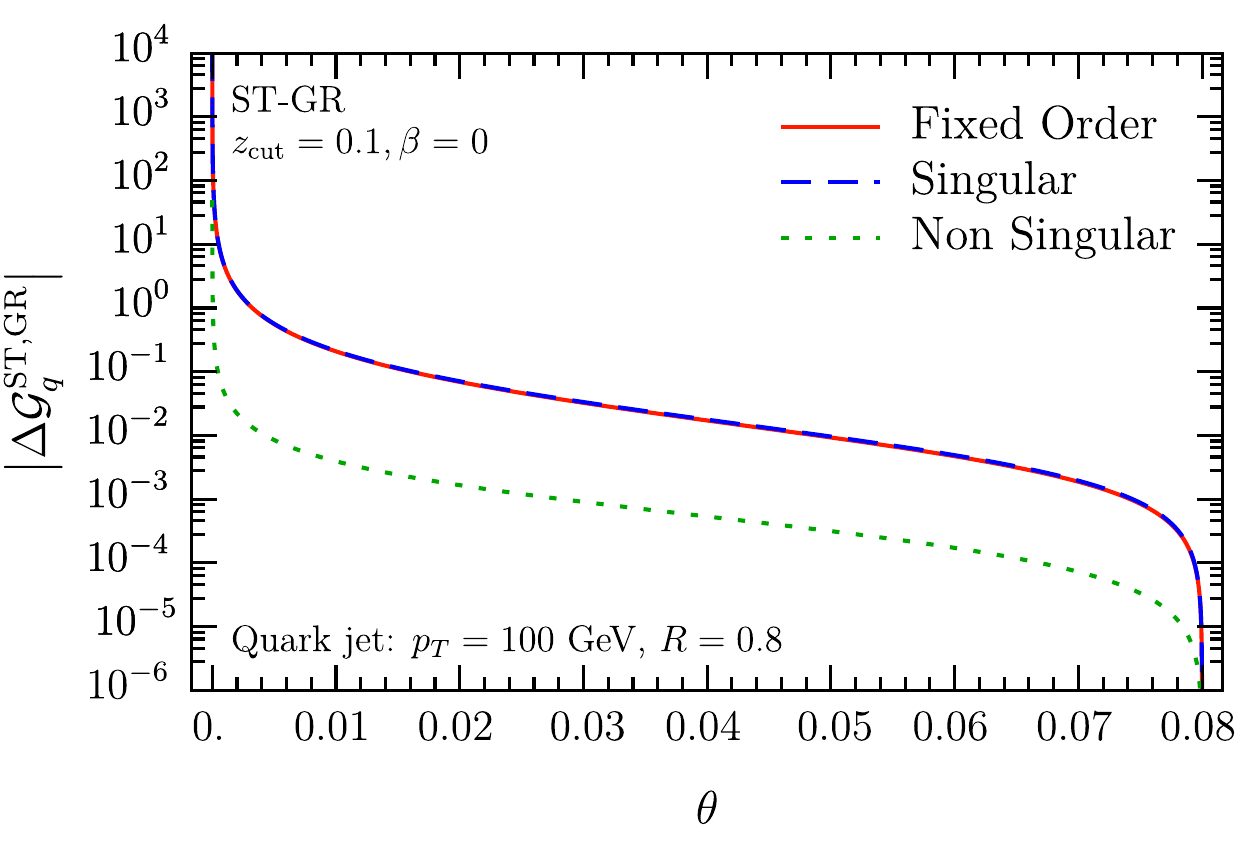} \hfill 
     \includegraphics[width=0.48\textwidth]{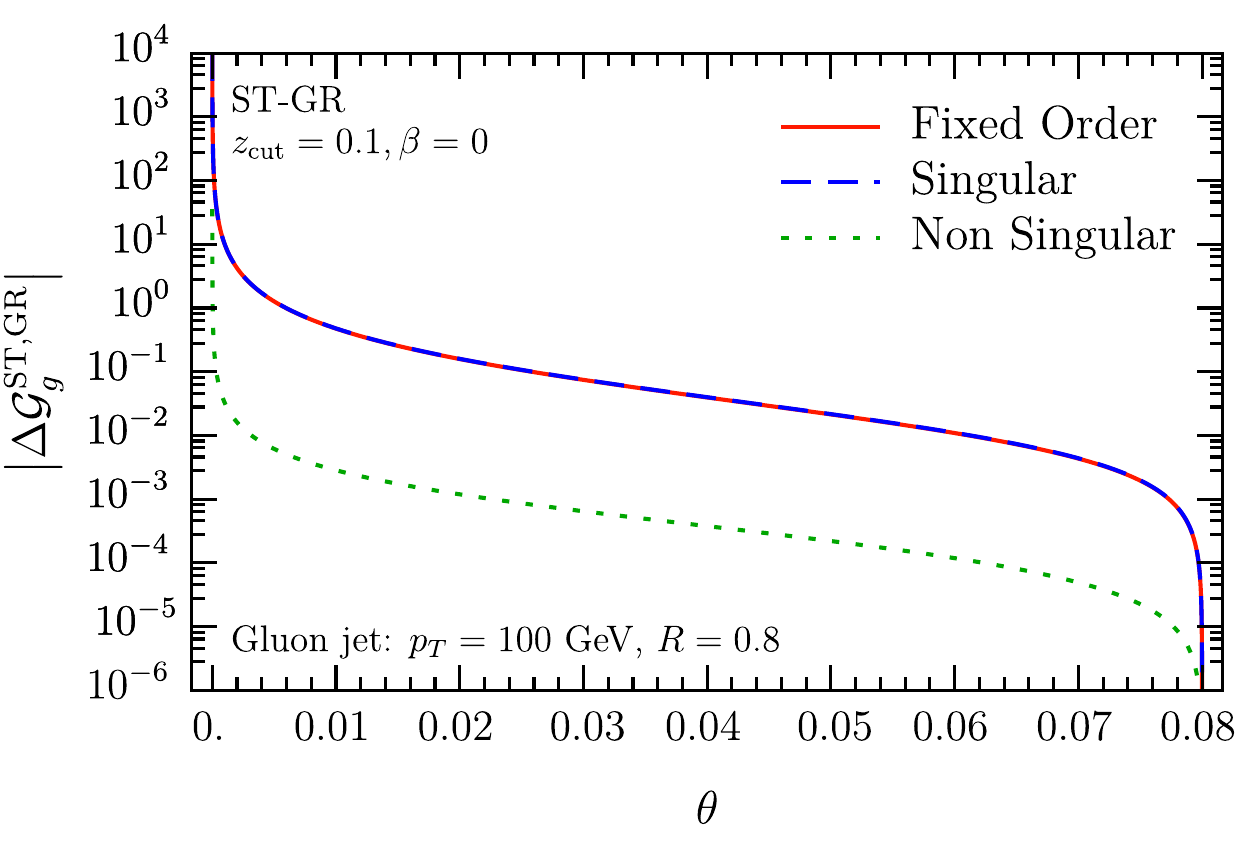} \hfill \phantom{.} \\
    \caption{Testing the refactorization of the jet function $\Delta \cG^{\rm ST,GR}$ where we include both regions $A$ and $B$, see eqs.~\ref{eq:refact_ST_GR_A} and~\eq{refact_ST_GR_B} at $\ord{\al_s}$. We show the full fixed order result (solid red), the singular distributions (dashed blue) and the difference between the two which corresponds to the nonsingular pieces (dotted green) is the difference. The left (right) panel corresponds to quark (gluon) jets.~\label{fig:refact_ST_GR}}
\end{figure}

Similar to the other angles considered above, we finish this section by comparing the singular and nonsingular terms at order $\al_s$, as shown in~\fig{refact_ST_GR}. Specifically, we compare the singular terms to the full fixed order expression for both quark (left) and gluon jets (right). Here we merged the contributions from regime $A$ and $B$ and integrate out the $R_g$ dependence. We observe that the singular terms dominate over the entire range of $\theta$. 

\section{Numerical implementation}
\label{sec:implement}

In this section we present details of the numerical implementation relevant for the results presented below. We start by describing how the resummation of $k_\perp/(p_T R)$ is carried out and matched to fixed order in~\sec{resummation}. In addition, we discuss scale choices and how we obtain perturbative QCD scale uncertainties for each angle. A discussion of nonperturbative effects can be found in~\sec{nonp}.

\subsection{Resummation, matching and uncertainties}
\label{sec:resummation}

The refactorization formulas in secs.~\ref{sec:fac_ST_WTA}, \ref{sec:fac_GR_WTA}  and \ref{sec:fac_ST_GR} are written in terms of convolutions in $\vec k_\perp$ since the individual contributions to the transverse momentum from each mode are added up vectorially. By going to impact parameter $b_\perp$ space, these equations become multiplicative which thus avoids the convolution structure and makes the implementation more straightforward. The same reasoning applies to the RG evolution equations, for which the convolution integrals  also become multiplicative in $b_\perp$ space. We then implement the $b_\perp$ space version of the formulas in~\sec{fac} and carry out the resummation, after which we perform a numerical inverse Fourier transform to obtain or predictions for the angle in $\kp$ space in the resummation region. We note that the natural scales of the different functions in $b_\perp$ space can be obtained from eqs.~\eqref{eq:can_ST_WTA},  \eqref{eq:can_GR_WTA}, \eqref{eq:can-ST-GR-A} and \eqref{eq:can-ST-GR-B}  with the replacement $\kp\rightarrow \mu_b=2 e^{-\gamma_E}/b_\perp$. 

Instead of using profile scales to merge the resummation of $k_\perp/(p_T R)$ with the fixed order (FO) calculations for $\tilde \cG$, we perform the matching by directly interpolating between the two results
\begin{align}
\tilde \cG_{\rm match} =   g(\theta)\, \tilde \cG_{\rm FO} + [1- g(\theta) ]\, \tilde \cG_{\rm resum} .
\end{align}
The function $g$ is $0$ in the resumation region, $1$ in the fixed order region and it smoothly interpolates between both regimes. For our numerical implementation we choose the following double quadratic function
\begin{align} \label{eq:g_match}
g(x) &= \left\{
\begin{tabular}{ll}
$0$ & $\phantom{x_1 \leq}\, x \leq x_1 $
\\
$\frac{(x-x_1)^2}{(x_2-x_1)(x_3-x_1)}$ & $x_1 \leq x \leq x_2$
\\
$1 - \frac{(x-x_3)^2}{(x_3-x_1)(x_3-x_2)}$ & $x_2 \leq x \leq x_3$
\\
$1$ & $x_3 \leq x$ \,.
\end{tabular}\right.
\end{align}
The choice of the transition points $x_1$ and $x_3$ depends on the angle we consider. For each angle, we determine the transition point between the resummation and fixed order region by comparing the singular and nonsingular terms at fixed order as shown in figs.~\ref{fig:refact_ST_WTA}, \ref{fig:refact_GR_WTA} and~\ref{fig:refact_ST_GR}. 

For the angle between the standard and WTA jet axes, we conclude from the the left panel of fig.~\ref{fig:refact_ST_WTA} that the refactorization of the cross section is valid until $\theta \sim 0.35 $, where the singular and the nonsingular (power corrections) become of comparable size. This indicates we have reached the region where one should use the fixed-order expression. Likewise, for the gluon (right panel) we see the resummation is necessary for almost the entire range that is shown in the figure, and we can transition to fixed-order result at slightly higher values, here $\theta \sim 0.38 $.

For the angle between the WTA and groomed jet axes we need to perform two transitions. The refactorization formula \eqref{eq:refact_GR_WTA} is expected to accurately describe the full fixed-order expression in the limit $ \kp  \ll \zc p_T R$, implying that it will only reproduce the terms proportional to $\Theta (\kp < \zc p_T R)$ in $\Delta \cG$. Once we go to values of $\kp$ that are larger than $\zc p_T R$, the fixed-order expression becomes equal to the ungroomed case, where the singular terms still dominate. This explains why in fig.~\ref{fig:refact_GR_WTA} there is a sudden change in the behavior of the power corrections at $\theta= \zc R$, which corresponds to the transition between the theta functions of the fixed order expression. Therefore, we need to switch from the groomed resummed case to the ungroomed resummed result at $\theta=\zc R$. We then proceed to transition from the ungroomed resummed to the fixed order result just as in the case of the angle between the standard and WTA jet axes.

By examining fig.~\ref{fig:refact_ST_GR}, we notice that the refactorization for the angle between the standard and groomed jet axes is particularly good throughout the whole range of $\theta$. Thus we transition to the fixed-order expression only at the very end, requiring that the cross section vanishes at $\theta=\zc R$.
 
The predictions for the angle between the standard and WTA axes and the groomed and WTA axes are computed at NLL$'$ accuracy. The perturbative QCD scale uncertainties are obtained by varying all scales simultaneously up and down by a factor of 2, by individually varying them by a factor of 2, and then taking the envelope. The angle between the standard and groomed axes is calculated at NLL accuracy. At this accuracy the scale variations give rather large uncertainties which may be up to 50\% in some kinematic regions, because there is no (partial) cancellation between the evolution kernels and fixed-order ingredients as the scales are varied. We therefore only show the central curve for this angle. However, the general features of the prediction, such as the angle at which the distribution peaks, do not change much. In order to integrate out the dependence on the soft drop groomed radius $R_g$, we freeze the running of the coupling constant at the scale $0.5$~GeV. We explored the dependence of our results on this cutoff scale by varying it by factors of 2 and found that the impact on the final numerical results is very small, especially compared to the QCD scale uncertainty at NLL accuracy.

\subsection{Nonperturbative effects}
\label{sec:nonp}

For all observables considered in this work, we work in impact parameter $b_\perp$ space and perform a numerical Fourier inverse transformation. For large $b_\perp$, we enter the nonperturbative regime and therefore we adopt the so-called $b_*$-prescription~\cite{Collins:1984kg}, modifying the scale $\mu_b=2 e^{-\gamma_E}/b_*$ with $b_*=b_\perp/\sqrt{1+(b_\perp/b_\perp^{\rm max})^2}$. Here $b_{\rm max}$ is chosen such that $b_*$ avoids the Landau pole for all values of $b_\perp$ and approaches $b_\perp$ at low values of $b_\perp$. We follow ref.~\cite{Makris:2017arq} in making the approximation that the dominant nonperturbative hadronization effects are due to the nonperturbative component of the rapidity anomalous dimension, as it is multiplied by a large logarithm of the ratio of the respective rapidity scales of the $b_\perp$-dependent collinear(-soft) and soft functions. The all-orders result of the rapidity anomalous dimension is given by
\begin{equation}\label{eq:rapidityanomalous}
\gamma_{\nu, i}^{S}(\mu)=-2 \int_{\mu_{b}}^{\mu} {\rm d} \ln \mu^{\prime}\,\Gamma_{\mathrm{cusp}}^{i}\left[\alpha_{s}\left(\mu^{\prime}\right)\right]+\gamma_{f}^{i}(\mu)-g_{K}\left(b_\perp, b_\perp^{\max }\right) \,,
\end{equation}
where the non-cusp part $\gamma_f^i$ vanishes to the accuracy at which we are working. The choice of $\mu_b$ as the lower bound of the integral in~\eq{rapidityanomalous} is compensated for by introducing the nonperturbative model function $g_K$, which needs to be determined by a fit to experimental data. For the two variables $g_2,\, b_\perp^{\rm max}$ which parametrize our nonperturbative model, we use the fitted results of ref.~\cite{Konychev:2005iy} where
\begin{equation}\label{eq:non-perturbative-model}
g_K(b_\perp,b_\perp^{\rm max})=g_2(b_\perp^{\rm max}) b_\perp^2\,,
\end{equation}
with $b_\perp^{\rm max}=1.5$~GeV$^{\rm -1}$ and $g_2(b_\perp^{\rm max})=0.18 \text{ GeV}^2$. For the gluon case, we include an additional factor of $C_A/C_F$ in $g_2$. Note that this parametrization of $g_K$ vanishes in the limit $b_\perp\to 0$. Other extractions of these or related nonperturbative parameters can be found for example in refs.~\cite{Landry:2002ix,Su:2014wpa,Bacchetta:2017gcc,Bertone:2019nxa}.  For example, for the angle between the standard and the groomed jet axis, we include in regime $A$ the following nonperturbative exponent in $b_\perp$ space
\begin{equation}
\exp\Big[-g_K(b_\perp,b_*)\frac{1}{1+\beta}\ln\frac{z_{\rm cut}p_T R}{\mu_b}\Big]\,.
\end{equation}
We note that the nonperturbative component in this case vanishes in the limit $\beta\to\infty$, which is consistent with the expectation that the the two axes are aligned in the limit that the grooming condition is removed. Similarly in regime $B$, the nonperturbative component vanishes when $\theta_g=1$ which corresponds to the case where the standard and groomed jet axis are aligned.

\begin{figure}[t]
     \hfill \includegraphics[width=0.48\textwidth]{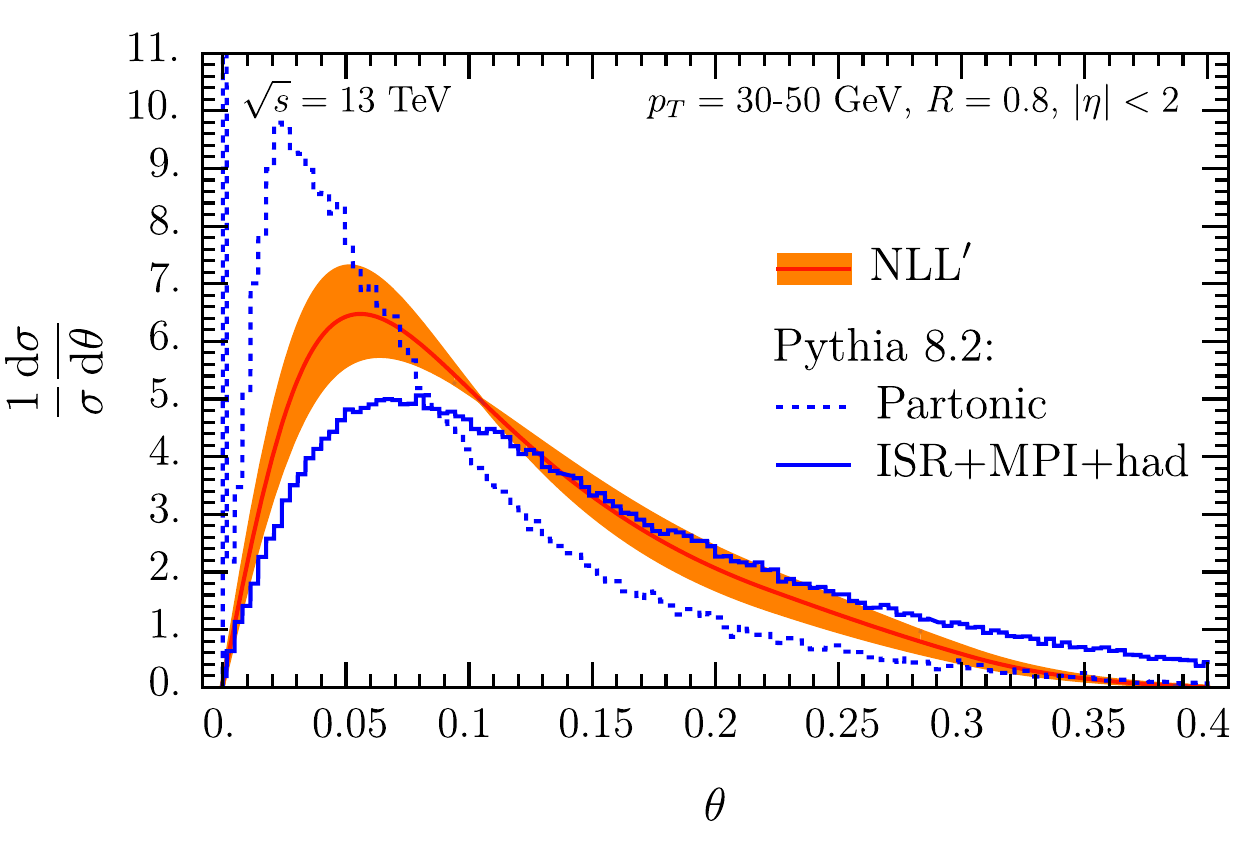} \hfill 
     \includegraphics[width=0.48\textwidth]{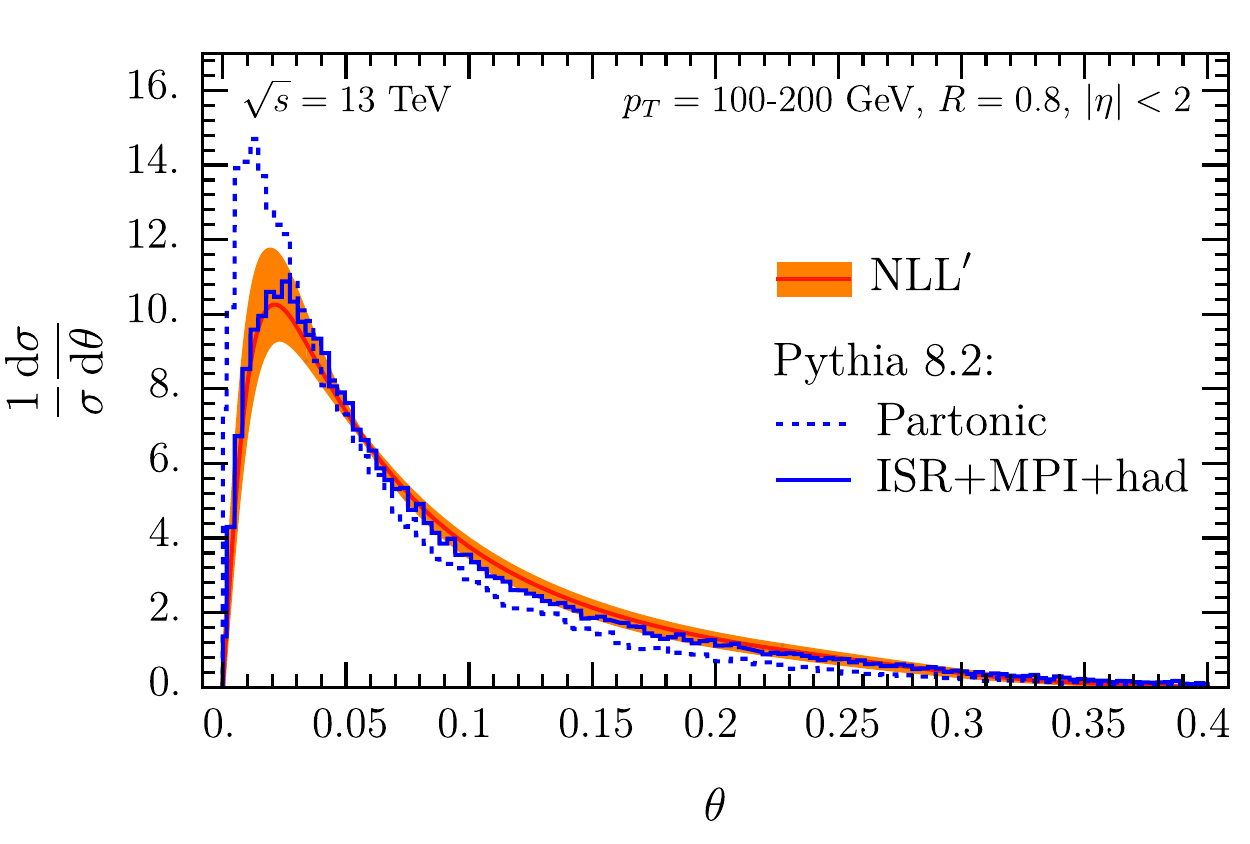} \hfill \phantom{.} \\
          \hfill \includegraphics[width=0.48\textwidth]{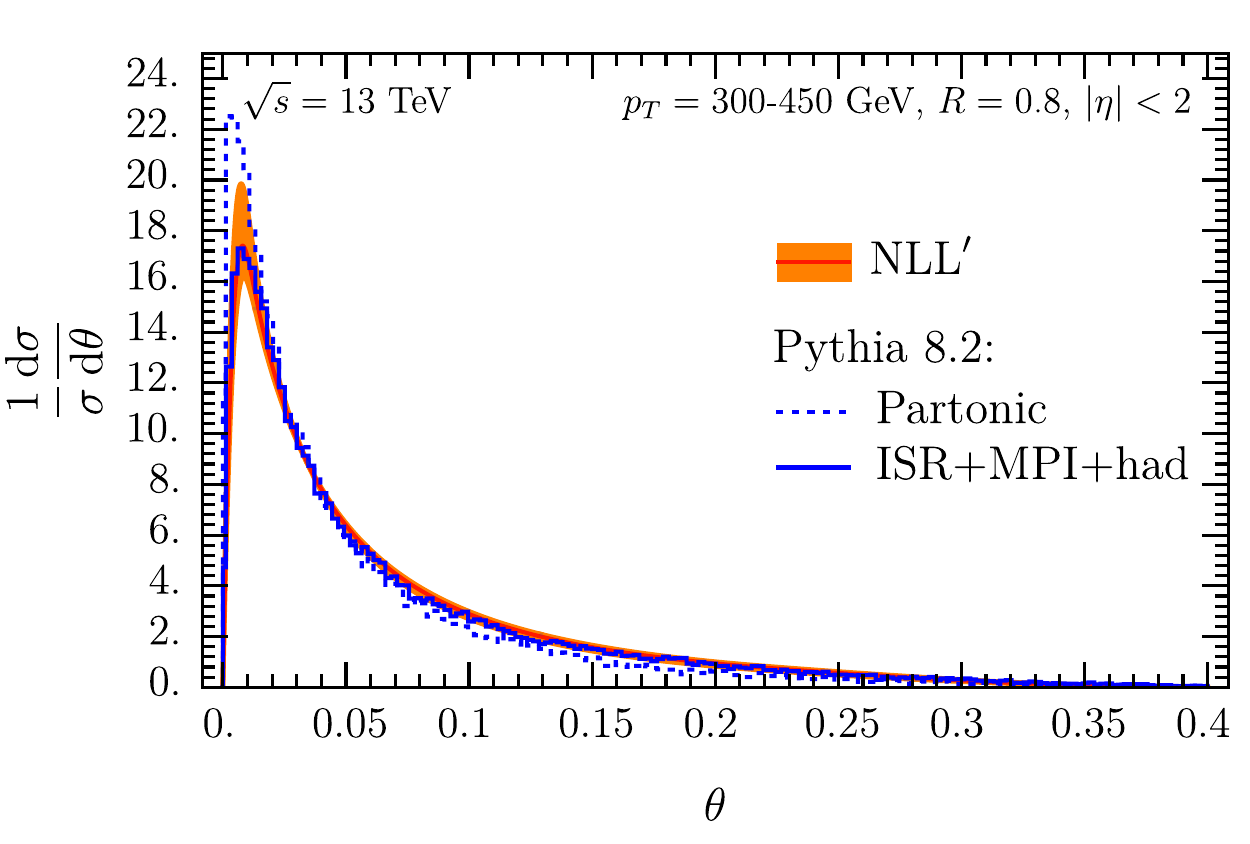} \hfill 
     \includegraphics[width=0.48\textwidth]{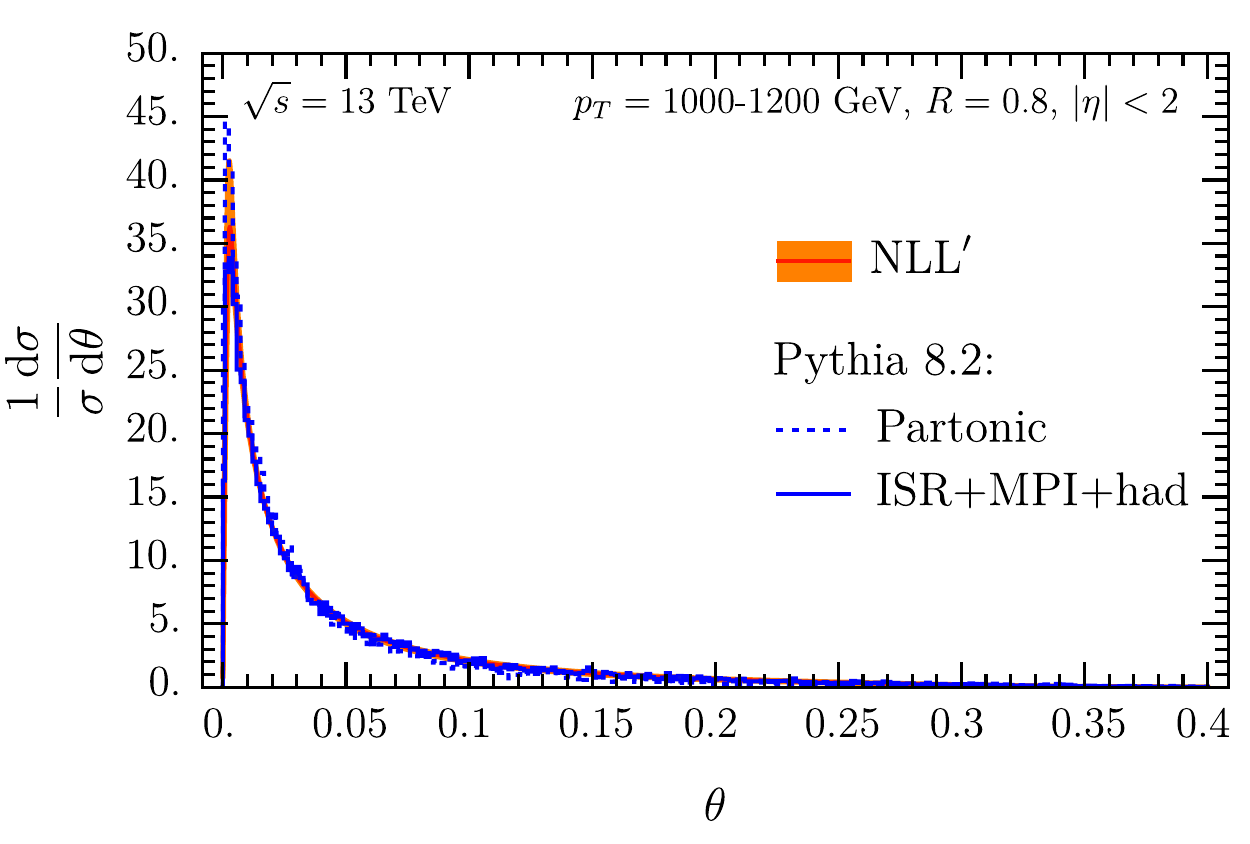} \hfill \phantom{.} 
    \caption{Numerical results for the angle between the WTA and the standard jet axis at NLL$'$ accuracy (orange curve and band) for four jet transverse momentum intervals as indicated in the figure. We choose proton-proton collisions at $\sqrt{s}=13$~TeV where jets are identified with the anti-k$_T$ algorithm with $R=0.8$ at central rapidity with $|\eta|<2$. For comparison, we show \Pythia~8.2 results at parton level (dotted blue) and when ISR, MPI and hadronization are included (solid blue).~\label{fig:WTA-ST-results}} 
\end{figure}

\begin{figure}[t]
     \hfill \includegraphics[width=0.48\textwidth]{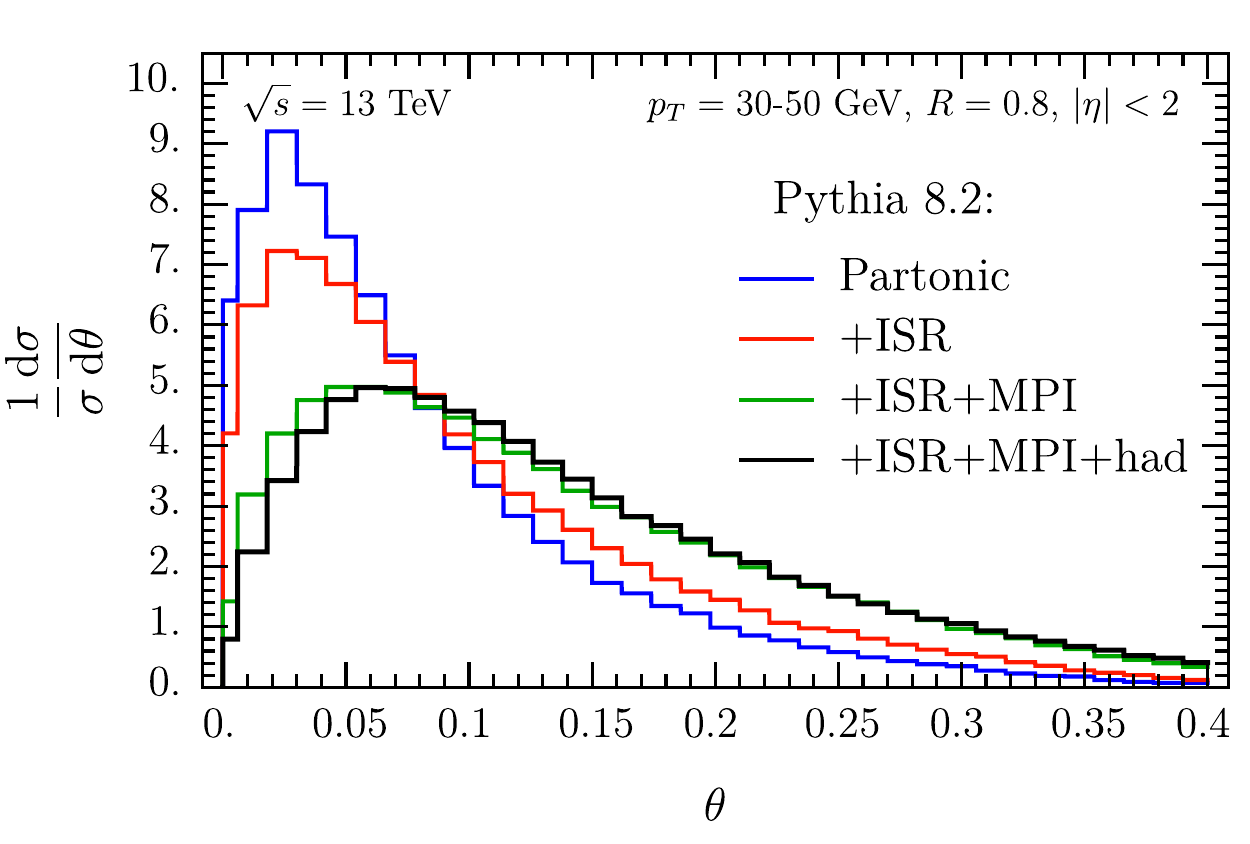} \hfill 
     \includegraphics[width=0.48\textwidth]{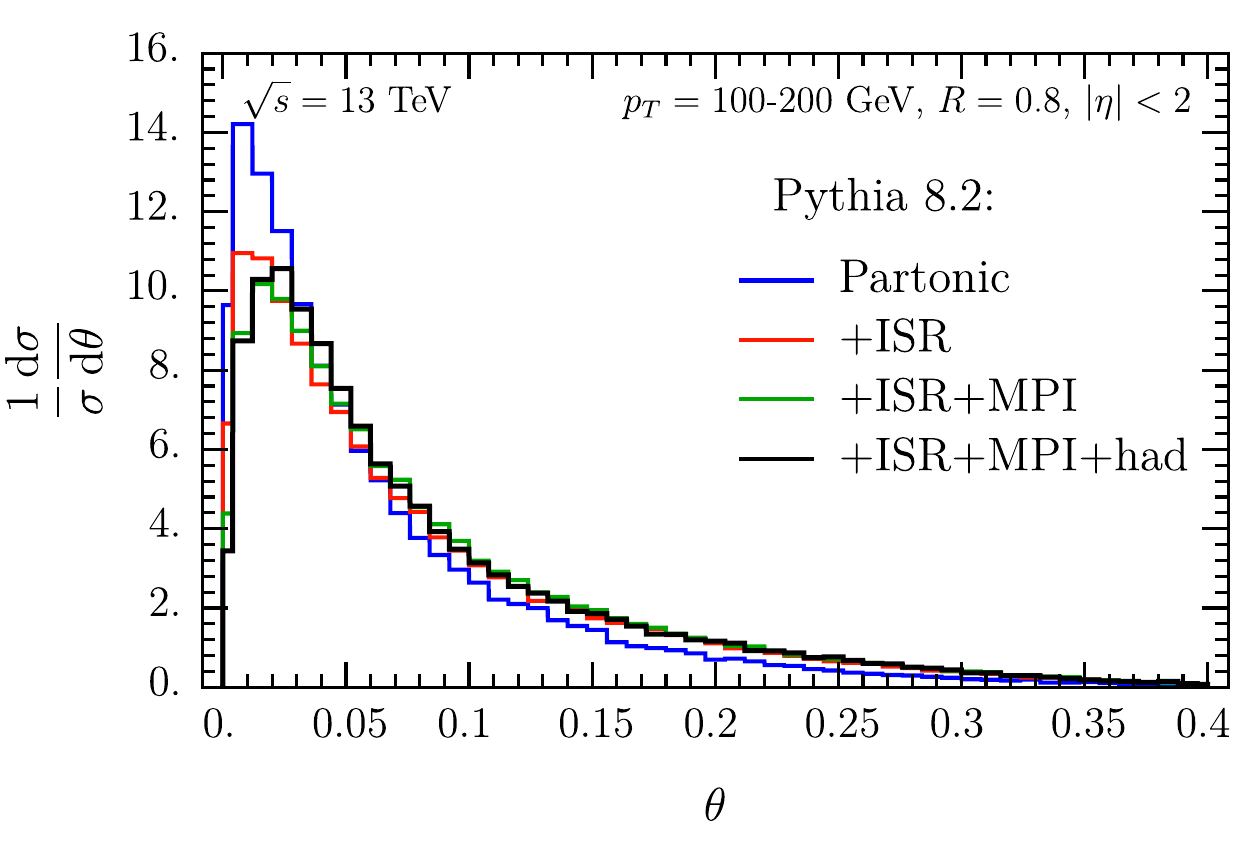} \hfill \phantom{.} \\
    \caption{\Pythia results for the angle between the standard and WTA jet axis, for the kinematics corresponding to the upper row in~\fig{WTA-ST-results}. Shown are the partonic result (blue), including ISR (red), MPI (green) and also hadronization (black).~\label{fig:Pythiaonly}}
\end{figure}

\begin{figure}[t]
\begin{center}
\includegraphics[width=0.65\textwidth]{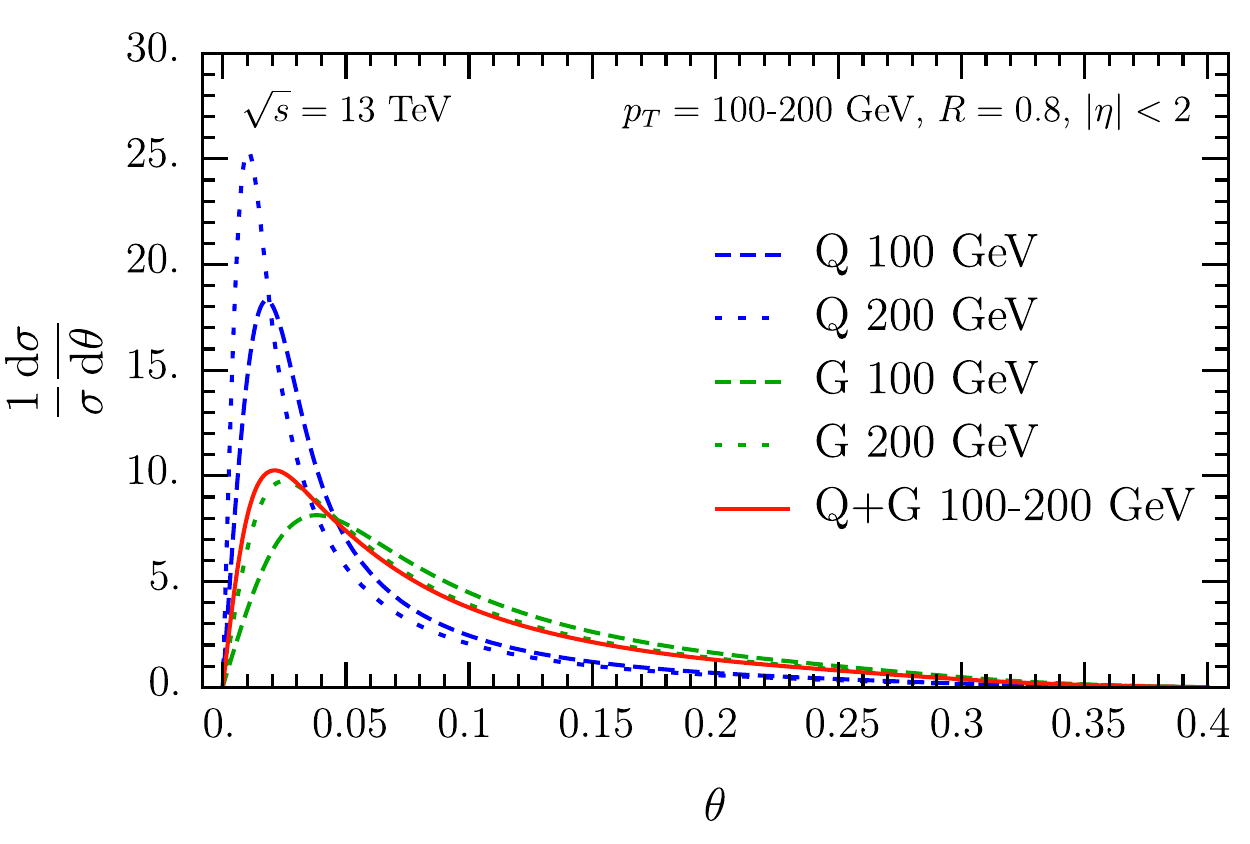}
\caption{Quark (blue) and gluon (red) distributions for the angle between the standard and the WTA axis for $p_T=100$~GeV (dashed) and 200~GeV (dotted). The solid line shows the $p_T$ integrated result summed over quarks and gluons which corresponds to the central curve in the upper right panel of~\fig{WTA-ST-results}.~\label{fig:quarkgluon}} 
\end{center}
\end{figure}

\section{Results for the LHC}
\label{sec:results}

\begin{figure}[t]
\begin{center}
\hfill \includegraphics[width=0.48\textwidth]{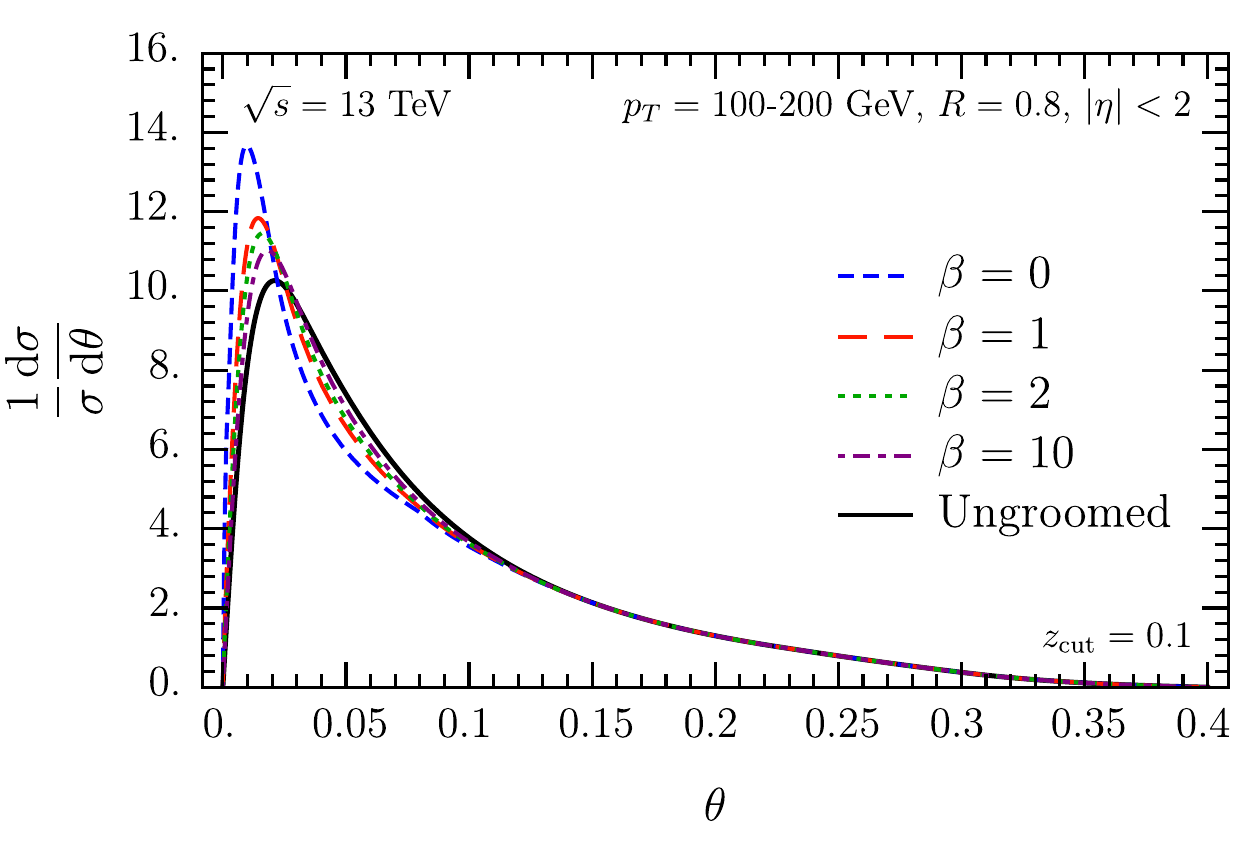} \hfill 
     \includegraphics[width=0.48\textwidth]{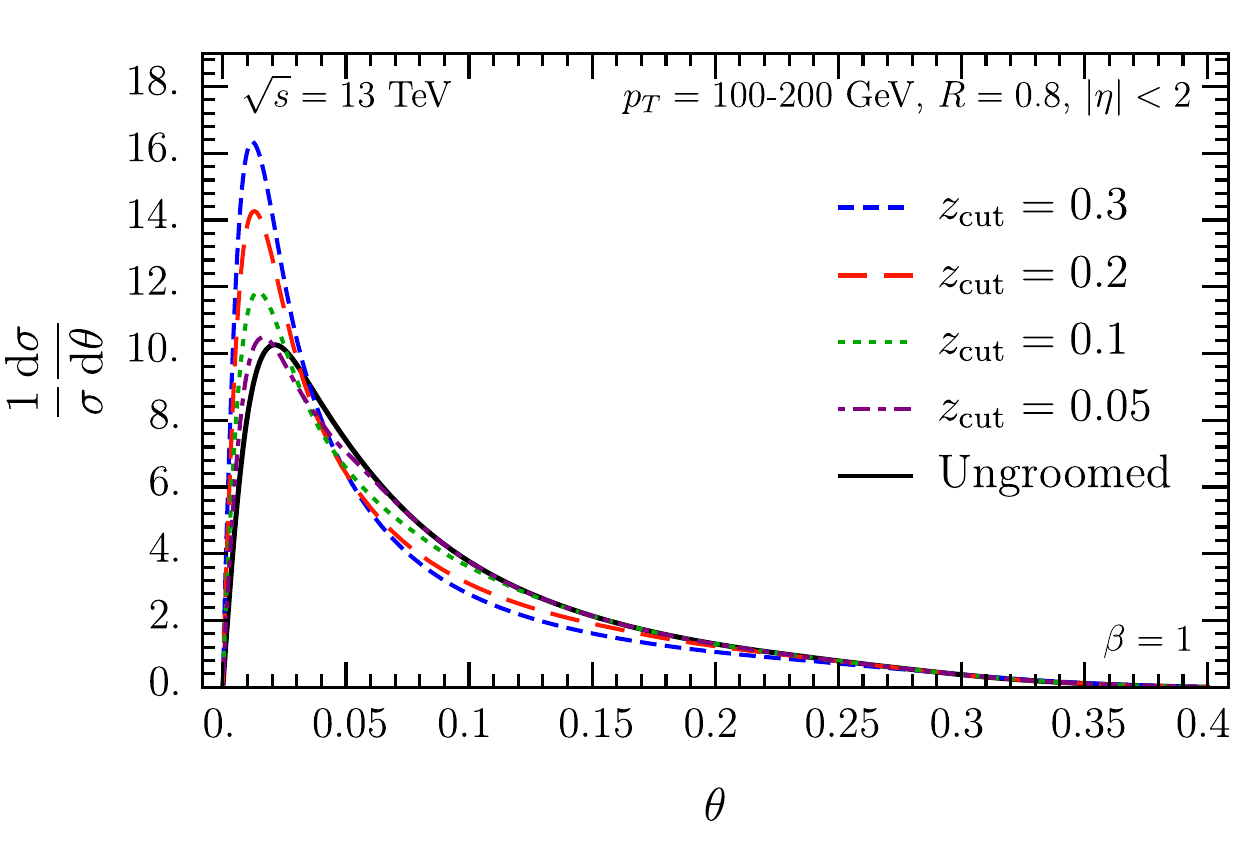} \hfill \phantom{.} \\
\caption{Left: Numerical results for the angle between the WTA and the soft drop groomed jet axis with $z_{\rm cut}=0.1$ and different values of $\beta=0,\,1\,,\,2,\,10$. Right: Results for a fixed value of $\beta=1$ but different values of $z_{\rm cut}=0.05,\, 0.1,\,0.2,\,0.3$. In both panels we show the ungroomed result (solid black) for comparison.~\label{fig:WTA-GR-beta}} 
\end{center}
\end{figure}

In this section, we present numerical results for the angles between the different jet axes as introduced in the previous sections. We start with the angle between the standard and the WTA axes, see~\sec{ST_WTA}. Throughout this section, we use the CT14 NLO PDF set~\cite{Dulat:2015mca}. In~\fig{WTA-ST-results}, we show the perturbative results at NLL$'$ accuracy including a QCD scale uncertainty band, which is obtained by varying all scales and taking the envelope, as discussed in~\sec{resummation} above. We consider inclusive jets in proton-proton collisions at $\sqrt{s}=13$~TeV at central rapidities $|\eta|<2$ where jets are reconstructed using the anti-k$_T$ algorithm with $R=0.8$. We choose four exemplary jet transverse momentum intervals between $p_T=30-1200$~GeV as indicated in each panel of the figure. As expected, we observe the typical Sudakov suppression for small angles. As the jet $p_T$ is increased, the distribution becomes narrower and peaks at small values of $\theta$. For comparison, we also show \Pythia~8.2~\cite{Sjostrand:2014zea} results at parton level and including Initial State Radiation (ISR), Multi Parton Interactions (MPI) and hadronization effects. Note that the spike seen in the lowest $\theta$ bin of the \Pythia results at parton level is due to the shower cutoff. Overall, we find very good agreement between our perturbative results and the \Pythia~8.2 simulations. Only for the lowest jet transverse momentum bin $p_T=30-50$~GeV, we observe that the \Pythia results are not within the displayed uncertainty band. This is the case with the largest contribution from ISR and MPI, which are not (or only partially) contained in our calculation. Note that in this (and subsequent) figures we show results for fairly small angles. To access these experimentally it would be natural to  consider track-based measurements, since these have vastly superior angular resolution. On the theory side this could be included in the calculation by using ref.~\cite{Chang:2013rca}.

We explore ISR, MPI and hadronization in more detail within \Pythia~8.2 in~\fig{Pythiaonly}, for the two jet transverse momentum intervals $p_T=30-50$~GeV (left) and $p_T=100-200$~GeV (right) (corresponding to the upper row in~\fig{WTA-ST-results}). All three contributions are sources of additional soft radiation in the jet. The standard jet axis is sensitive to this additional soft radiation whereas the WTA axis is insensitive, leading to the broadening of the distribution seen in the figure. We observe that in the lower $p_T$ interval in~\fig{Pythiaonly} ISR and MPI are the dominant effects, which are relevant over the entire displayed range of $\theta$. For the higher $p_T$ interval, ISR is the most important effect. The correction due to hadronization is only relevant for small values of $\theta$ for both $p_T$ intervals considered here, in agreement with our model in \sec{nonp}. 

Next, we investigate differences between quark and gluon jets. We separately show the quark (blue) and gluon (red) distributions in~\fig{quarkgluon} for $p_T=100$~GeV (dashed) and 200~GeV (dotted). The $p_T$ integrated result where we include appropriate quark/gluon fractions is shown by the solid red curve. Gluons radiate more, and therefore the standard and WTA axes are further separated, leading to a broader distribution for gluon jets compared to quark jets.

\begin{figure}[t]
     \hfill \includegraphics[width=0.48\textwidth]{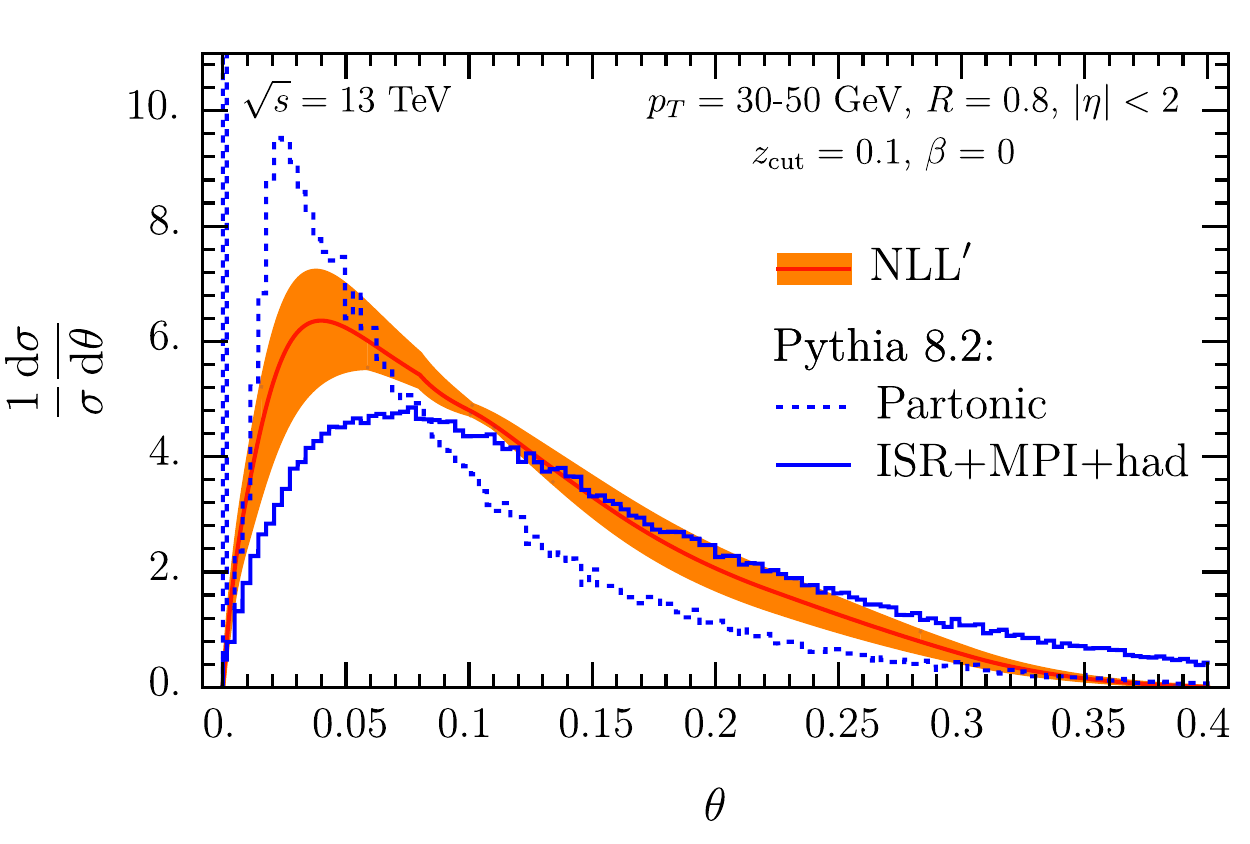} \hfill 
     \includegraphics[width=0.48\textwidth]{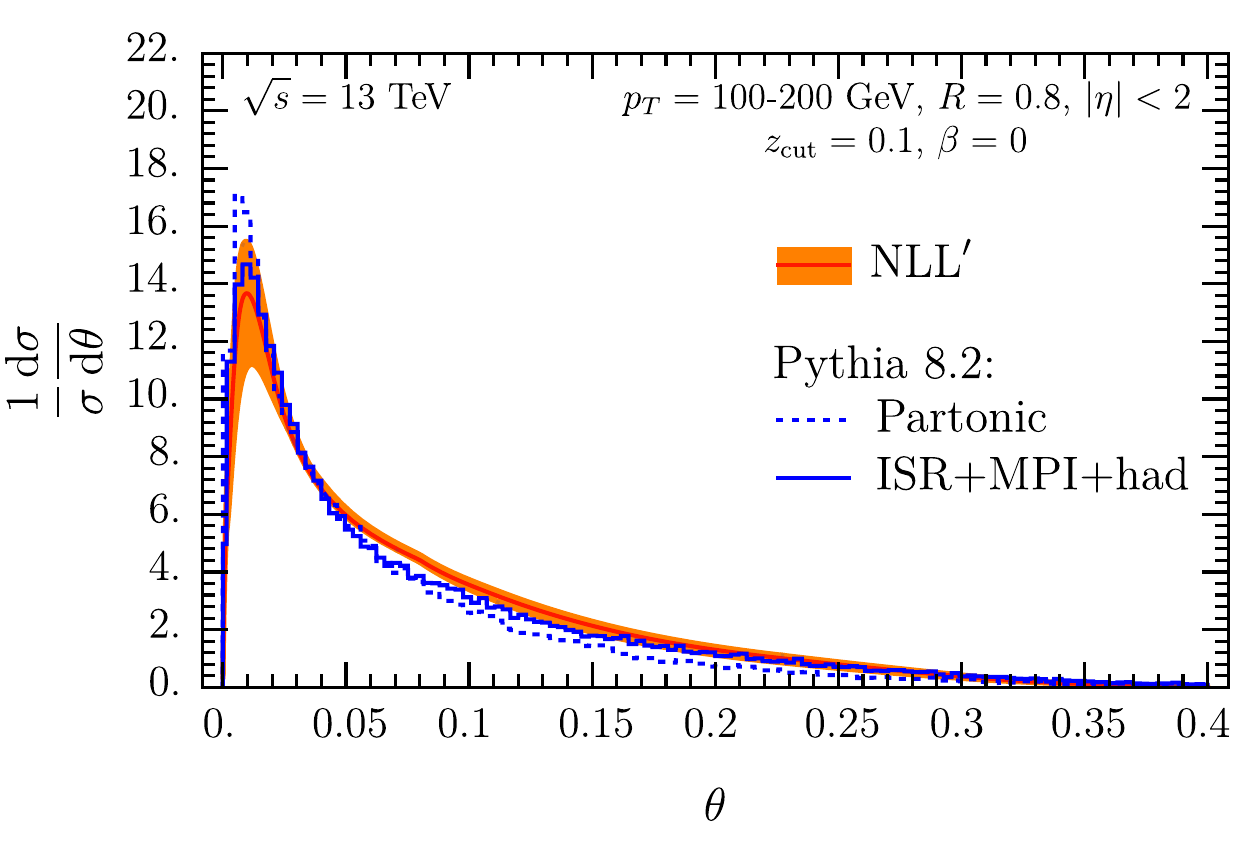} \hfill \phantom{.} \\
          \hfill \includegraphics[width=0.48\textwidth]{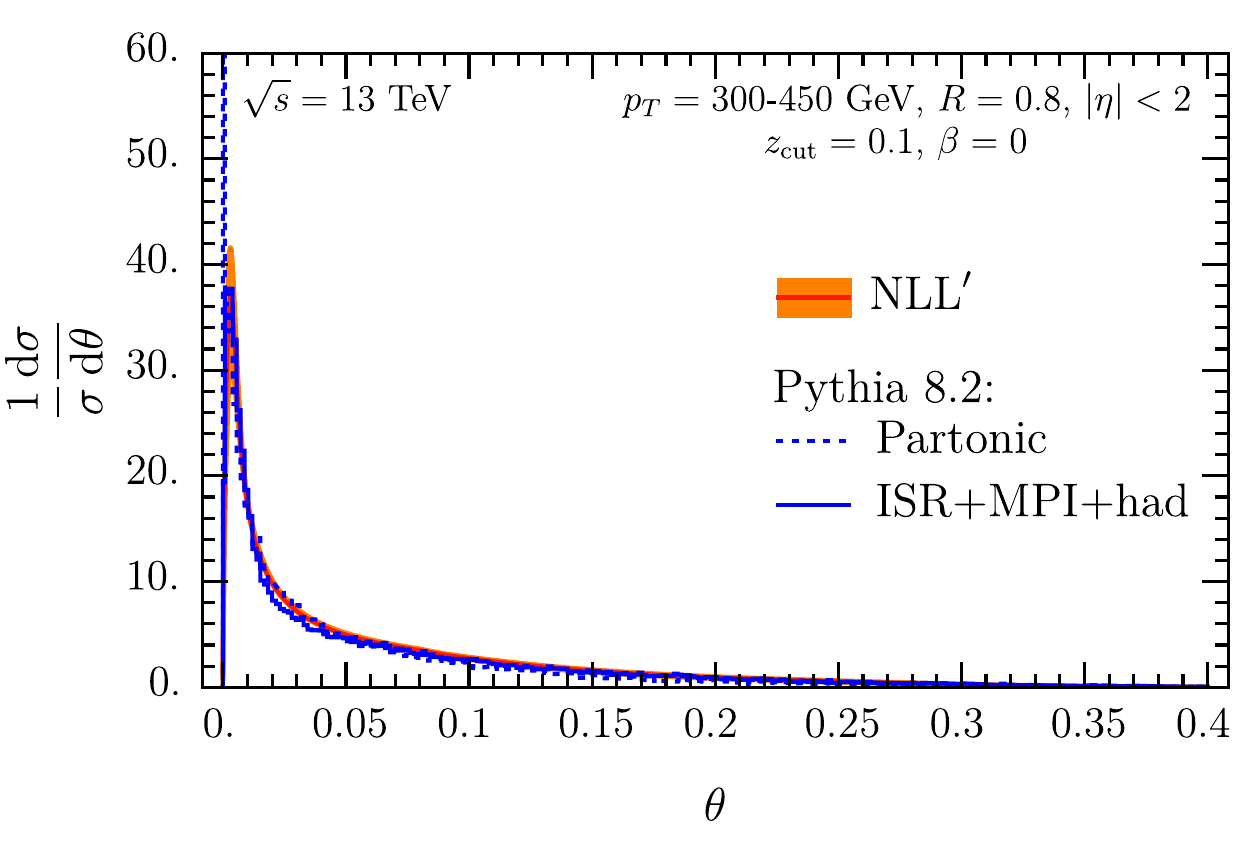} \hfill 
     \includegraphics[width=0.48\textwidth]{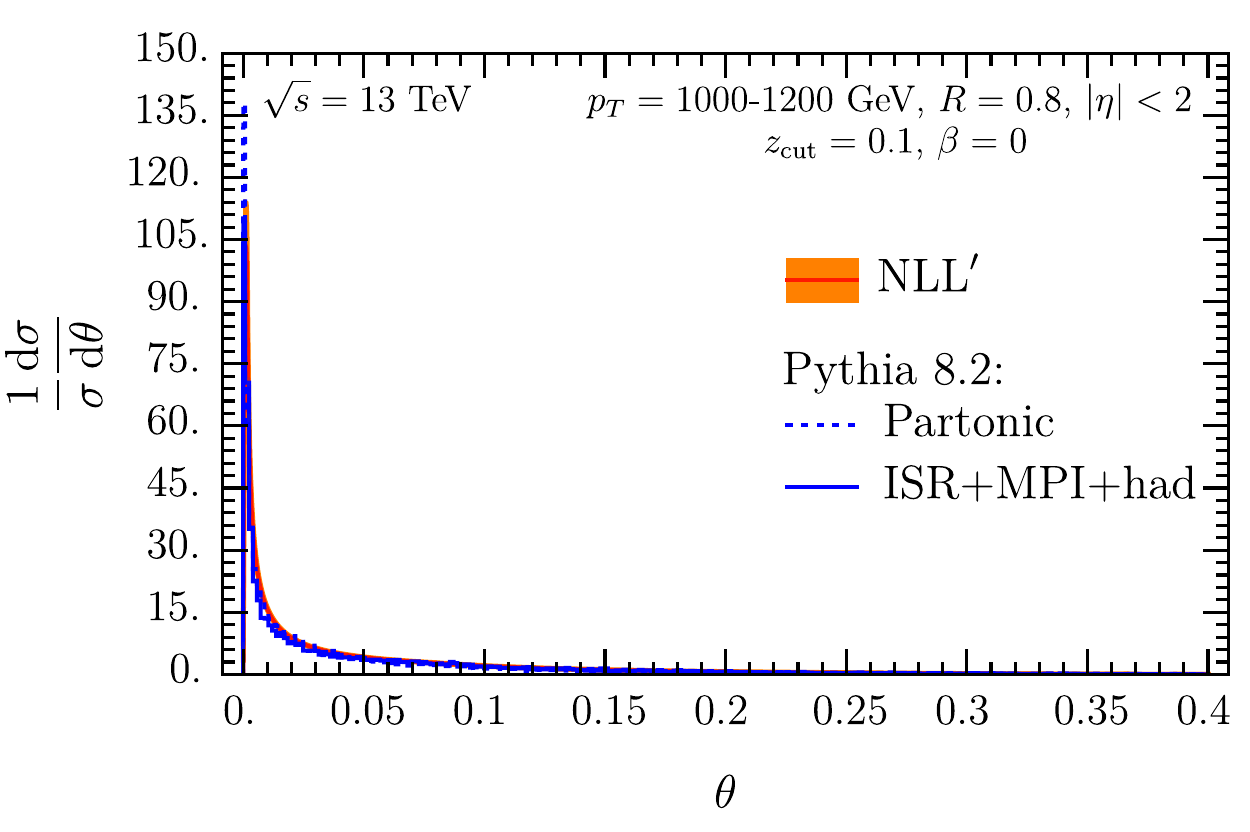} \hfill \phantom{.} 
    \caption{Numerical results for the angle between the WTA and the soft drop groomed jet axis with $z_{\rm cut}=0.1$ and $\beta=0$ at NLL$'$ accuracy. We consider the same kinematics as in~\fig{WTA-ST-results} and compare to the corresponding \Pythia~8.2 results.~\label{fig:WTA-GR-results}} 
\end{figure}

\begin{figure}[t]
     \hfill \includegraphics[width=0.48\textwidth]{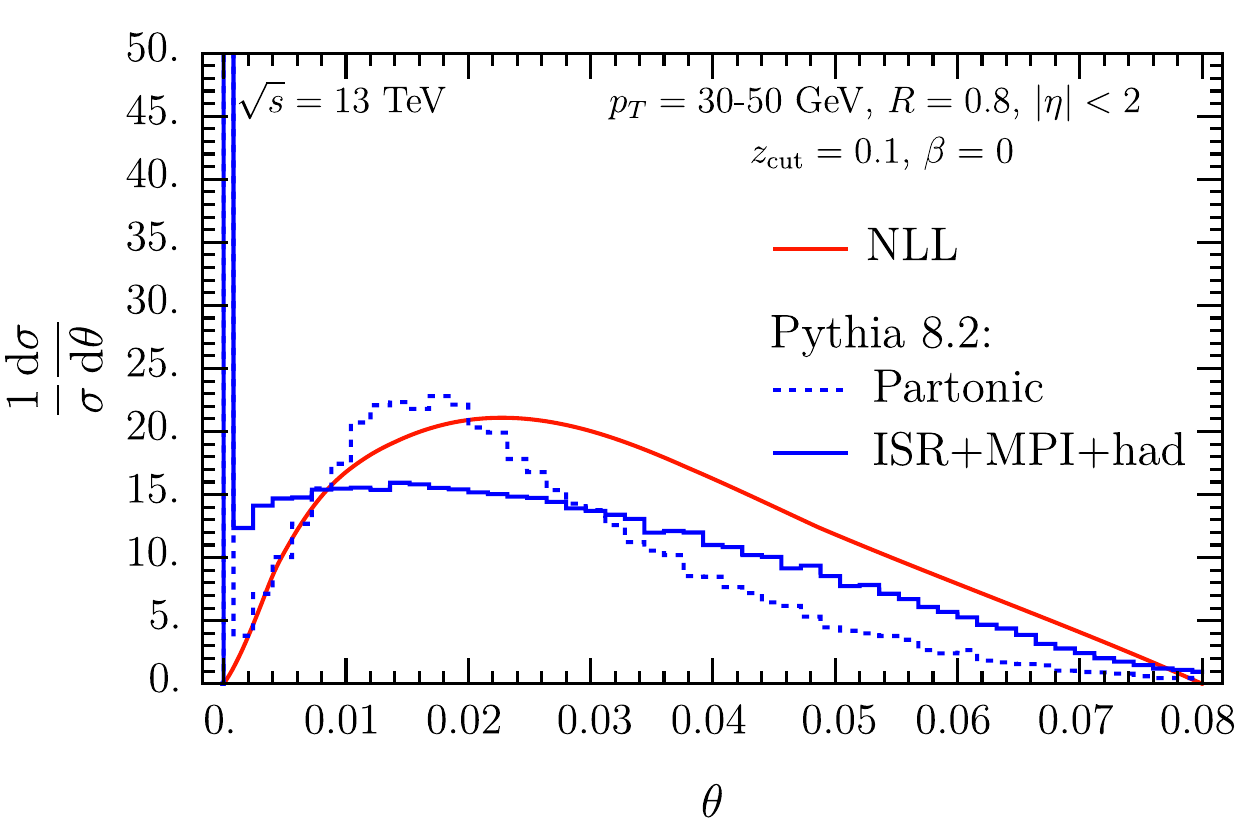} \hfill 
     \includegraphics[width=0.48\textwidth]{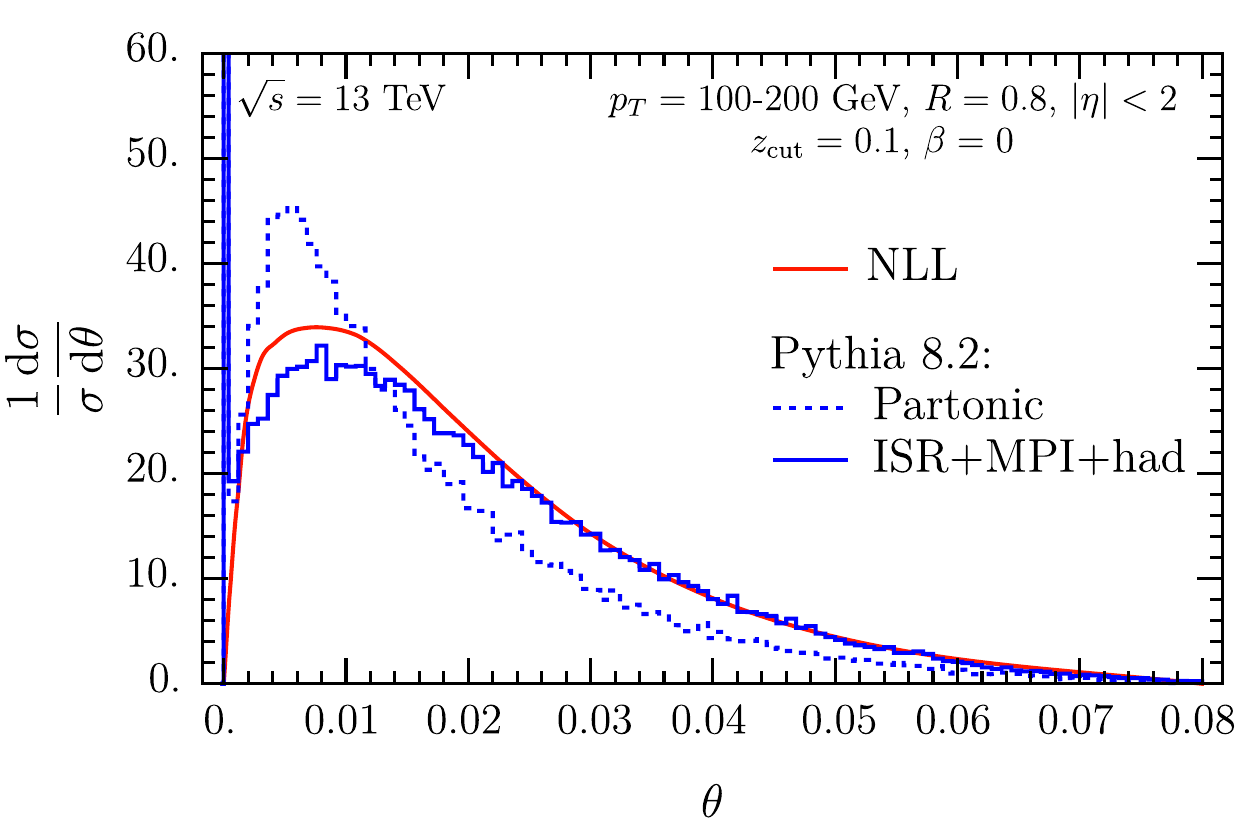} \hfill \phantom{.} \\
          \hfill \includegraphics[width=0.48\textwidth]{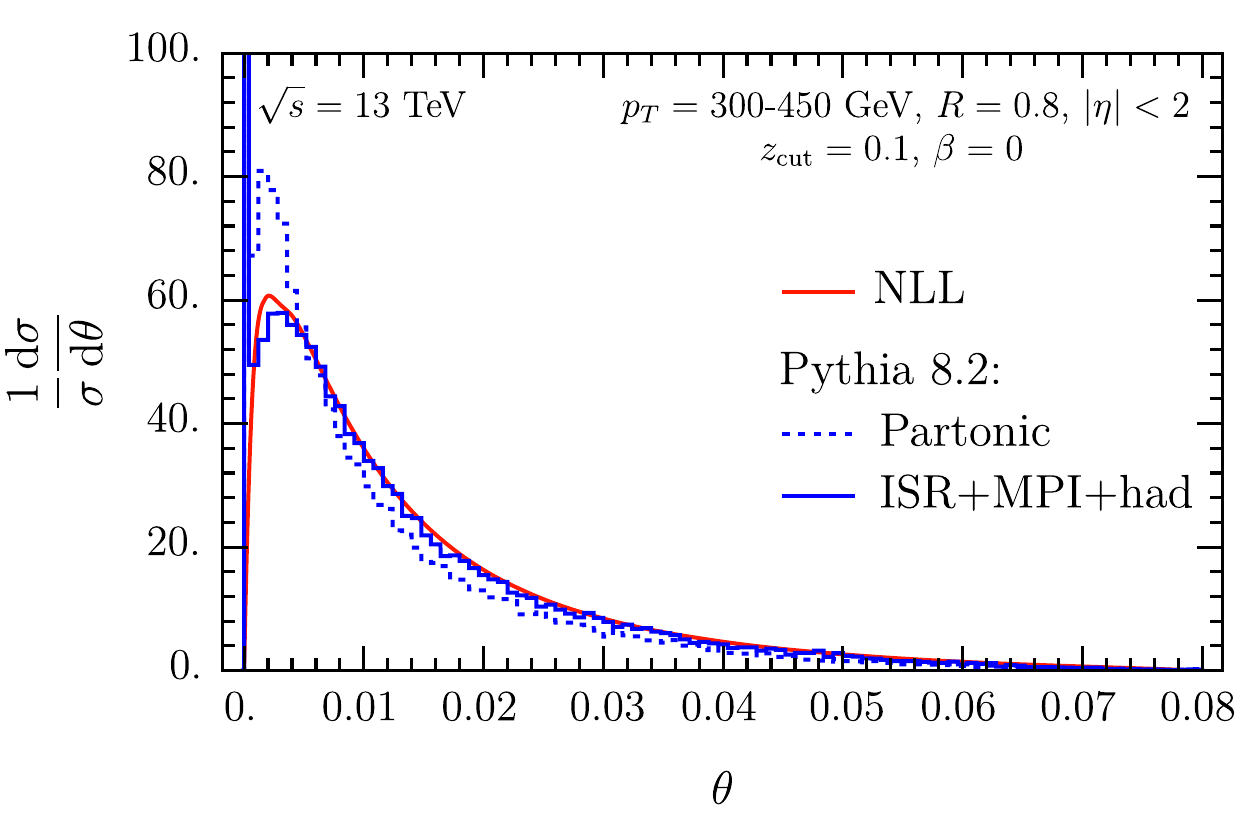} \hfill 
     \includegraphics[width=0.48\textwidth]{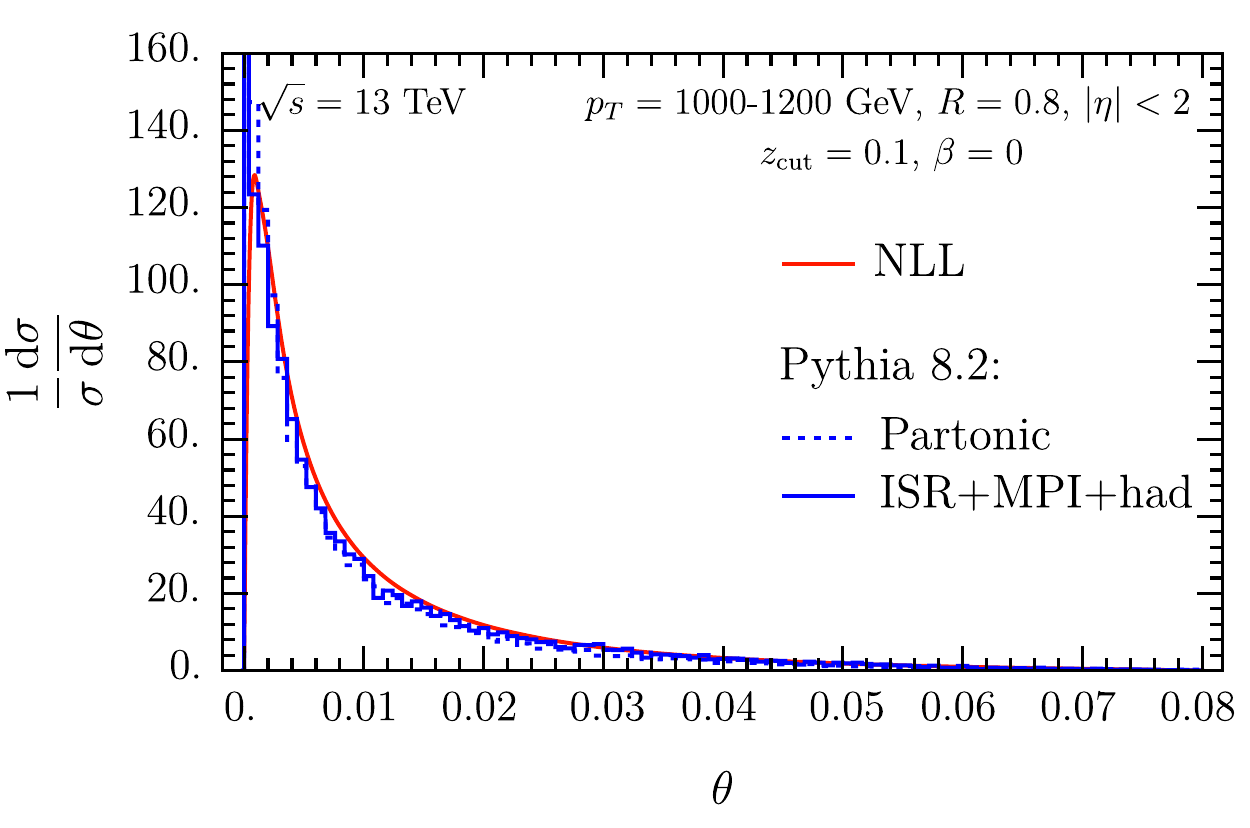} \hfill \phantom{.} 
    \caption{Numerical results for the angle $\theta$ between the standard and the soft drop groomed jet axis with $z_{\rm cut}=0.1$ and $\beta=0$ at NLL accuracy. We consider the same LHC kinematics as in~\fig{WTA-ST-results} and compare to the corresponding \Pythia~8.2 results.~\label{fig:ST-GR-results}} 
\end{figure}

We now consider the angle between the WTA and the soft drop groomed jet axes. First, we study the impact of soft drop grooming  in~\fig{WTA-GR-beta}, choosing the same kinematical setup as in~\fig{WTA-ST-results} with the jet transverse momentum interval of $p_T=100-200$~GeV as a representative example. The ungroomed result is shown for reference. For the soft drop grooming parameters we choose $z_{\rm cut}=0.1$ and $\beta=0,\,1,\,2$ (left panel) and $z_{\rm cut}=0.05,\,0.1,\,0.2,\,0.3$ and $\beta=1$ (right panel), and the corresponding curves are shown at NLL$'$ accuracy. From the left panel, we observe that the ungroomed result is approached for large values of $\beta$, which is expected as the ungroomed result is recovered in the limit $\beta\to \infty$. Similarly, the right panel shows that for smaller values of $z_{\rm cut}$ the ungroomed result is approximated. We also note that the transition point between the groomed and the ungroomed distribution depends on $z_{\rm cut}$ but it is independent of $\beta$. This is consistent with the perturbative results presented in~\sec{GR_WTA} above. In general, we observe that the grooming leads to a distribution that peaks at smaller values of the angle $\theta$. This can be understood in the sense that soft drop grooming removes soft radiation from the jet and therefore the groomed axis and the WTA axis, which is insensitive to soft radiation by construction, are closer in angle than the WTA and the standard jet axis. The size of the peak at small values of $\theta$ is enhanced when more aggressive grooming parameters are chosen, i.e. small values of $\beta$ or large values of $z_{\rm cut}$. 

Next, we show the comparison of our perturbative results to \Pythia~8.2 simulations. In~\fig{WTA-GR-results}, we show the comparison of our perturbative results for $\zc=0.1$ and $\bt=0$, including the nonperturbative model as described in~\sec{nonp}, and the \Pythia results at parton level and including ISR, MPI and hadronization. We choose the same kinematics as in~\fig{WTA-ST-results}. In the highest jet $p_T$ bins, the difference between the two calculations is very small and the cross section for the angle between the WTA and the groomed jet axes has a narrow peak at very small values $\theta\lesssim 0.01$ which implies that the two jet axes almost agree. In the lowest $p_T$ bin there is still a small discrepancy between the perturbative results and \Pythia due to ISR, MPI and hadronization. We observe that soft drop grooming does not significantly reduce the size of these effects at low $p_T$, compared to the ungroomed case shown in \fig{WTA-ST-results}.

\begin{figure}[t]
\begin{center}
\hfill \includegraphics[width=0.48\textwidth]{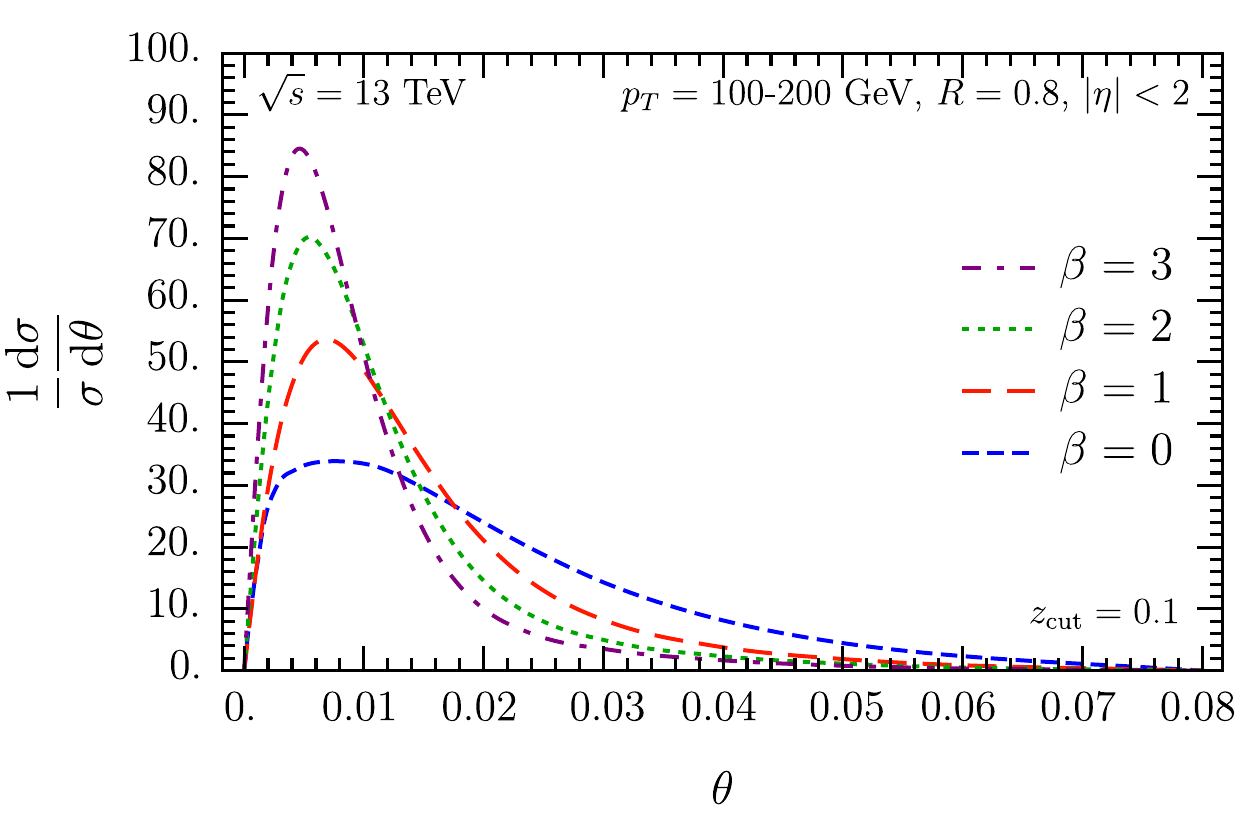} \hfill 
     \includegraphics[width=0.48\textwidth]{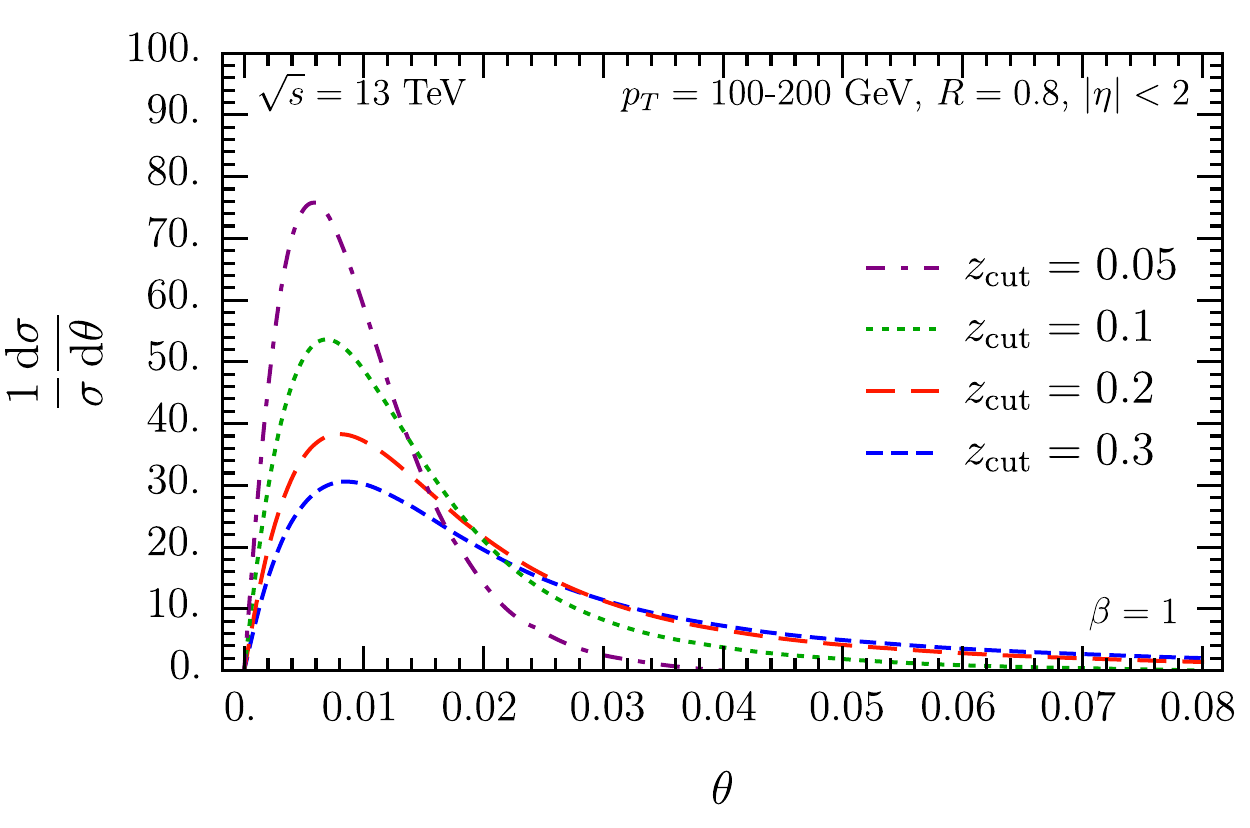} \hfill \phantom{.} \\
\caption{The angle between the standard and soft drop groomed jet axis for the same kinematics as in the upper right panel of~\fig{ST-GR-results} but for different values of $\beta=0-3$ (left), and for different values of $\zc$ (right).~\label{fig:ST-GR-beta}} 
\end{center}
\end{figure}

 Further, we present numerical results for the angle between the standard and the soft drop groomed jet axes. Our numerical results at NLL accuracy are shown in~\fig{ST-GR-results} along with \Pythia results, for the same LHC kinematics as in~\fig{WTA-ST-results}. We integrated out the dependence on the soft drop groomed radius $R_g$, but note that in principle it is possible to directly obtain double differential results from our numerical setup. This may be advantageous if it is experimentally necessary to impose an additional cut on the soft drop groomed radius $R_g$ as is sometimes the case for groomed jet substructure observables~\cite{Sirunyan:2017bsd,Sirunyan:2018gct,Acharya:2019djg}. Even though this observable is more sensitive to soft physics, the agreement between \Pythia and our perturbative results is nevertheless good. We observe that the perturbative results vanish for $\theta\to 0$, whereas the \Pythia results show a spike in the leftmost bin. In \Pythia the spike corresponds to jets where no branch gets groomed away and, hence, the standard and soft drop groomed jet axis are exactly aligned. We note that in general the difference between the two \Pythia curves shown in~\fig{ST-GR-results} is larger than for the other two angles considered above. This is expected as the angle between the standard and groomed jet axes is very soft sensitive making this observable a promising candidate to tune parton shower event generators.

In~\fig{ST-GR-beta}, we show the dependence of the angle between the standard and groomed axes on the grooming parameter $\beta$ for $p_T=100-200$~GeV which corresponds to the upper right panel of~\fig{ST-GR-results}. We choose four exemplary $\beta$ values 0,~1,~3,~4 (different dashing and colors) but the same $z_{\rm cut}=0.1$. In the limit $\beta\to\infty$, the grooming is removed and the ungroomed jet is recovered, which implies that the two jet axes considered here are aligned. This manifests itself in our numerical results by the fact that the curves shown in~\fig{ST-GR-beta} eventually approach a delta function at $\theta=0$ for large values of $\beta$. Note that we obtain the correct limit numerically because the nonperturbative exponent in $b_\perp$ space approximates unity for $\beta\to\infty$ (regime A), see the discussion in~\sec{nonp}. Similarly, we consider four different values of $\zc$, suggesting that in the limit $\zc \to 0$ the distribution also approaches a delta function, even though our formalism is strictly speaking not valid in this limit.

\begin{figure}[t]
\begin{center}
     \includegraphics[width=0.48\textwidth]{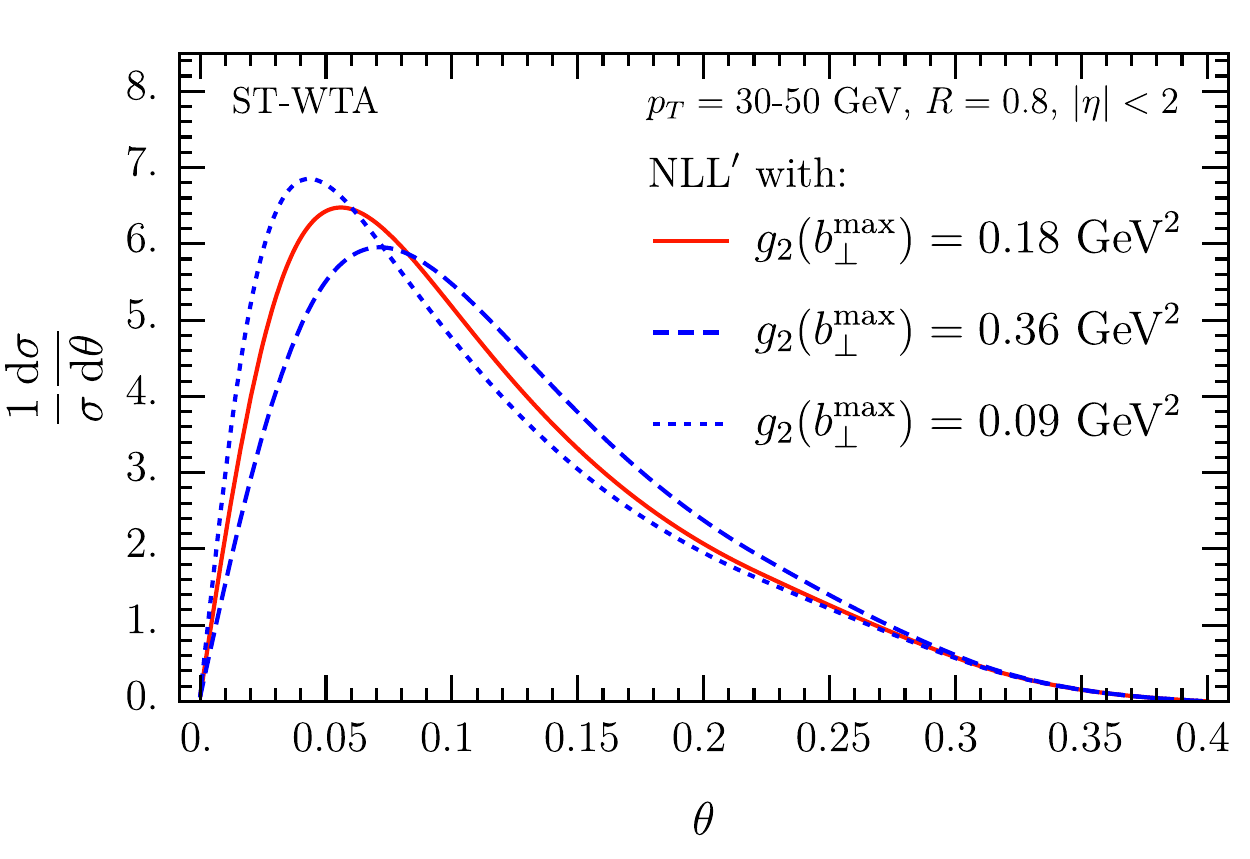}\hfill 
     \includegraphics[width=0.48\textwidth]{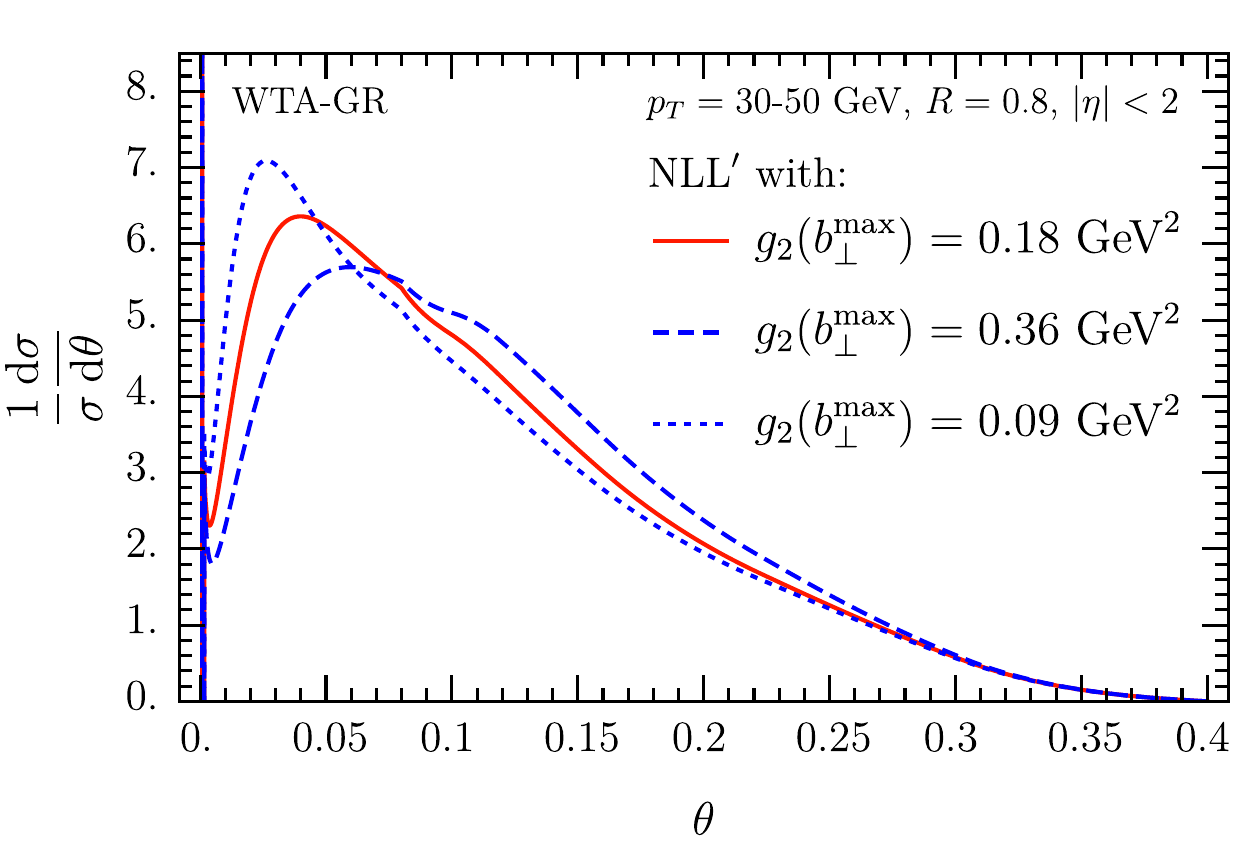}  \phantom{.} \\
     \hfill \includegraphics[width=0.48\textwidth]{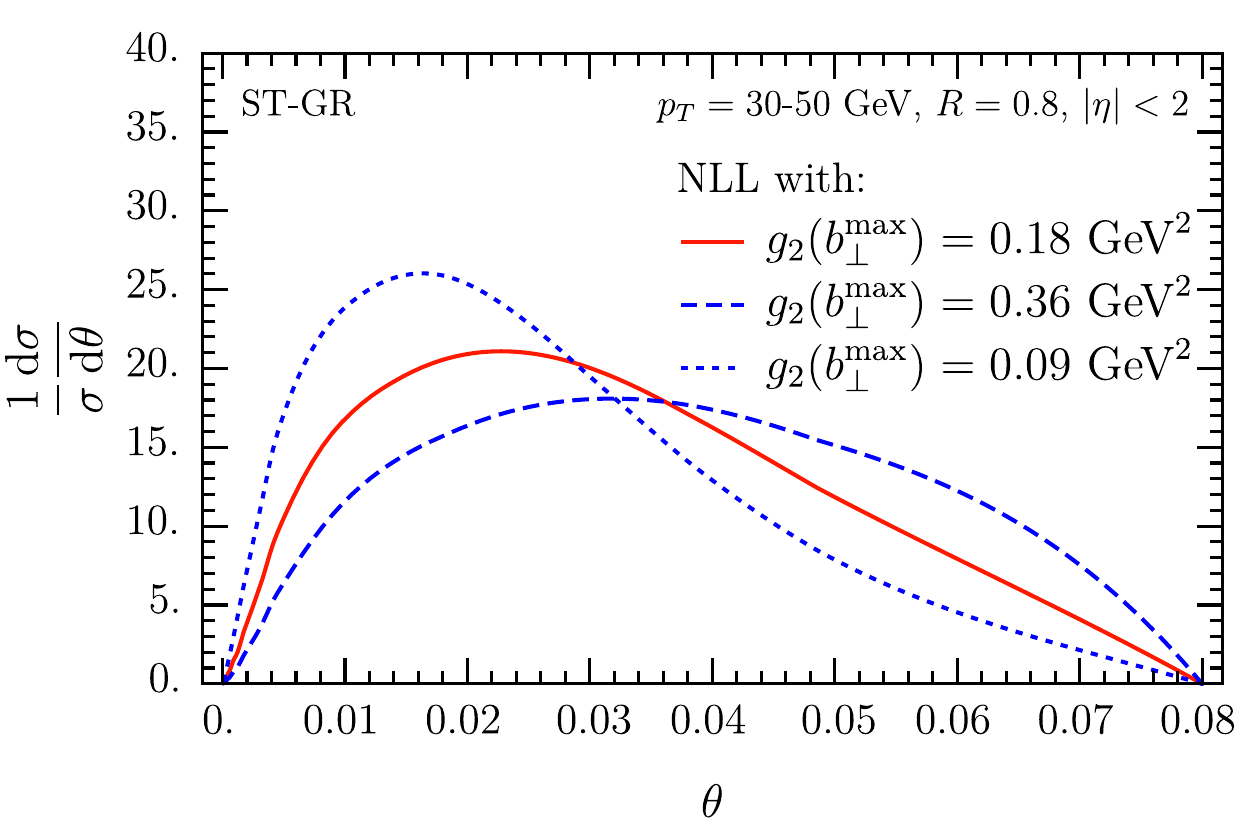} \hfill \phantom{.}

\caption{ The nonperturbative sensitivity of the angle between the WTA and ST axis (upper left), WTA and GR axis (upper right) and ST and GR axis (bottom), for the kinematics corresponding to the upper left panel of \fig{WTA-ST-results}. Shown are our NLL$'$ (or NLL) result with the default $g_2(b_\perp^{\rm max})=0.18$ GeV$^2$ (orange), as well as double (blue dashed) and half (blue dotted) this value.}
\label{fig:non-pert}
\end{center}
\end{figure}

 We next examine the sensitivity of each observable to nonperturbative physics. In \fig{non-pert} we focus on the lowest $p_T$-bin, $30-50$~GeV, where the effect of nonperturbative physics is largest. At larger jet transverse momentum, the nonperturbative effects arise at smaller angles and are therefore strongly Sudakov suppressed by the resummation. To gauge the nonperturbative sensitivity, we vary the model parameter $g_2(b_\perp^{\rm max})$ defined in \eq{non-perturbative-model} between our canonical choice of $g_2(b_\perp^{\rm max})=0.18 \text{ GeV}^2$ to twice and half its value. The angle between the winner-take-all axis and either the standard or the groomed axis has the least sensitivity to our choices of nonperturbative parameters. The effect of the nonperturbative model does not change the spectrum at large angles, so the nonperturbative sensitivity is most apparent in the peak of the distribution. Furthermore, the variation of $g_2(b_\perp^{\rm max})$ is within the perturbative uncertainty estimated from scale variations. In stark contrast, the angle of the standard to groomed axes displays a large sensitivity, and at no point in the spectrum do we see any turning off of the nonperturbative physics.


\section{Conclusions}
\label{sec:conclusions}

In this work we presented a first calculation of the angles between different jet axes. We considered three different jet axes: The standard jet axis, the soft drop groomed jet axis and the jet axis using a winner-take-all recombination scheme. Our studies were motivated by the different soft sensitivity of these different jet axes. The winner-take-all scheme yields a jet axis which is insensitive to soft radiation at leading power and also the soft drop groomed jet axis has a reduced sensitivity to soft physics compared to the standard jet axis. By considering the angles between different axes, the soft radiation pattern inside reconstructed jets can be studied. Within Soft Collinear Effective Theory, we performed calculations at next-to-leading logarithmic (NLL$'$ or NLL) accuracy, where large logarithms of the angle between the the standard and groomed vs.~the winner-take-all axes are resummed to all orders, including the contribution of non-global logarithms in the leading color approximation. We presented numerical results for relevant LHC kinematics at $\sqrt{s}=13$~TeV and compared our results to \Pythia~8.2 simulations. Overall we found very good agreement except for jets with very low jet transverse momentum, where different nonperturbative and power-suppressed effects play an important role. 

The angle between the standard and soft drop groomed jet axes is particularly soft sensitive, as shown in~\fig{non-pert}, as it is a measure of the radiation which is groomed away by the soft drop algorithm. The corresponding factorization theorem depends on the soft drop groomed jet radius $R_g$. We resummed large logarithms of both the angle between the jet axes and the groomed radius $R_g$, integrating over $R_g$ after the resummation is performed to obtain a distribution for the angle between the axes. The all-order structure of non-global logarithms for this observable is nontrivial, limiting the accuracy of our calculation in this case. The soft sensitivity of the angle between the standard and groomed jet axis makes this observable very well suited to tune parton shower event generators, however, further theoretical work must be done to extend the resummation to NLL$'$ including non-global correlations, before such studies can be conducted.

We included nonperturbative effects in impact parameter space by introducing a model function, which is related to the nonperturbative component of the rapidity anomalous dimension relevant for transverse momentum resummation. This was crucial to obtain a sensible $\beta$-dependence for the angle between the standard and groomed jet axis. Therefore, the observables considered here can provide important constraints on this universal nonperturbative quantity. 

We expect that the observables considered here will have important applications in proton-proton as well as heavy-ion collisions at the LHC and RHIC. In addition to the aforementioned tuning of parton shower Monte Carlo programs, or studying the nonperturbative contribution to the rapidity anomalous dimension, it could provide valuable insight into the effect of the medium in heavy ion collisions. Furthermore, for jets with a large radius parameter, these axes could be sensitive to the color flow of a collision, similar to the pull~\cite{Gallicchio:2010sw,Larkoski:2019urm}. As we noted, a measurement using charged particle tracks would be necessary to access the small angles we consider. Thus, another future direction is to directly incorporate the effect of a track-based measurement in our calculations. 

\acknowledgments
We thank Y.~Chen, R.~Elayavalli, Z.-B.~Kang, M.~LeBlanc, K.~Lee, Y.-J.~Lee, J.~Mulligan, B.~Nachman, M.~Ploskon and J.~Roloff for helpful discussions. This work is supported by the U.S.~Department of Energy under Contract No.~DE-AC02-05CH11231, the LDRD Program of Lawrence Berkeley National Laboratory, the National Science Foundation under Grant No.~ACI-1550228 within the JETSCAPE Collaboration, by the ERC grant ERC-STG-2015-677323, and the D-ITP consortium, a program of the Netherlands Organization for Scientific Research (NWO) that is funded by the Dutch Ministry of Education, Culture and Science (OCW), the Department of Energy under Contract DE-AC52-06NA25396 at LANL and through the LANL/LDRD Program via a Feynman Distinguished Fellowship. This material is based upon work supported by the U.S. Department of Energy, Office of Science, Office of Nuclear Physics, within the framework of the TMD Topical Collaboration.

\appendix

\section{Anomalous dimensions}
\label{app:anom}

The one-loop splitting functions are given by
\begin{align} \label{eq:split}
P_{qq}(z) &= C_F\,\Big((1+z^2) \cL_0(1-z) + \frac32 \de(1-z)\Big) \,,
&
P_{gq}(z) &= C_F\,\f{1+(1-z)^2}{z}
\,, \nn \\
   P_{gg}(z) &= 2C_A \bigg[ z \cL_0(1-z) + \frac{1-z}{z}+z(1-z)\bigg] + \frac{\beta_0}{2}\,\de(1-z)\,,
   &
   P_{qg}(z) &= T_F \big[z^2+(1-z)^2\big]
\,,\end{align}
with
\begin{align} \label{eq:beta_0}
  \beta_0 = \frac{11}{3} C_A - \frac{4}{3} T_F n_f
\,.\end{align}
Here we list all relevant anomalous dimensions in $b_\perp$-space 
\begin{align}\label{eq:anomdim}
 \ga^{\tilde H}_{q}(p_{T} R,\mu)
  &= \f{\as C_F}{\pi}  \Big[2 \ln\Big(\f{p_T R}{\mu} \Big)  - \frac32\Big]
, \nn \\
\ga^{\tilde H}_{g}(p_{T} R,\mu)
  &= \f{\as}{\pi} \Big[2 C_A \ln\Big(\f{p_T R}{\mu} \Big) - \frac12 \beta_0 \Big]
, \nn \\
\ga^C_q(\mu,\nu/p_T) 
&= \frac{\al_s C_F}{\pi}\, 
  \Big[2  \ln \Big(\frac{\nu}{2p_{T}}\Big) + \frac32\Big]
\,, \nn \\
\ga^C_g(\mu,\nu/p_T) 
&= \frac{\al_s}{\pi}\, 
  \Big[2C_A  \ln \Big(\frac{\nu}{2p_{T}}\Big) + \frac12 \beta_0\Big]
\,, \nn \\
\ga^S_i(\mu,\nu R)
&= \frac{2\al_s C_i}{\pi}\, \ln \Big( \frac{2\mu}{\nu R}\Big)
\,, \nn \\
\ga^\nu_i(b_\perp,\mu)
&= \frac{2\al_s C_i}{\pi}\,\ln \Big(\frac{\mu_b}{\mu}\Big)
\,, \nn \\
\ga^{S^{\notin {\rm gr}}}_i(z_{\rm cut}p_T R,\mu)
&= \frac{2\al_s C_i}{\pi}\frac{1}{1+\beta}\, \ln \Big(\frac{\mu}{z_{\rm cut}p_T R}\Big)
\,, \nn \\
\gamma^{\CS}_i(p_T R, \zc, \bt, \mu, \nu/p_T)
&= \f{2 \as C_i }{\pi} \f{\beta}{1+\beta} \ln \bigg[\f{\mu}{\zc^{-1/\beta}p_T R }\Big(\f{2 p_T}{\nu} \Big)^{\f{1+\beta}{\beta}} \bigg] 
\,, \nn \\
 \ga^{C^{\in {\rm gr}}}_{q}(\theta_g^c p_{T} R,\mu)
  &= \f{\as C_F}{\pi} \Big[2 \ln\Big(\frac{\mu}{\theta_g^c p_T R}\Big) + \frac32\Big]
\,, \nn \\
\ga^{C^{\in {\rm gr}}}_{g}(\theta_g^c p_{T} R,\mu)
  &= \f{\as}{\pi} \Big[2C_A \ln \Big(\frac{\mu}{\theta_g^c p_T R}\Big) + \frac12 \beta_0 \Big]
\,, \nn \\
\ga^{{\cal S}_G}_i(z_{\rm cut} (\theta_g^c)^{1+\beta} p_T R,\mu,\beta)
&=-\f{2\as C_i}{\pi} \f{1}{1+\beta} \ln \bigg(\f{\mu }{ \zc (\theta_g^c)^{1+\beta} p_T R} \bigg)
\,, \nn \\
\ga^{{\cal S}_X}_i(z_{\rm cut} p_TR,\beta,\mu,\nu R)
&=  -\frac{2 \alpha_{s} C_{i}}{\pi} \frac{\beta}{1+\beta} \ln \bigg[\Big(z_{\rm cut} p_{T} R\Big(\frac{2}{\nu R}\Big)^{1+\beta}\Big)^{\frac{1}{\beta}} \mu\bigg]
\,, \nn \\
\ga^{{\cal S}_K}_i(\mu,\nu \tg R)
&= -\f{2\as C_i}{\pi} \ln \Big(\f{2 \mu }{\nu \tg R} \Big)
\,,\end{align}
where $C_i = C_F$ ($C_A$) for $i=q$ ($i=g$). We achieve full NLL$'$ accuracy by including the two-loop cusp anomalous dimension which amounts to multiplying all $\ln \mu$ and $\ln \nu$ terms in~\eq{anomdim} by
\begin{align}
   1 + \frac{\al_s}{4\pi} \Bigl[\Bigl( \frac{67}{9} -\frac{\pi^2}{3} \Bigr)\,C_A  -
   \frac{20}{9}\,T_F\, n_f \Bigr]
\,.\end{align}

\bibliographystyle{JHEP}
\bibliography{bibliography}

\end{document}